\documentclass[aps,pra,twocolumn,superscriptaddress]{revtex4}
\usepackage{graphicx}% Include figure files
\usepackage{dcolumn}% Align table columns on decimal point
\usepackage{bm}% bold math
\usepackage{physics}
\usepackage{amsmath}
\usepackage{mathtools}
\usepackage{amssymb}
\usepackage{array}
\usepackage{color}
\usepackage{xcolor}
\usepackage{nicefrac}

\usepackage{tikz}
\usetikzlibrary{calc,tikzmark} 

\usepackage[export]{adjustbox}
\usepackage[colorlinks=true, breaklinks=true, linkcolor=blue, citecolor=blue, urlcolor=blue]{hyperref}

\newcolumntype{P}[1]{>{\centering\arraybackslash}p{#1}}
\newcommand{\bonnpi}{Physikalisches Institut, University of Bonn, Nussallee 12, 53115 Bonn, Germany}
\newcommand{\geneva}{Department of Quantum Matter Physics, University of Geneva, Quai Ernest-Ansermet 24, 1211 Geneva, Switzerland}
\newcommand{\pks}{Max Planck Institute for the Physics of Complex Systems, N\"othnitzer Str.~38, 01187 Dresden, Germany}

\usepackage{color}
\usepackage{ifthen}
\usepackage{booktabs}
\usepackage{multirow}

\begin{document}

\title{Steady state diagram of interacting fermionic atoms coupled to dissipative cavities
}

\date{\today}
\author{Luisa Tolle}
\affiliation{\bonnpi}
\author{Ameneh Sheikhan}
\affiliation{\bonnpi}
\author{Thierry Giamarchi}
\affiliation{\geneva}
\author{Corinna Kollath}
\affiliation{\bonnpi}
\author{Catalin-Mihai Halati}
\affiliation{\geneva}
\affiliation{\pks}

\begin{abstract}
We investigate fermionic atoms subjected to an optical lattice and coupled to a high finesse optical cavity with photon losses. A transverse pump beam introduces a coupling between the atoms and the cavity field. We explore the steady state phase diagram taking fluctuations around the mean-field of the atoms-cavity coupling into account. Our approach allows us to investigate both one- and higher-dimensional atomic systems.
The fluctuations beyond mean-field lead to an effective temperature which changes the nature of the self-organization transition. 
We find a strong dependence of the results on the atomic filling, in particular when contrasting the behavior at low filling and at half filling. 
At low filling the transition to a self-organized phase takes place at a critical value of the pump strength. In the self-organized phase the cavity field takes a finite expectation value and the atoms show a modulation in the density. Surprisingly, at even larger pump strengths a strongly non-monotonous behavior of the temperature is found and hints towards effects of cavity cooling at many-body resonances. Additionally multiple self-organized stable solutions of the cavity field and the atoms occur, signaling the presence of a fluctuation-induced bistability, with the two solutions having different effective temperatures previously discussed in Ref.~\cite{TolleHalati2024}. In contrast, at half filling a bistable region arises at the self-organization transition already neglecting the fluctuations. The presence of the fluctuations induce an effective temperature, as at lower fillings, and change the behavior of the transition and the steady states drastically. 
We analyze the properties of the occurring steady states of the coupled atoms-cavity system. 
\end{abstract}
\nopagebreak
{\let\clearpage\relax\maketitle}

\section{Introduction}

The study of non-equilibrium dynamics in many particle systems is a major challenge across multiple fields. A prominent platform are many-body quantum systems dissipatively coupled to external environments.
While typically the coupling of the system of interest to external degrees of freedom is detrimental for the preparation of many-body entangled quantum states, it turns out that dissipation can be used constructively as an avenue to obtain complex quantum phenomena
\cite{DiehlZoller2008, VerstraeteCirac2009, WeimerBuchler2010, BarreiroBlatt2010, MuellerZoller2012}. 
An avenue to introduce driving and dissipation in a controlled manner are hybrid systems, in which quantum matter is coupled to bosonic quantum fields like photons or phonons. 
This has been explored in several platforms, such as ultracold atoms in optical cavities \cite{RitschEsslinger2013, MivehvarRitsch2021}, electrons in solid state cavities \cite{FornDiazSolano2019, SchlawinSentef2022}, trapped ions coupled to phonons \cite{TomzaJulienne2019, MonroeYao2021}, or superconducting circuits \cite{LEHURSchiro2016, BlaisWallraff2021}.

The recent experimental achievements of ultracold atoms in optical cavities allow one to access the strong coupling regime and gain precise control over the collective matter-light coupling \cite{RitschEsslinger2013, MivehvarRitsch2021}. 
Bosonic atoms have been experimentally coupled to cavities, either as three-dimensional Bose-Einstein condensates \cite{BaumannEsslinger2010, KroezeLev2018, VaidyaLev2018, KesslerHemmerich2021, FerriEsslinger2021, DreonDonner2022, KongkhambutKessler2022}, or confined to an external optical lattice structure \cite{LandigEsslinger2016, HrubyEsslinger2018, KlinderHemmerich2015, KlinderHemmerich2015b}.
In their fermionic counterpart both the steady state self-organization phase transition to density wave states \cite{HelsonBrantut2023, ZwettlerBrantut2025b, ZhangWu2021} and the out-of-equilibrium dynamics have been experimentally investigated \cite{ZwettlerBrantut2025, WuWu2023}.
Coupling fermionic atoms to optical cavities has been a rich playground also theoretically \cite{LarsonLewenstein2008b, SunLiu2011,PiazzaStrack2014b, ChenZhai2014, KeelingSimons2014, ChenYu2015, WolffKollath2016, MarijanovicDemler2024, TolleHalati2024, OrtunoGonzalezChitra2025}. 
In particular, the dissipative dynamics has been proposed for realization of exotic quantum phases \cite{DongPu2014,  FanJia2018, ColellaChiofalo2018, ColellaRitsch2019, SchlawinJaksch2019, ZhengWang2020, NieZheng2023, MivehvarPiazza2017}, topological effects \cite{KollathBrennecke2016, SheikhanKollath2016, SheikhanKollath2016b, Mendez-CordobaQuiroga2020, PanGuo2015}, and pairing in superfluids \cite{SchlawinJaksch2019b}.
While a large part of the theoretical atoms-cavities studies have relied on a mean-field treatment of the cavity field \cite{RitschEsslinger2013, MivehvarRitsch2021}, a growing body of the literature is showing the importance of including the quantum fluctuations of the atoms-cavity coupling term in order to correctly capture the steady states and the dissipative dynamics \cite{GopalakrishnanGoldbart2010, PiazzaStrack2014, WallRey2016, DamanetKeeling2019, LinLode2019, HalatiKollath2020, HalatiKollath2020b, BezvershenkoRosch2021, HalatiKollath2022, LinkDaley2022, ChiriacoChanda2022, JagerBetzholz2022, LentrodtEvers2023, HalatiKollath2025, TolleHalati2024, HalatiJager2025, MuellerStrunz2025, SchmitJaeger2025, OrsoDeuar2025}.

In this work, we compute the steady state phase diagram for interacting fermionic atoms confined to optical lattices and transversely coupled to the field of a dissipative optical cavity over a wide range of parameters. In particular, we focus on the case of low fillings and on the commensurate half filling case.  We employ a commonly used mean-field approach and additionally, the many-body adiabatic elimination method that takes into account the quantum fluctuations of the 
atoms-cavity coupling. This approach was originally developed for bosonic atoms \cite{BezvershenkoRosch2021} in order to capture the mixed state character of the atomic steady states naturally occurring in the presence of dissipation, when the cavity field fluctuations are taken into account.
We compare the findings of the two approaches and point out similarities and differences in the properties of the steady state diagram. Our work here complements the investigation of Ref.~\cite{TolleHalati2024} in which we focused on the existence of a fluctuation-induced bistability at low filling. We briefly summarize the previous findings of the fluctuation induced bistability at low filling of Ref.~\cite{TolleHalati2024} for completeness in Sec.~\ref{sec:fluctuation_induced_bistability}.
Additionally to the previous work, we analyze in detail the self-organization transition and the cavity induced cooling due to many-body resonances. We derive analytical formulas for the critical coupling in different regimes. 
Further, we analyze the self-organization transition at half filling a case not considered in the previous work. 
For the commensurate half-filling case the transition exhibits a bistability analog to the optical bistability \cite{RitschEsslinger2013}, which has a very different character than the fluctuation-induced bistability present at low filling.
We uncover that the bistability at half filling is already present in the zero-temperature mean-field approach. However, we show that important deviations occur when fluctuations are included, in particular, when the cavity energy scales dominate.

The paper is organized as follows. 
We begin by introducing our model of interacting fermions on an external optical lattice that globally couples to a dissipative cavity mode via transverse driving with a pump laser (Sec.~\ref{sec:setup}). Then we proceed to introduce the analytical methods, we derive to determine the steady state phase diagram (Sec.~\ref{sec:methods}). We employ the \emph{many-body adiabatic elimination technique} that takes into account perturbatively the quantum fluctuations of the atoms-cavity coupling (Sec.~\ref{sec:mbae}). The fluctuations determine a self-consistent effective temperature of the atoms. We provide a stability condition for the thermal steady state solutions (Sec.~\ref{sec:stability}).
Further, we obtain perturbative results for vanishing kinetic energy in the thermodynamic limit, which is also applicable in higher dimensions (Sec.~\ref{sec:pertubation_J}). In order to facilitate the understanding of the phases we obtain in the complex atoms-cavity coupled system, we provide an overview of the ionic Hubbard model (Sec.~\ref{sec:ionic_hubbard}), since the effective Hamiltonian in our mean-field approach takes this form with a self-consistently determined staggered potential. We then proceed to present our results for the steady state phase diagrams across the self-ordering transition at both low filling (Sec.~\ref{sec:quarter_filling}) and commensurate half filling (Sec.~\ref{sec:half_filling}). Special focus is put on cooling processes. We extend the concept of cavity-cooling (Sec.~\ref{sec:cavity_cooling}) to the many-body regime (Sec.~\ref{sec:many_body_cooling}), and show that an efficient transfer of energy between the photons and the atoms, mediated by the fluctuations, determines many of the features present in the phase diagram, including the \emph{fluctuation-induced bistability} (Sec.~\ref{sec:fluctuation_induced_bistability}) we introduced in Ref.~\cite{TolleHalati2024}.
Here we contrast the fundamentally different nature of the bistability at low fillings and the special case of commensurate fillings (Sec.~\ref{sec:self_organization_transition_half_filling}).

\section{Setup and Model\label{sec:setup}}

\begin{figure}[ht]
    \includegraphics[width=0.45\textwidth]{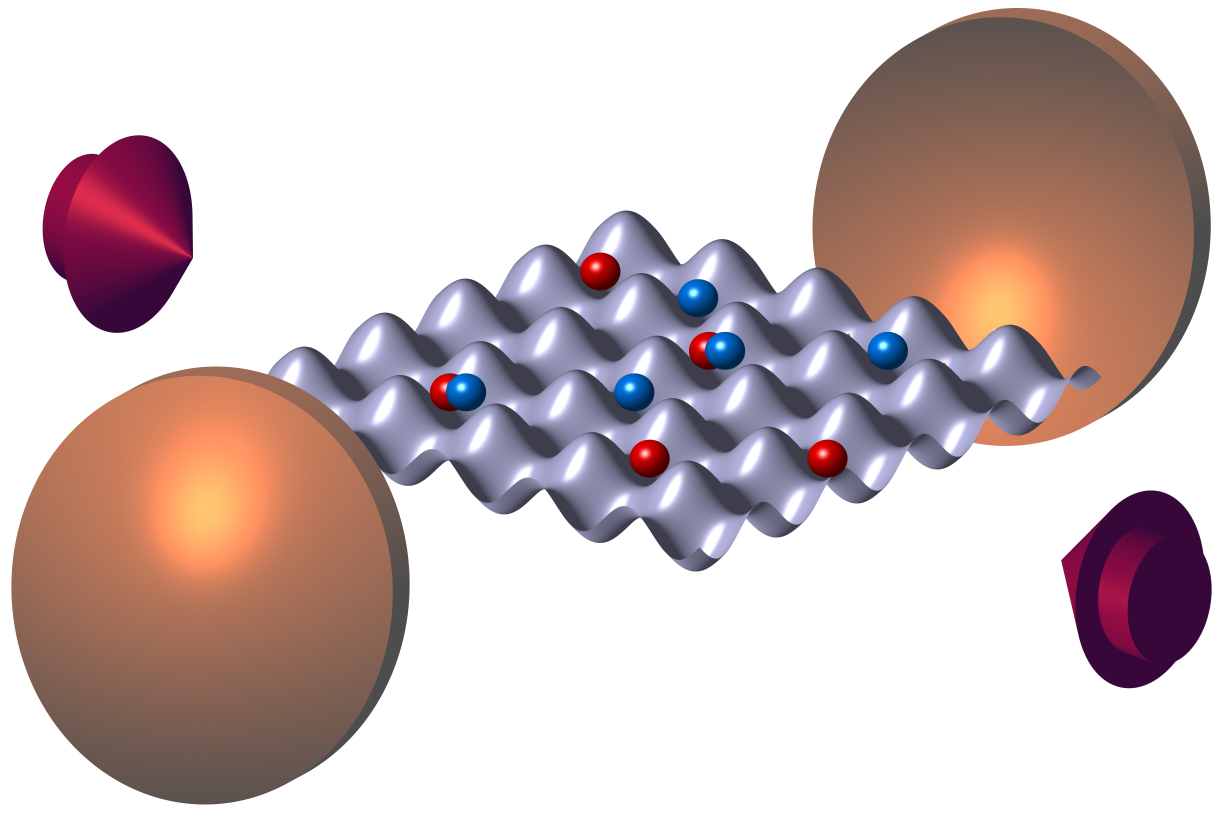} 
    \vspace{-10pt}
    \caption{Fermi-Hubbard system coupled globally to a single-mode cavity with strength $g$, by pumping with a retro-reflected transverse laser beam. The two fermion species tunnel with amplitude $J$ and have an on-site interaction $U$. Photons with a detuning $\delta$ from the pump laser leak through the cavity mirrors at rate $\varGamma$.}
    \label{fig:model_2d}
\end{figure}

We consider ultracold atoms in an optical lattice coupled to the field of an optical cavity by a transverse pump beam, setup sketch in Fig.~\ref{fig:model_2d}. The dynamics of the system is captured by a Lindblad master equation
\cite{MaschlerRitsch2008}
\begin{equation}
\label{eq:Lindblad_equation_model}
\frac{d}{dt}\hat{\rho}
=-\frac{i}{\hbar}[\hat{H},\hat{\rho}]+ \frac{\varGamma}{2}\left(2\hat{a}\hat{\rho} \hat{a}^\dagger-\hat{a}^\dagger \hat{a}\hat{\rho}-\hat{\rho}\hat{a}^\dagger \hat{a}\right).
\end{equation}
The density matrix $\hat{\rho}$ contains both the photonic and atomic degrees of freedom.
The cavity experiences photon losses described by the dissipator with the amplitude $\varGamma$ and the photon annihilation jump operator $\hat{a}$.
The atoms-cavity system is described by the following Hamiltonian  \cite{RitschEsslinger2013, MivehvarRitsch2021,TolleHalati2024}
\begin{align}
\label{eq:Hamiltonian}
\hat{H}&=\hat{H}_\text{FH}+\hat{H}_\text{cav}+\hat{H}_\text{ac}, \\
\hat{H}_\text{FH}&= -J\sum_{\langle j,l\rangle, \sigma}\big(\hat{c}_{j\sigma}^\dagger \hat{c}_{l\sigma}+\text{H.c.}\big)+ U\sum_{j}\hat{n}_{j\uparrow}\hat{n}_{j\downarrow}, \nonumber \\
\hat{H}_\text{cav}&=\hbar\delta \hat{a}^\dagger\hat{a}, \nonumber \\
\hat{H}_\text{ac}&=-\frac{\hbar g}{\sqrt{L^d}}(\hat{a}+\hat{a}^\dagger)\hat{\Delta},~\hat{\Delta}=\sum_{\substack{j\in A,\sigma}}\hat{n}_{j\sigma}-\sum_{\substack{l\in B,\sigma}}\!\hat{n}_{l\sigma}. \nonumber
\end{align}
The ultracold atoms we consider are interacting fermionic atoms subjected to additional optical lattices. These lattices can confine the geometry to be one-, two- or three-dimensional. For sufficiently strong optical lattices, the atoms can be described by the Hubbard model $\hat{H}_\text{FH}$ with $\hat{c}_{j \sigma}$ fermionic operators for the atoms on site $j$ and spin $\sigma\!\in\!\{\uparrow,\downarrow\}$. The local density operator is $\hat{n}_{j,\sigma}\!\equiv\!\hat{c}_{j,\sigma}^\dagger \hat{c}_{j,\sigma}$. 
The Hubbard model contains a tunneling process between neighboring sites $\langle j,l\rangle$ with amplitude $J$ and a repulsive on-site interaction of strength $U\!>\!0$. 
The second term $\hat{H}_\text{cav}$ describes the cavity mode in the rotating frame of the pump beam, where $\delta$ is the detuning between the cavity and pump beam.
The atoms-cavity coupling $\hat{H}_\text{ac}$ is realized via a standing-wave transverse pump beam far-detuned from the atomic resonance. The large atomic detuning allows the adiabatic elimination of the excited internal atomic states. The form of the atoms-cavity coupling stems from the commensurability of the cavity mode with twice the periodicity of the lattice spacing, which creates a bipartite lattice (with sublattices $A,B$). This results in a coupling of the cavity field to the atomic density imbalance between the two sublattices, $\hat{\Delta}$, with the effective coupling strength $g$ \cite{MaschlerRitsch2008, RitschEsslinger2013}.
We present results for systems of dimension $d\!=\!1,2$, with $L$ the number of sites along each dimension of the lattice and $N$ the total number of atoms.

\section{Methods\label{sec:methods}}

In this section, we begin by introducing the mean-field elimination of the cavity mode, reducing the system to an effective atomic model (Sec.~\ref{sec:mf}). Then we go beyond this approximation by including fluctuations in the atoms-cavity coupling with our many-body adiabatic elimination approach (Sec.~\ref{sec:mbae}). Further, a stability condition for the solutions from the previously obtained equations under variations of the steady state cavity field is derived (Sec.~\ref{sec:stability}). 
We conclude the methods section by performing a perturbative calculation in the weak tunneling regime (Sec.~\ref{sec:pertubation_J}), going to the thermodynamic limit.

\subsection{Mean-field elimination of the cavity mode \\($T\!=\!0$ MF)\label{sec:mf}}

An often employed approach for dealing with systems of particles coupled to cavities is to perform a mean-field decoupling of the coupling term $\hat{H}_\text{ac}$ and adiabatically eliminate the dynamics of the cavity mode \cite{RitschEsslinger2013, MivehvarRitsch2021}. Within the mean-field approach the coupling is given by
\begin{align}
\label{eq:Hac_MF}
\hat{H}_\text{ac}^\text{MF}&=-\frac{\hbar g}{\sqrt{L^d}}\big[\langle\hat{a}+\hat{a}^\dagger\rangle\hat{\Delta}+(\hat{a}+\hat{a}^\dagger)\langle\hat{\Delta}\rangle\big]. 
\end{align}
One assumes that the photons are in their steady state which is determined by
\begin{equation}
\label{eq:time_dependence_a}
\frac{\partial}{\partial t}\langle\hat{a}\rangle=i\frac{g}{\sqrt{L^d}}\langle\hat{\Delta}\rangle-\Big(i\delta+\frac{\varGamma}{2}\Big)\langle\hat{a}\rangle=0.
\end{equation}
The solution of this equation results in a coherent state for the cavity
\begin{align}
\label{eq:cav_MF}
\lambda\equiv\frac{1}{\sqrt{L^d}}\langle \hat{a}+\hat{a}^\dagger\rangle=\frac{2g\delta}{\delta^2+(\varGamma/2)^2}\frac{\langle \hat{\Delta}\rangle}{L^d}.
\end{align}
By considering the coherent state for the cavity, the following effective Hamiltonian for the atoms arises
\begin{align}
\label{eq:Heff_atoms}
\hat{H}_\text{eff}=\hat{H}_\text{FH}-\hbar g\lambda\hat{\Delta}.
\end{align}
For our model, Eq.~(\ref{eq:Hamiltonian}), the effective Hamiltonian is given by the ionic Hubbard model, whose properties we discuss in Sec.~\ref{sec:ionic_hubbard}. The value of the staggered potential is $\hbar g\lambda$.
A common approach is to take the self-consistently determined ground state of the effective Hamiltonian $\hat{H}_\text{eff}$ \cite{RitschEsslinger2013, MivehvarRitsch2021}, which in the following we dub the zero-temperature mean-field method ($T\!=\!0$ MF). However, this is an arbitrary choice since any eigenstate of $\hat{H}_\text{eff}$ is a steady state solution of the atomic equations of motion \cite{BezvershenkoRosch2021, TolleHalati2024, HalatiPhD}.
Thus, in order to determine the contributions of each eigenstate of $\hat{H}_\text{eff}$ to the steady state, one needs to go beyond the mean-field approach.

We note that by performing the mean-field treatment the weak $\mathbb{Z}_2$ symmetry of the model, $(\hat{a},\hat{\Delta})\!\to\!(-\hat{a},-\hat{\Delta})$, is broken, resulting in two steady states for which the cavity field has different signs. In the following, we show only the solution with $\langle \hat{a}\rangle\!\geq\!0$, corresponding to the sublattice $A$ having a lower energy.

In the next section we consider the quantum fluctuations on top of the mean-field decoupling in order to determine the steady state of the coupled atoms-cavity system.

\subsection{Many-body adiabatic elimination technique for atoms-cavity coupled systems\label{sec:mbae}}

In recent years, the crucial role of fluctuations in the coupling of quantum matter and quantum light has been the focus of several theoretical works \cite{MivehvarRitsch2021, GopalakrishnanGoldbart2010, PiazzaStrack2014, WallRey2016, DamanetKeeling2019, HalatiKollath2020, HalatiKollath2020b, BezvershenkoRosch2021, HalatiKollath2022, ChiriacoChanda2022, JagerBetzholz2022, LinkDaley2022, LentrodtEvers2023, HalatiKollath2025, TolleHalati2024, MuellerStrunz2025, HalatiJager2025, SchmitJaeger2025, OrsoDeuar2025}. 
Here, we employ a method developed in Ref.~\cite{BezvershenkoRosch2021} in the framework of the many-body adiabatic elimination technique \cite{Garcia-RipollCirac2009, ReiterSorensen2012, Kessler2012, PolettiKollath2013, ZanardiCamposVenuti2014, SciollaKollath2015, LangeRosch2018}. 
This approach perturbatively includes the fluctuations in the atoms-cavity coupling on top of the mean-field treatment (see Sec.~\ref{sec:mf}).
This method was developed and benchmarked against numerically exact results for bosonic atoms \cite{BezvershenkoRosch2021, HalatiKollath2022, HalatiPhD} and extended to fermionic atoms in Ref.~\cite{TolleHalati2024}.

In the following, we briefly present how to employ the many-body adiabatic elimination approach to include the effects of the fluctuations $\delta\hat{H}_\text{ac}$ in the atoms-cavity coupling \cite{TolleHalati2024}
\begin{align}
\label{eq:H_fluct}
\delta\hat{H}_\text{ac}=-\hbar g\left(\frac{\hat{a}+\hat{a}^\dagger}{\sqrt{L^d}}-\lambda\right)\hat{\Delta}.
\end{align}

The first step for performing the perturbation theory for density matrices within the many-body adiabatic elimination framework is to determine the decoherence free and lowest decaying subspaces of the unperturbed Liouvillian $\mathcal{L}_0$. 
For the considered case, $\mathcal{L}_0$ is given by
\begin{align}
\label{eq:Free_Liouvillian}
\mathcal{L}_0\hat{\rho}=-&\frac{i}{\hbar}\big[\hat{H}_\text{FH}+\hat{H}_\text{cav}+\hat{H}_\text{ac}^{MF},\hat{\rho}\big]\\
&+\frac{\varGamma}{2}\left(2\hat{a}\hat{\rho}\hat{a}^\dagger-\hat{a}^\dagger \hat{a}\hat{\rho}-\hat{\rho}\hat{a}^\dagger \hat{a}\right).\nonumber
\end{align}
The eigenvalues of $\mathcal{L}_0$, defined by $\mathcal{L}_0 \hat{\rho}_\lambda\!=\!(-\lambda^R\!+\!i\lambda^I)\hat{\rho}_\lambda$, are complex numbers with $\lambda^R\!\geq\!0$. If the dissipation is the dominating energy scale, these lie in bands separated by gaps of order $\mathcal{O}(\varGamma)$. The bands broaden and eventually merge as $\varGamma$ decreases relative to the other energies.
We define $\Lambda_\alpha$ as the subspace of right eigenvectors sharing the same $\lambda^R_\alpha$, with $\Lambda_0$ the decoherence free subspace for which $\lambda^R_0\!=\! 0$ holds. 
The right eigenstates spanning the decoherence free subspace, determined by $\mathcal{L}_0 \hat{\rho}^0\!=\!\lambda_0 \hat{\rho}^0$, are given by
\begin{equation}
\begin{aligned}
\label{eq:ansatz}
\hat{\rho}^0=\hat{\rho}_{\text{cav}}^0\otimes \hat{\rho}_{\text{eff}}^0, \quad \text{with}\quad&\hat{\rho}_{\text{cav}}^0=\ket{\alpha(\Delta)}\bra{\alpha(\Delta)}\\ \qquad& \hat{\rho}_{\text{eff}}^0=\ket{n_1(\lambda)}\bra{n_2(\lambda)}.
\end{aligned}
\end{equation}
The atomic part is determined by eigenstates $\ket{n(\lambda)}$ of $\hat{H}_\text{eff}$ with energy $E_n(\lambda)$ and $\lambda$ is self-consistently determined as in Eq.~(\ref{eq:cav_MF}). The cavity state $\hat{\rho}^0_{\text{cav}}$ is given by a coherent state with the field 
\begin{align}
\label{eq:cavity_sc}
\frac{\alpha(\Delta)}{\sqrt{L^d}}=\frac{g}{\delta-i\varGamma/2} \frac{\langle \hat{\Delta}\rangle}{L^d},
\end{align}
where the expectation value is calculated with respect to $\hat{H}_\text{eff}$.
Thus, the decoherence free subspace $\Lambda_0$ is spanned by the states $\ket{0,n}\!\bra{0,m}$. These are product states of the coherent photonic state $\ket{0}\equiv\ket{\alpha}$ and the eigenstates of the effective atomic Hamiltonian $\ket{n}$ with $\hat{H}_\text{eff}\ket{n}\!=\!E_n\ket{n}$.
We obtain the following eigenvalues for the unperturbed Liouvillian $\mathcal{L}_0$
\begin{align}
\label{eq:eigenstates0}
&\mathcal{L}_0\ket{0,n}\!\bra{0,m}=-\frac{i}{\hbar}(E_{n}-E_{m})\ket{0,n}\!\bra{0,m}.
\end{align}

A general state in the decoherence free subspace $\Lambda_0$ can be written as 
\begin{equation}
\label{eq:rho_0}
\hat{\rho}^0\! =\! \ket{\alpha(\Delta)}\bra{\alpha(\Delta)}\!\otimes\!\sum_{n_1,n_2}\!c(n_1,n_2)\ket{n_1(\lambda)}\bra{n_2(\lambda)},
\end{equation}
with the coefficients $c(n_1,n_2)$ the entries of the reduced atomic density matrix. The atomic eigenstates of the Hamiltonian $\hat{H}_\text{eff}$ depend self-consistently on the cavity field via the relation given in Eq.~(\ref{eq:cav_MF}). 

The other most relevant subspace $\Lambda_1$ is the slowest decaying subspace and is spanned by the states $\ket{1,n}\!\bra{0,m}$ and $\ket{0,n}\!\bra{1,m}$, where the photonic state is given by $\ket{1}\equiv (\hat{a}^\dagger\!-\!\alpha^*)\!\ket{0}$. The corresponding eigenvalues are
\begin{align}
\label{eq:eigenstates1}
& \mathcal{L}_0\ket{1,n}\!\bra{0,m}=\left[-\frac{i}{\hbar}(E_{n}-E_{m}+\!\hbar\delta)-\frac{\varGamma}{2}\right]\ket{1,n}\!\bra{0,m} \nonumber\\
& \mathcal{L}_0\ket{0,n}\!\bra{1,m}=\left[-\frac{i}{\hbar}(E_{n}-E_{m}-\hbar\delta)-\frac{\varGamma}{2}\right]\ket{0,n}\!\bra{1,m}.\nonumber
\end{align}
In particular, the subspaces that we consider contain excitations of the cavity field, decaying exponentially on a time scale $\mathcal{O}(\varGamma)$.

The perturbation, given in Eq.~(\ref{eq:H_fluct}), can take the system out of the decoherence free subspace of $\mathcal{L}_0$. Thus, to determine an effective equation of motion within the decoherence free subspace we adiabatically eliminate the states from the higher subspaces $\Lambda_{\alpha\neq0}$ \cite{Garcia-RipollCirac2009,ReiterSorensen2012,Kessler2012,PolettiKollath2013, SciollaKollath2015}. We obtain
\begin{equation}
\pdv{t}\hat{\rho}^{0} =\mathcal{L}_0\hat{\rho}^{0}+ \frac{1}{\hbar^2} \hat{P}_0 \Big[ \delta\hat{H}_\text{ac},(\mathcal{L}_0)^{-1} \hat{P}_1 \big[\delta\hat{H}_\text{ac},\hat{\rho}^0\big]\Big],
\label{eq:ae}
\end{equation}
where $\hat{P}_0$ and $\hat{P}_1$ are the projectors to $\Lambda_0$ and $\Lambda_1$, respectively.
This equation captures the long-time behavior of the system beyond the fast decay on the time scale $\mathcal{O}(\varGamma)$, including the steady state. However, the approach does not ensure that the solutions are positive semi-definite \cite{LiKoch2014}.

To explicitly obtain the steady state of Eq.~(\ref{eq:ae}) one would need to insert the state given in Eq.~(\ref{eq:rho_0}) and solve self-consistently for the entries of the atomic density matrix.
However, the number of coefficients scales with the square of the atomic Hilbert space dimension. In order to reduce the complexity of the system of equations obtained and to restrict the solutions to physical density matrices, we make the assumption that we can describe the atomic system with a thermal state \cite{BezvershenkoRosch2021, HalatiPhD, LangeRosch2018, TolleHalati2024}. 
This is justified when the thermalization time of the atomic system is shorter than the timescale of the scattering from photon fluctuations. 
For an arbitrary system, one should consider a generalized Gibbs ensemble state for the atoms in order to take into account all symmetries \cite{HalatiKollath2022, LangeRosch2017, LangeRosch2018}. However, if the dynamics is confined to a single symmetry sector, or the effective model describes a chaotic system we can assume an atomic thermal state
\begin{equation}
\label{eq:rho_T}
\hat{\rho}_\text{at}=\frac{1}{Z}e^{-\beta\hat{H}_\text{eff}(\lambda)},
\end{equation}
with $\beta\!=\!1/k_B T$ the inverse temperature and $Z$ the partition function. 
In Ref.~\cite{BezvershenkoRosch2021} this approach was validated in a wide parameter regime by a comparison to results obtained with numerically exact results obtained with matrix product state methods for bosonic atomic.  
For the fermionic lattice model, when the cavity field has a finite occupation the effective Hamiltonian $\hat{H}_\text{eff}$, Eq.~(\ref{eq:Heff_atoms}) is the ionic Hubbard model. 
This model was shown to obey Gaussian orthogonal ensemble (GOE) statistics in the presence of the staggered potential \cite{DeMarcoKollath2022} associated with a chaotic dynamics. Since this level statistics \cite{DeMarcoKollath2022} is similar to the case of interacting bosons \cite{KolovskyBuchleitner2004, KollathLaeuchli2010}, we expect the thermal Ansatz for the fermions to lead to reliable results for the steady states above the self-organization transition and for higher dimensions both below and above the transition. 
In contrast, in one-dimension below the transition, the effective model is the Hubbard model, which is an integrable model. For this special case, more reliable results could be obtained using a generalized Gibbs Ansatz, taking into account the conserved quantities. However, due to the non-local nature of the conservation laws in many situations the dynamics of the Hubbard model is still well described neglecting these, such that we expect that typical quantities still show the correct features. 

Importantly, the fluctuations in the atoms-cavity coupling, Eq.~(\ref{eq:H_fluct}), determine the effective temperature at which the atomic system thermalizes. The temperature causes the admixture of excited states of the effective Hamiltonian in the density matrix of the system, in contrast to the $T\!=\!0$ MF method.
We determine the effective inverse temperature $\beta$ and the cavity field parameter $\lambda$ from the mean-field relation, Eq.~(\ref{eq:cav_MF}), and the steady state condition of the energy transfer 
\begin{equation}
\label{eq:energy_transfer}
\frac{\partial}{\partial t}\langle\hat{H}_{\text{eff}}\rangle=0,
\end{equation}
given by \cite{BezvershenkoRosch2021, TolleHalati2024}
\begin{equation}
\label{eq:energy_transfer_spectral}
 \frac{\partial}{\partial t}\langle\hat{H}_{\text{eff}}\rangle\propto\sum_{n,m} |\Delta_{nm}|^2 \frac{e^{\!-\beta E_{m}} (E_n\!-\!E_m)\varGamma}{\left(E_n\!-\!E_m\!+\!\hbar\delta\right)^2\!+\!\left(\hbar\varGamma/2\right)^2},
\end{equation}
with $\Delta_{nm}\!=\!\bra{n} \hat{\Delta}\ket{m}$. 
We discuss the nature of the excited states of the ionic Hubbard model in Sec.~\ref{sec:ionic_hubbard}. The eigenenergies and states are calculated via exact diagonalization for small system sizes and perturbatively for the approximation introduced in Sec.~\ref{sec:pertubation_J}. 
We can also write the energy transfer with the aid of the self-consistently computed retarded susceptibility of the operator $\hat{\Delta}$
\begin{equation}
\label{eq:energy_transfer_susceptibility}
    \frac{\partial}{\partial t}\langle\hat{H}_{\text{eff}}\rangle\propto\int\! d\omega \frac{\hbar\omega \text{Im}\left[\chi_T(\omega)\right]}{1-e^{-\beta\hbar\omega}}\frac{\varGamma/(2\pi)}{(\omega+\delta)^2+(\varGamma/2)^2},
\end{equation}
where 
\begin{equation}
\label{eq:susceptibility}\chi_T(\omega)\!=\!-\frac{i}{\hbar}\int_{0}^{\infty} dt e^{i(\omega+i\epsilon) t}\big\langle\big[\hat{\Delta}(t),\hat{\Delta}(0)\big]\big\rangle
\end{equation}
is evaluated for the thermal state $\hat{\rho}_\text{at}$. 
The self-consistently determined temperature can stabilize to different values as we vary the parameters throughout the phase diagram, as we discuss in Sec.~\ref{sec:results}. In particular this effect is a crucial ingredient for the many-body cooling mechanisms and the fluctuation-induced bistability \cite{TolleHalati2024}. 
In the normal phase fluctuations in the atoms-cavity coupling are present despite the vanishing steady state cavity field. 

\subsubsection{Large temperature limit}
\label{sec:large_dissipation}
In certain regimes we can derive an approximate value of the effective temperature which does not depend on the microscopic details of the atomic model to which the cavity is coupled.
In this subsection, we derive the approximate expression of the effective temperature in the limit of large dissipation $\varGamma,\varGamma^2/\delta\!\gg\! J/\hbar,U/\hbar,g\lambda$ or large detuning $\hbar\delta\!\gg\!U,J,g\lambda$, which are directly linked to a large self-consistently determined effective temperature.

In either of these limits, the susceptibility $\text{Im}[\chi_T(\omega)]$, Eq.~(\ref{eq:energy_transfer_susceptibility}), decays rapidly for $\hbar\omega\! >\!J,U,\hbar g\lambda$, thus we can make the approximation
\begin{align}
    \delta_\varGamma(\omega\!+\!\delta)&\equiv\!\frac{\varGamma/2\pi}{\delta^2(1\!+\!\omega/\delta)^2\!+\!(\varGamma/2)^2}\\&\!\approx\!\frac{\varGamma/2\pi}{\delta^2\!+\!(\varGamma/2)^2}\!-\!\frac{\omega\delta\varGamma/\pi}{\big[\delta^2\!+\!(\varGamma/2)^2\big]^2}\nonumber.
\end{align}
At the same time we take the high temperature limit $\beta\!\to\!0$ and approximate the factor $\big(1\!-\!e^{-\beta\hbar\omega}\big)^{-1}$ appearing in the integrand of the energy transfer equation, Eq.~(\ref{eq:energy_transfer_susceptibility}), with the two lowest orders of Taylor expansion $1/(\beta\hbar\omega)\!+\!1/2$.
Of the product $$\delta_\varGamma(\omega\!+\!\delta)\big(1\!-\!e^{-\beta\hbar\omega}\big)^{-1}\!\approx\!\delta_\varGamma(\omega\!+\!\delta)\bigg(\frac{1}{\beta\hbar\omega}\!+\!\frac{1}{2}\bigg)$$ only even functions in the frequency contribute in the integral since $\omega\text{Im}[\chi_T(\omega)]$ is an even function and we integrate over a symmetric interval. The leading order terms are constant in $\omega$
\begin{equation}
    \frac{\varGamma/4\pi}{\delta^2\!+\!(\varGamma/2)^2}-\frac{\delta\varGamma/\pi}{\hbar\beta\big[\delta^2\!+\!(\varGamma/2)^2\big]^2}.
\end{equation}
Fulfilling the steady state condition $\frac{\partial}{\partial t}\langle\hat{H}_\text{eff}\rangle\!=\!0$, we require the integrand to vanish and solve for $\beta$, resulting in
\begin{equation}
    \label{eq:beta_approx_large_dissipation}
   \beta\!\approx\!\frac{4\delta/\hbar}{\delta^2\!+\!(\varGamma/2)^2}.
\end{equation}
This result agrees with the result of inverse effective temperature, $\beta$, for bosons \cite{BezvershenkoRosch2021} and is consistent with results based on the semiclassical limit for non-interacting atoms \cite{AsbothVukics2005, SchuetzMorigi2014, PiazzaStrack2014}.
This relation shows that by increasing the separation of energy scales between the cavity and the atoms, i.e.~increasing $\hbar\delta$ and $\hbar\varGamma$, the atoms are driven into states with large effective temperatures.

\subsection{Stability analysis of the solutions \label{sec:stability}}

In this section, we show how to decide if self-consistent solutions obtained from Eq.~(\ref{eq:cav_MF}) and Eq.~(\ref{eq:energy_transfer}) are stable by analyzing their behavior in the presence of perturbations acting on the cavity field. For unstable results, small fluctuations are amplified and grow exponentially.
The stability analysis follows the extension to mixed atomic states at finite temperature, shown in Ref.~\cite{TolleHalati2024}, of previously done work for the stability of atoms-cavity systems described by pure states, e.g.~in Refs.~\cite{Tian2016, HalatiKollath2017}.

For the determination of the stability, we need to consider the dependence on the cavity field $\lambda$ of the following quantities: the eigenenergies of the atomic effective Hamiltonian $E_n(\lambda)$; the inverse temperature $\beta(\lambda)$; matrix elements of the imbalance operator $\Delta_{nm}(\lambda)$.
The equations of motion of the quadratures of the cavity field, $\lambda\!=\!\langle\hat{a}^\dag\!+\!\hat{a}\rangle/\sqrt{L^d}$ and $p\!=\!i\langle \hat{a}^\dag\!-\!\hat{a}\rangle/\sqrt{L^d}$, are given by
\begin{equation*}
\frac{d\lambda}{dt}=-\frac{\varGamma}{2}\lambda+\delta p,\qquad\frac{dp}{dt}=-\delta\lambda-\frac{\varGamma}{2} p+\frac{2g}{L^d}\langle \hat{\Delta}\rangle.
\end{equation*}

In the next step we find the stationary solutions of these equations
\begin{equation*}
\lambda^{(s)}=\frac{2\delta g}{\delta^2+(\varGamma/2)^2}\frac{\langle \hat{\Delta}\rangle}{L^d}^{(s)},\quad p^{(s)}=\frac{\varGamma g}{\delta^2+(\varGamma/2)^2}\frac{\langle \hat{\Delta}\rangle}{L^d}^{(s)},
\end{equation*}
where the expectation value $\langle ...\rangle^{(s)}$ is computed with respect to the self-consistent thermal density matrix. The stability of the stationary solutions is determined from considering linear fluctuations around the stationary values
\begin{equation*}
    \lambda=\lambda^{(s)}+\tilde{\lambda},\quad
p=p^{(s)}+\tilde{p}, \quad\langle \hat{\Delta}\rangle=\langle \hat{\Delta}\rangle^{(s)}+\frac{\differential \langle \hat{\Delta}\rangle^{(s)}}{\differential \lambda^{(s)}}\tilde{\lambda}.
\end{equation*}
The dynamics of the fluctuations is given by
\begin{equation}
\label{eq:dynamics_linear_fluctuations}
\frac{d\tilde{\lambda}}{d t}\!=\!-\frac{\varGamma}{2}\tilde{\lambda}+\delta \tilde{p},\quad\frac{d\tilde{p}}{d t}\!=\!-\delta\tilde{\lambda}-\frac{\varGamma}{2} \tilde{p}+\frac{2g}{L^d}\frac{\differential \langle \hat{\Delta}\rangle^{(s)}}{\differential \lambda^{(s)}}\tilde{\lambda}.
\end{equation}

The differential of the atomic imbalance expectation value in a thermal state $$\langle\hat{\Delta}\rangle^{(s)}\!=\!\frac{1}{Z(\lambda^{(s)})}\sum_n\!\text{e}^{-\beta(\lambda^{(s)})E_n(\lambda^{(s)})}\Delta_{nn}(\lambda^{(s)})$$ is given by
\begin{align}
\frac{\differential \langle \hat{\Delta}\rangle^{(s)}}{\differential \lambda^{(s)}}&=\frac{1}{Z}\sum_n \textstyle e^{-\beta E_n}\left[\frac{\partial \Delta_{nn}}{\partial \lambda}-\Delta_{nn}\left(\frac{\partial \beta}{\partial \lambda}E_n+\beta\frac{\partial E_{n}}{\partial \lambda}\right)\right]\nonumber\\
+\frac{1}{Z^2}&\sum_{n,m}\textstyle e^{-\beta (E_n+E_m)}\Delta_{nn}\left(\frac{\partial \beta}{\partial \lambda}E_m+\beta\frac{\partial E_{m}}{\partial \lambda}\right).
\end{align}

For stable stationary solutions small perturbations decay back to the stationary state.
To obtain a requirement for states that fulfill this condition we calculate the Jacobian eigenvalues $k_{\pm}$ for the system of equations Eqs.~(\ref{eq:dynamics_linear_fluctuations}), where stable solutions require $\Re{k_{\pm}}\!<\!0$. 
For a positive detuning $\delta\!\geq\!0$, this gives the condition for stability of the thermal steady states under fluctuations of the cavity field
\begin{widetext}
\begin{align}
\label{eq:condition}
\frac{\delta^2+(\varGamma/2)^2}{2\delta g}>\sum_n \frac{\text{e}^{-\beta E_n}}{L^dZ}\left[\frac{\partial \Delta_{nn}}{\partial \lambda}-\Delta_{nn}\left(\frac{\partial\beta}{\partial \lambda}E_n+\beta\frac{\partial E_{n}}{\partial \lambda}\right)\right]
+\sum_{n,m}\frac{\text{e}^{-\beta(E_n+E_m)}}{L^{2d}Z^2}\Delta_{nn}\left(\frac{\partial\beta}{\partial \lambda}E_m+\beta\frac{\partial E_{m}}{\partial \lambda}\right).
\end{align}
\end{widetext}

For the results we present in Sec.~\ref{sec:results}, we numerically evaluate the derived condition in Eq.~(\ref{eq:condition}) for the solutions of Eq.~(\ref{eq:cav_MF}) and Eq.~(\ref{eq:energy_transfer}) for both the small system exact-diagonalization and in perturbative approach in the thermodynamic limit. 
In practice, we employ a five-point finite differences scheme $f'(\lambda)=[-f(\lambda+2\epsilon)+8f(\lambda+\epsilon)-8f(\lambda-\epsilon)+f(\lambda-2\epsilon)]/12\epsilon$ with a small step size $\epsilon$ of order $\mathcal{O}(10^{-4})$ to obtain numerically robust results. 

For completeness we state the stability condition at the $T\!=\!0$ MF level which is
\begin{equation}
    \label{eq:stability_condition_MF}
    \frac{\delta^2+(\varGamma/2)^2}{4\delta g}>   \frac{\differential \langle \hat{\Delta}\rangle^{(s)}}{\differential \lambda^{(s)}}.
\end{equation}
Here, an interpretation is straightforward since the presented observables do not implicitly depend on each other via the self-consistently determined temperature.
If the effective coupling $\hbar g\lambda\!=\!\frac{4\hbar\delta g^2}{\delta^2+(\varGamma/2)^2}$ is slightly increased the slope of the line $g\lambda\langle \hat{\Delta}\rangle^{(s)}/\hbar g\lambda$ will decrease. Starting from a stable solution, the solution is shifted to slightly larger values of $\langle \hat{\Delta}\rangle^{(s)}$ and $\hbar g\lambda$. Contrarily for an unstable solution solution shifts towards to the trivial solution of $\hbar g\lambda\!=\!\langle \hat{\Delta}\rangle^{(s)}\!=\!0$. At finite temperature the many additional terms in Eq.~(\ref{eq:condition}) make the interpretation more complicated.

\subsection{Perturbative solutions in the regime of weak tunneling\label{sec:pertubation_J}}

In order to gain analytical insights into the self-consistent steady state solutions and obtain results for the thermodynamic limit and higher dimensional systems, we perturbatively solve Eq.~(\ref{eq:cav_MF}) and Eq.~(\ref{eq:energy_transfer}) in the limit of small kinetic energy \cite{TolleHalati2024}.
This approach can be generalized for situations in which the dominant contribution to the atomic Hamiltonian stems from terms that commute with the atomic operator to which the cavity is coupled to.
In our case this corresponds to the situation in which the hopping amplitude $J$ is the smallest energy scale in the system and the on-site interaction $U$ dominates.
We can perform the calculation in the thermodynamic limit, allowing us to understand the finite size effects present in the exact-diagonalization approach.

We split the effective Hamiltonian, Eq.~(\ref{eq:Heff_atoms}), into a part containing density operators $\tilde{H}_0$ and the perturbative kinetic part $\hat{H}_\text{kin}$
\begin{align}
\label{eq:Hamiltonian_perturbation}
\tilde{H}_0&= U\sum_{j}\hat{n}_{j\uparrow}\hat{n}_{j\downarrow}-\hbar g\lambda\hat{\Delta}-\mu\sum_{j,\sigma}\hat{n}_{j\sigma}, \\
\hat{H}_\text{kin}&=-J\sum_{\langle j,l\rangle, \sigma}\big(\hat{c}_{j\sigma}^\dagger \hat{c}_{l\sigma}+\text{H.c.}\big), \nonumber 
\end{align}
where in $\tilde{H}_0$ we added a chemical potential term in order to fix the atomic filling.

We can rewrite the susceptibility $\chi_T(\omega)$, Eq.~(\ref{eq:susceptibility}), in terms of the perturbation $\hat{H}_\text{kin}$ as
\begin{align}
\label{eq:Susceptibility}
\chi_T(\omega)&=\frac{-i}{\hbar}\!\int_{0}^{\infty} dt e^{i(\omega+i\epsilon)t}\big\langle\big[\hat{\Delta}(t),\hat{\Delta}(0)\big]\big\rangle\\
&=\frac{-i}{\hbar^3\omega^2}\!\int_{0}^{\infty} dt e^{i(\omega+i\epsilon)t} \nonumber\\
&\qquad\times\big\langle\big[\big[\hat{H}_\text{kin},\hat{\Delta}\big](t),\big[\hat{H}_\text{kin},\hat{\Delta}\big](0)\big]\big\rangle\nonumber,
\end{align}
where $\epsilon$ is small. We used the time-translational symmetry $\big[\hat{\Delta}(t),\hat{\Delta}(0)\big]\!=\!\big[\hat{\Delta}(0),\hat{\Delta}(-t)\big]$.
Within this expression by computing the expectation values with respect to a thermal state of the unperturbed Hamiltonian $\tilde{H}_0$ we obtain a perturbative result for $\chi_T(\omega)$ of order $\mathcal{O}(J^2)$.

For the one-dimensional system in the thermodynamic limit the expectation value calculated with respect to the two-site unit cell is given by \cite{TolleHalati2024}
\begin{align}
&\big\langle\big[\big[\hat{H}_\text{kin},\hat{\Delta}\big](t),\big[\hat{H}_\text{kin},\hat{\Delta}\big](0)\big]\big\rangle\!\\
&\quad=\frac{16iJ^2}{Z_\text{1D}}\bigg\{2\sin(2g\lambda t)\big(e^{\beta\mu}+e^{\beta(3\mu-U)}\big)\sinh(\beta\hbar g\lambda)\nonumber\\
&\qquad-\!\sum_{\mathclap{p=\pm 1}}\textstyle p\sin\big[(2g\lambda-pU/\hbar)t\big]e^{2\beta\mu}\big(1-e^{\beta(p2\hbar g\lambda-U)}\big)\bigg\}\nonumber.
\end{align}
Using this result to compute the susceptibility, Eq.~(\ref{eq:Susceptibility}), we can derive the energy transfer, Eq.~(\ref{eq:energy_transfer_susceptibility}), up to order $\mathcal{O}(J^2)$. Together with the mean-field self consistency condition, Eq.~(\ref{eq:cav_MF}), and the condition fixing the total particle number, they lead to the following system of equations \cite{TolleHalati2024}
\begin{widetext}
\begin{subequations}
	\label{eq:system_of_equations_1D}
	\begin{align}
	\label{eq:EOM_1D}
	 \frac{\partial}{\partial t}\langle \hat{H}_\text{eff}\rangle&\propto
	J^2\bigg\{\frac{\text{e}^{\beta\mu}+e^{\beta(3\mu-U)}}{2g\lambda}\left[\frac{-e^{\beta\hbar g\lambda}}{(2g\lambda+\delta)^2+(\varGamma/2)^2}+\frac{e^{-\beta\hbar g\lambda}}{(2g\lambda-\delta)^2+(\varGamma/2)^2}\right]\\
	&+\sum_{\mathclap{p=\pm 1}}\frac{e^{2\beta\mu}}{2g\lambda-pU/\hbar}\left[\frac{-e^{\beta(2\hbar g\lambda-pU)}}{(2g\lambda-pU/\hbar+\delta)^2+(\varGamma/2)^2}+\frac{1}{(2g\lambda-pU/\hbar-\delta)^2+(\varGamma/2)^2}\right]\bigg\}=0\nonumber,
	\end{align}
	\begin{align}
	\label{eq:pc_1D}
	\left\langle \hat{n}\right\rangle&= \frac{1}{Z_{\text{1D}}}\left[\cosh(\beta\hbar g\lambda)\left(e^{\beta\mu}+e^{\beta(3\mu-U)}\right)+e^{2\beta\mu}\left(2+e^{-\beta U}\cosh(2\beta\hbar g\lambda)\right)+e^{\beta(4\mu-2U)}\right],\\
	\lambda&=\frac{1}{Z_{\text{1D}}}\frac{4g\delta}{\delta^2+(\varGamma/2)^2}\left[\sinh(\beta\hbar g\lambda)\label{eq:sc_1D}\left(e^{\beta\mu}+e^{\beta(3\mu-U)}\right)+e^{\beta(2\mu-U)}\sinh(2\beta\hbar g\lambda)\right],
	\end{align}
with the partition function on the two-site unit cell
	\begin{align}
	\label{eq:partition_function_1D}
	Z_{\text{1D}}=&1+4\cosh(\beta\hbar g\lambda)\big(e^{\beta\mu}+e^{\beta(3\mu-U)}\big)+2e^{2\beta\mu}\left[2+e^{-\beta U}\cosh(2\beta\hbar g\lambda)\right]+e^{\beta(4\mu-2U)}.
	\end{align}
\end{subequations}
\end{widetext}
By solving these coupled self-consistent equations we determine the steady state in the perturbative approach.
The stability of the steady states against small perturbations in the cavity field can also be determined using the condition derived in Eq.~(\ref{eq:condition}) and evaluated up to $\mathcal{O}(J^2)$. 

We can obtain similar results for atomic systems confined to a higher dimensional lattice structure. 
As we consider kinetic processes up to the order $\mathcal{O}(J^2)$, one can decompose the commutators in the integrand of the susceptibility $\chi_T(\omega)$, Eq.~(\ref{eq:Susceptibility}), to independent contributions along the different lattice direction. Tunneling processes in multiple directions are contained only at higher orders. The susceptibility for $d\!>\!1$, thus, only differs from the one-dimensional case by a rescaled tunneling amplitude, depending on the lattice geometry and the number of nearest neighbors (the connectivity).
For the steady state solution of Eq.~(\ref{eq:EOM_1D}), the dimension in which the fermionic atoms are confined thus does not play a role. However, the rates at which the steady state is approached do depend on the lattice geometry.
Thus, for the parameters regimes discussed in Sec.~\ref{sec:results} which show the agreement between the small system exact-diagonalization calculations and the perturbation theory calculation our results are valid for any dimensionality of the atomic system.

To confirm this, we explicitly calculated the weak tunneling perturbation equations on the $2\times2$ unit cell for the case of a two-dimensional square lattice and recover the same system of equations as for one dimension.

\section{Properties of the ionic Hubbard model}
\label{sec:ionic_hubbard}

In the previous sections we have seen that following the mean-field decoupling of the atoms and the cavity field the effective atomic Hamiltonian $\hat{H}_\text{eff}$ is given by the ionic Hubbard model with a self-consistently determined potential imbalance.
Thus, we want to revise briefly in the following the properties of the ground state phase diagram and of the energy spectrum which are important for the self-organization transition in the cavity considered. 

\subsection{Ionic Hubbard Hamiltonian\label{sec:ionic_hubbard_hamiltonian}}
The Hamiltonian of the ionic Hubbard model is given by 
\begin{align}
    \label{eq:Hamiltonian_ionic_hubbard}
    \hat{H}_\text{IH}\!=&\! -J\!\sum_{\langle j,l\rangle, \sigma}\!\big(\hat{c}_{j\sigma}^\dagger \hat{c}_{l\sigma}\!+\text{H.c.}\big)+ U\!\sum_{j}\!\hat{n}_{j\uparrow}\hat{n}_{j\downarrow}\nonumber \\&-\eta\Big(\sum_{\substack{j\in A,\sigma}}\!\hat{n}_{j\sigma}\!-\!\sum_{\substack{l\in B,\sigma}}\!\hat{n}_{l\sigma}\Big).
\end{align}
The ground state of the ionic Hubbard model is governed by the competition between the kinetic energy, the offset between lattice sites of different sublattices $2\eta$, and the on-site interaction $U$. 
At finite interaction strength the nature of the ground state phases becomes non-trivial and have been explored using many different many-body approaches.
Previous studies have tackled this model in both one dimension \cite{TorioCeccatto2001, WilkensMartin2001, ManmanaSchonhammer2004, LegezaSolyom2006, OtsukaNakamura2005, GoJeon2011} and higher dimension \cite{GidopoulosTosatti2000}, or its excitations \cite{HafezTorbatiUhrig2014} and finite temperature effects \cite{KimJeon2014}. 
Some works have focused on the predicted exotic bond-order wave phase at commensurate filling \cite{FabrizioNersesyan1999, ZhangLin2003, BatistaAligia2004, LoidaKollath2017}, that we will briefly comment on later.

We first discuss the effect of the ionic potential term in the non-interacting model ($U/J\!=\!0$), which is exactly solvable in any dimensions. 
The \emph{non-interacting} model with periodic boundary conditions can be diagonalized in momentum space on the reduced Brillouin zone, with $k\!\in\![-\pi/2,\pi/2]$ in one dimension. A generalization to higher dimensions is straightforward. 
One performs the Bogoliubov transformation defined by 
\begin{align*}
\label{eq:bogoliubov_transform}
    \hat{c}^{\phantom{\dagger}}_{k,\sigma}\!=\!u_k\hat{\gamma}^{\phantom{\dagger}}_{-,k,\sigma}\!+\!v_k\hat{\gamma}^{\phantom{\dagger}}_{+,k,\sigma},\quad &\hat{c}^{\phantom{\dagger}}_{k+\pi,\sigma}\!=\!-v_k\hat{\gamma}^{\phantom{\dagger}}_{-,k,\sigma}\!+\!u_k\hat{\gamma}^{\phantom{\dagger}}_{+,k,\sigma},
\end{align*}
with $\abs{u_k}^2\!+\! \abs{v_k}^2=1$, $\epsilon(k)\!=\!-2J\cos(k)$ and the parameters of the transformation given by
\begin{align}
    u_k,v_k\!=\!\sqrt{\frac{1}{2}\bigg(1\!\pm\!\frac{\epsilon(k)}{\sqrt{\epsilon(k)^2\!+\!\eta^2}}\bigg)}.
\end{align}
Using 
\begin{equation}
\label{eq:Ek}
   E(k,\eta)=\sqrt{\epsilon(k)^2\!+\!\eta^2} 
\end{equation}
the transformed Hamiltonian is then given by 
\begin{equation}
\label{eq:hamiltonian_momentum_space}
    \hat{H}_\text{IH}\!=\!\sum_{k,\sigma}E(k,\eta)(\hat{\gamma}^\dagger_{+,k,\sigma}\hat{\gamma}^{\phantom{\dagger}}_{+,k,\sigma}\!-\!\hat{\gamma}^\dagger_{-,k,\sigma}\hat{\gamma}^{\phantom{\dagger}}_{-,k,\sigma}).
\end{equation}
Due to the ionic potential, at the edges of the reduced Brillouin zone a band gap $\propto\!2\eta$ opens, splitting the band into the low energy band ($\hat{\gamma}^{\phantom{\dagger}}_{-,k,\sigma}$) and the high energy band ($\hat{\gamma}^{\phantom{\dagger}}_{+,k,\sigma}$).

The average filling can be calculated by 
\begin{equation}
\label{eq:n_U0_momentum}
    n\!=\!\sum_{k,\sigma}(u_k\!-\!v_k)^2\big(\langle\hat{\gamma}^\dagger_{-,k,\sigma}\hat{\gamma}^{\phantom{\dagger}}_{-,k,\sigma}\rangle\!+\!\langle\hat{\gamma}^\dagger_{+,k,\sigma}\hat{\gamma}^{\phantom{\dagger}}_{+,k,\sigma}\rangle\big),
\end{equation}
and the sublattice density imbalance as
\begin{align}
\label{eq:density_imbalance_momentum_space}
\langle\hat{\Delta}\rangle\!=\!\sum_{k,\sigma}&u_kv_k\big(\langle\hat{\gamma}^\dagger_{-,k,\sigma}\hat{\gamma}^{\phantom{\dagger}}_{-,k,\sigma}\rangle+\langle\hat{\gamma}^\dagger_{+,k,\sigma}\hat{\gamma}^{\phantom{\dagger}}_{+,k,\sigma}\rangle\big).
\end{align}
Here the expectation values are evaluated for the state one is interested in, so for example the ground state, or the finite temperature state. 
For the ground state one can use 
$\langle\hat{\gamma}^\dagger_{\pm,k,\sigma}\hat{\gamma}_{\pm,k,\sigma}\rangle\!=\!1$ for the filled momenta and $\langle\hat{\gamma}^\dagger_{\pm,k,\sigma}\hat{\gamma}_{\pm,k,\sigma}\rangle\!=\!0$ otherwise. 
We make use of these results to obtain analytical insights into the self-organization transition for atoms which have no direct interaction between themselves and and to obtain approximate results for strongly interacting atoms.

\subsection{Ground state phase diagram and excitations}

We discuss the phase diagram and low lying excitations of the ionic Hubbard model in one dimension. Since distinct phases arise at different fillings, we consider here two different fillings $n\!=\!N/L\!=\!\{1/2, 1\}$.

\begin{figure}[ht]
\begin{tikzpicture}
    \node[anchor=north west,inner sep=0pt] at (0,0){\includegraphics[width=0.45\textwidth,valign=c]{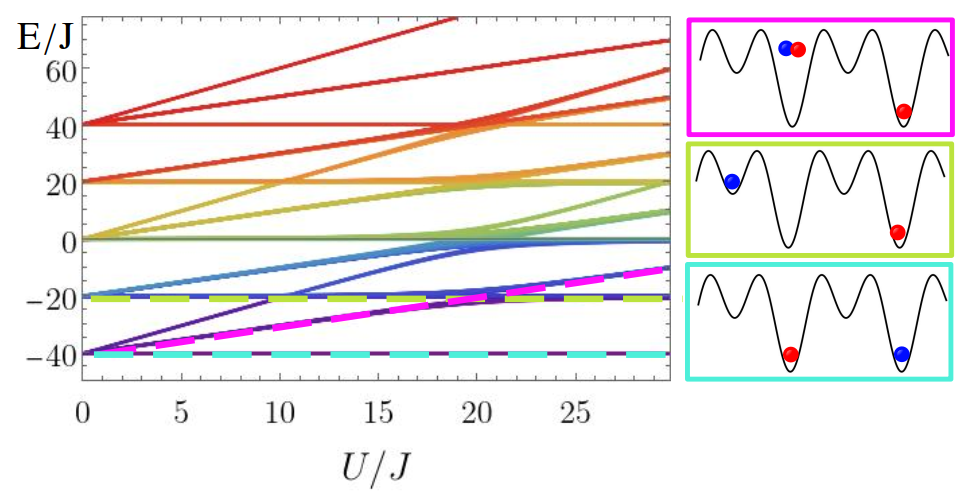}};
    \node[font=\normalfont] at (-1ex,-2ex) {(a)};
    \end{tikzpicture}
    \begin{tikzpicture}
    \node[anchor=north west,inner sep=0pt] at (0,0){\includegraphics[width=0.45\textwidth,valign=c]{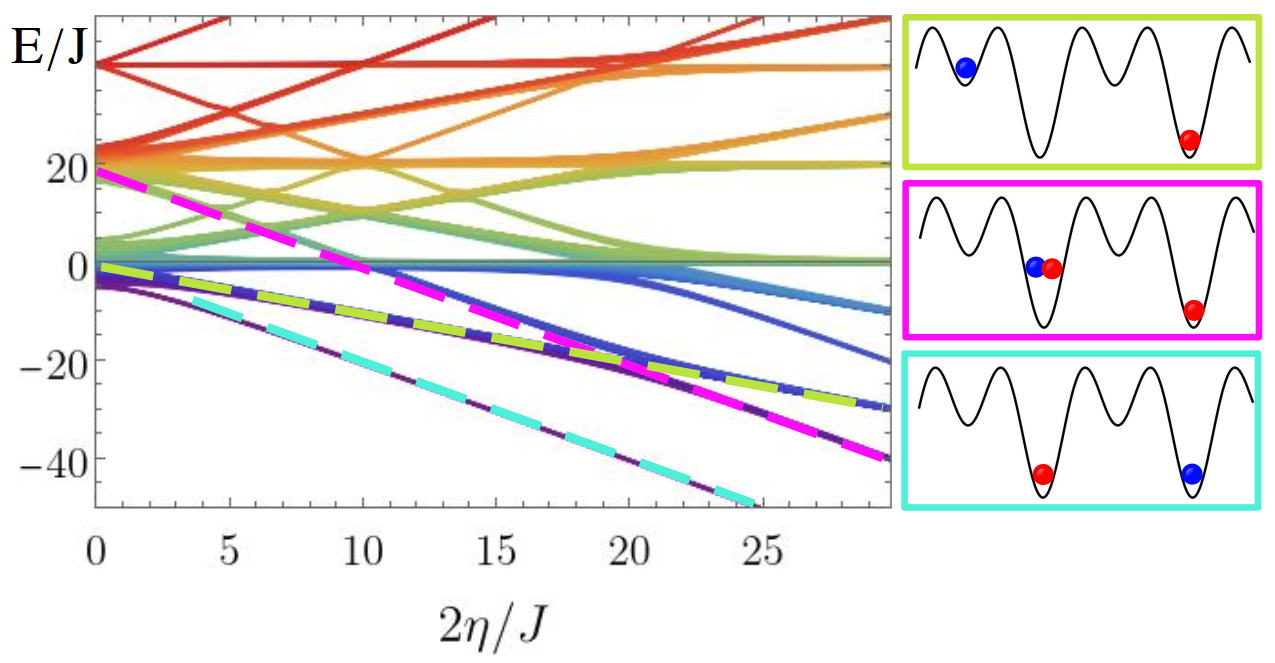}};
    \node[font=\normalfont] at (-1ex,-2ex) {(b)};
    \end{tikzpicture}
    \caption[]{Energy spectra of the ionic Hubbard model at quarter filling (solid lines). The parameters used are $L\!=\!8$, and in (a) $2\eta/J\!=\!20$, and in (b) $U/J\!=\!20$. We sketch the nature of ground states and excited states crossing around $2\eta\!\sim\!U$ (colored boxes) and mark the respective bands with dashed lines in the same color.}
    \label{fig:Espectrum_IH_L8N4_U20_etavar}
\end{figure}

At \emph{quarter filling} ($n\!=\!1/2$) the ground state of the non-interacting model ($U/J\!=\!0$), corresponds to the state in which the lower quasi-particle band is half-filled up to momenta $\abs{k}\!=\!\pi/4$. 
The ground state is a liquid and the low-lying excitations within the lower band are gapless. With increasing contribution of the ionic potential, the properties of the ground state change. The imbalance in the ground state increases with rising ionic potential.
Clear bands can be seen in the excitation spectrum in Fig.~\ref{fig:Espectrum_IH_L8N4_U20_etavar}~(a) at $U/J\!=\!0$ which correspond to excitations of the quasi-particles to the higher quasi-particle bands, for large $\eta/J\!\gg\!1$, the gaps between these bands grows with $\sim\! 2\eta/J$. 
At finite interaction strength the phases become more involved and many-body approaches are needed \cite{SchusterSchwingenschloegl2007, TorioNormand2006}.

Here we focus on revising the results needed in order to understand the self-organization transitions presented in Sec.~\ref{sec:self_organization_transition}. At finite interaction strength, the ground state of the one-dimensional Hubbard model at quarter filling is a liquid. This state has a gap $\Delta_c$ for charge excitations, whereas spin-excitations are gapless. At large interaction strength the charge gap is dominantly given by $\Delta_c\!\sim \!U$ corresponding in a simplified picture to the creation of a doubly occupied site.
As the potential imbalance $2\eta$ grows (e.g.~for $2\eta/J\!>\!U/J$), the imbalance of the density rises. The gap in the excited spectrum increases correspondingly as can be seen in Fig.~\ref{fig:Espectrum_IH_L8N4_U20_etavar}~(b). The nature of the second lowest band of excitations changes around $U/J\!\sim\!2\eta/J$ due to a crossing with the 3rd band. At this crossing the nature of the lowest charge excitations changes from states with a single particle occupation of a high-potential sublattice (light green box) to states with a double occupancy on the low-potential sublattice (magenta box).
We note that the spin excitations are always gapless.

\begin{figure}[ht]
    \includegraphics[width=0.23\textwidth,valign=c]{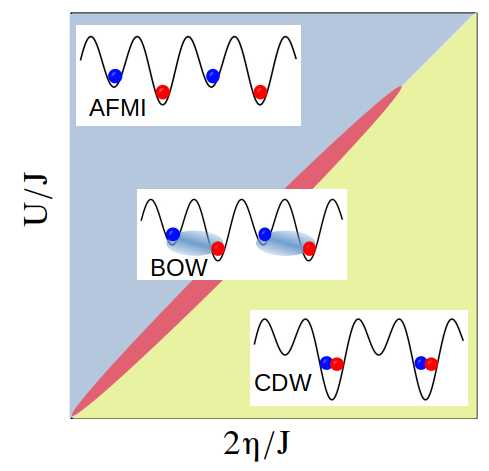}
    \caption[]{Sketch of the ground state phase diagram for the ionic Hubbard model at half filling.}
    \label{fig:phase_diagram_IH}
\end{figure}

\begin{figure}[h]
    \begin{tikzpicture}
    \node[anchor=north west,inner sep=0pt] at (0,0)
    {\includegraphics[width=0.45\textwidth,valign=c]{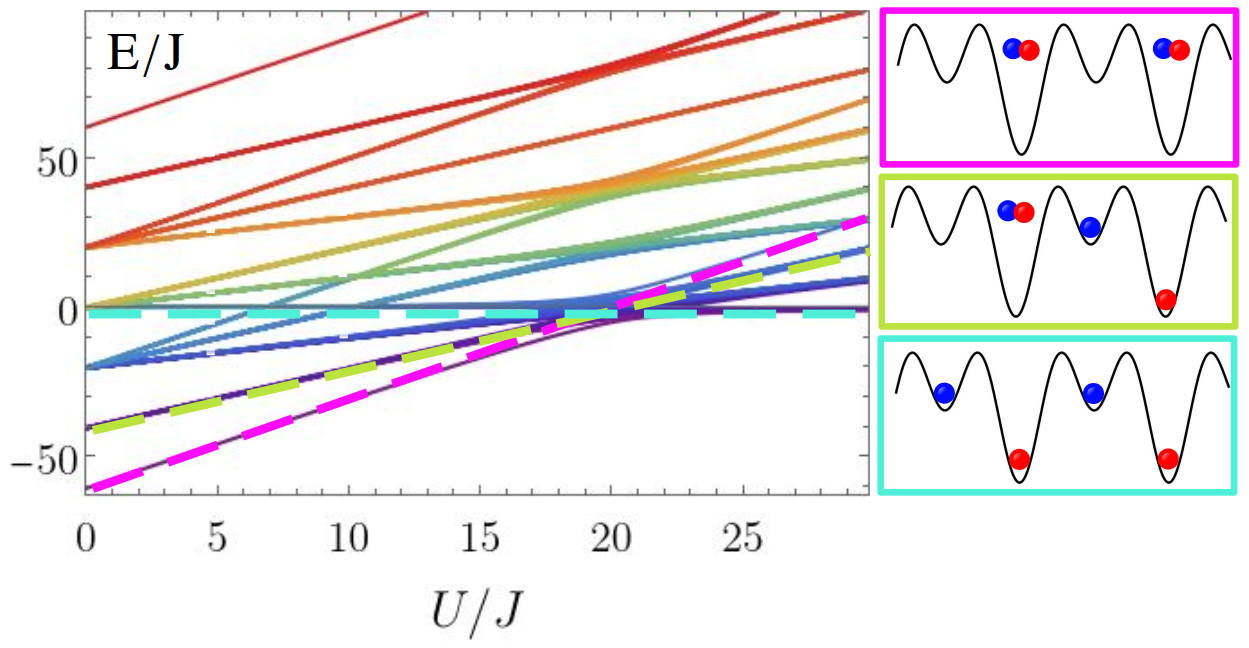}};
    \node[font=\normalfont] at (-1ex,-2ex) {(a)};
    \end{tikzpicture}
    \begin{tikzpicture}
    \node[anchor=north west,inner sep=0pt] at (0,0){\includegraphics[width=0.45\textwidth,valign=c]{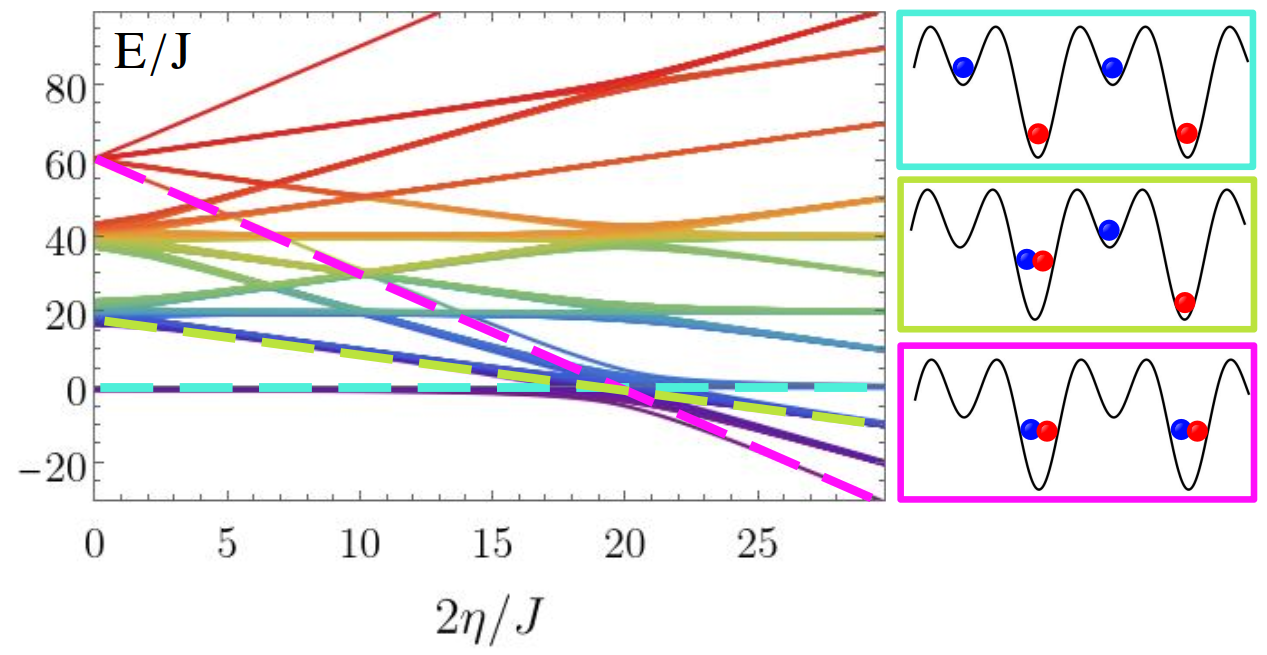}};
    \node[font=\normalfont] at (-1ex,-2ex) {(b)};
    \end{tikzpicture}
    \caption[]{Energy spectra of the ionic Hubbard model at half filling (solid lines). The parameters used are $L\!=\!6$, and in (a) $2\eta/J\!=\!20$, and in (b) $U/J\!=\!20$. We sketch the nature of the ground states and excited states crossing around $2\eta\!\sim\!U$ (colored boxes) and mark the respective bands with dashed lines in the same color.}
    \label{fig:Espectrum_IH_L6N6_U20_etavar}
\end{figure}

At \emph{half filling} ($n\!=\!1$) \cite{BagKrishnamurthy2015, ChattopadhyayGarg2019, KampfBrune2003, ManmanaSchonhammer2004} the ground state phase diagram of the model is very different from the low filling case (Fig.~\ref{fig:phase_diagram_IH}) due to commensurability effects. 
This distinct behavior can already be seen from the non-interacting case at finite ionic potential, where the lower quasi-particle band is completely filled for $n\!=\!1$, leading to a band insulator. At large $\eta$, the state can be described by a simplified picture where all lower potential sites are occupied by two particles, i.e.~forming a charge density wave. In Fig.~\ref{fig:Espectrum_IH_L6N6_U20_etavar}~(a) the excitation spectrum is plotted versus the interaction strength $U/J$. At $U/J\!=\!0$, the gap of the band insulator is seen which is proportional to $2\eta/J$.

In contrast, for vanishing ionic potential $\eta\!=\!0$ and at finite interaction strength, the system in one dimension is always in an antiferromagnetic Mott-insulating state (AFMI) (upper left in Fig.~\ref{fig:phase_diagram_IH}). In the large interaction limit a simple picture for this state is that every site is singly occupied with a charge gap $U/J$ as seen in Fig.~\ref{fig:Espectrum_IH_L6N6_U20_etavar}~(b). However, an entire band exists at low energies due to the gapless spin degrees of freedom. 
The Mott-insulating state survives also in the presence of the ionic potential (blue region in Fig.~\ref{fig:phase_diagram_IH}).
The finite charge gap becomes $\Delta_{c}\!\sim \!U\!-\!2\eta$ for $U\!-\!2\eta\gg J$, such that the higher value of the ionic potential $\eta/J$ the excitation comes down and crosses the ground state excitation [Fig.~\ref{fig:Espectrum_IH_L6N6_U20_etavar}~(b) around $2\eta/J\!\approx\!20$]. 
Far on the other side of the crossing the charge density wave (CDW) state dominates the low energy behavior (green region in Fig.~\ref{fig:phase_diagram_IH}). 
In the regime of this crossing of these two bands many more excitations are close to the ground state and a complex behavior is found.

In the intermediate regime $U\!\sim\!2\eta$, where the single particle band insulator gap becomes comparable to the Mott-gap, a narrow dimerized phase or bond-order wave (BOW) phase (red region in Fig.~\ref{fig:phase_diagram_IH}), characterized by a spontaneous dimerization of the kinetic energy, is predicted \cite{FabrizioNersesyan1999, ZhangLin2003, BatistaAligia2004, LoidaKollath2017}. It is bounded by two continuous transitions: a Kosterlitz-Thouless spin transition at $U_\text{cr,2}$ and a charge transition with vanishing charge gap at $U_\text{cr,1}\!<\!U_\text{cr,2}$. The dimerized phase vanishes at a multicritical point for very large $U/J,~2\eta/J$.
Due to the presence of these many different excitations, we expect the physics in this region to be very sensitive to a finite temperature.

\section{Results\label{sec:results}}

In this section, we present our results on the steady state phase diagram of the atoms-cavity coupled system. The transversely pumped coupled atoms-cavity open system exhibits a self-ordering transition \cite{RitschEsslinger2013, MivehvarRitsch2021}, which has been studied widely for bosonic and fermionic atoms \cite{AsbothVukics2005, LarsonLewenstein2008, LarsonLewenstein2008b, NiedenzuRitsch2011, PiazzaStrack2014, KeelingSimons2014, ChenZhai2014}. 
Our results investigate the self-organization transition for the considered system of interacting fermions coupled to the optical cavity. We analyze in detail the case of quarter filling (Sec.~\ref{sec:quarter_filling}), complementing our work presented in Ref.~\cite{TolleHalati2024}. Whereas the numerical results shown are obtained at $n\!=\!1/2$, our analytical findings can be generalized to lower fillings $n\!\leq\!1/2$.
Additionally, we present and analyze results at half filling $n\!=\!1$ (Sec.~\ref{sec:half_filling}). Fundamentally different behaviors are expected for the two fillings.

We begin the discussion (see Sec.~\ref{sec:quarter_filling}) by focusing on the regime around the self-ordering transition at $g\!\sim\! g_\text{cr}$ at quarter filling. We contrast the $T\!=\!0$ MF results to the approach which includes fluctuations in the atoms-cavity coupling (Sec.~\ref{sec:self_organization_transition}). 
Analytical approximations for the phase transition are derived in several limits. Then we emphasize features originating from the different cavity-cooling (Sec.~\ref{sec:cavity_cooling}) and many-body cooling mechanisms (Sec.~\ref{sec:many_body_cooling}), that are only possible due to the presence of fluctuations. 
At quarter filling, this includes the fluctuations-induced bistability (Sec.~\ref{sec:fluctuation_induced_bistability}), which has been the focus of Ref.~\cite{TolleHalati2024}. 
In Sec.~\ref{sec:self_organization_transition_half_filling}, we turn to the bistability occurring at half filling. In this case the self-organization transition is coupled with the onset of a bistabilitym which is already visible for the $T\!=\!0$ MF results.

We compare our results from solving of Eq.~(\ref{eq:cav_MF}) and Eq.~(\ref{eq:energy_transfer}) with exact diagonalization (ED) for finite systems in one dimension to those obtained in the weak-tunneling perturbation theory in the thermodynamic limit, Eqs.~(\ref{eq:system_of_equations_1D}). The latter is also applicable for higher dimensions of the atomic lattice.

\subsection{Low filling}

In this section, we present numerical data obtained at quarter filling ($n\!=\!1/2)$ and show analytical results that are derived for the more general case of low filling.
\label{sec:quarter_filling}

\subsubsection{Self-organization phase transition\label{sec:self_organization_transition}}

\begin{figure}[!hbtp]
    \begin{tikzpicture}
    \node[anchor=north west,inner sep=0pt] at (0,0){\includegraphics[width=0.22\textwidth]{Fig5a}};
    \node[font=\normalfont] at (1ex,-2ex) {(a)};
    \end{tikzpicture}
    \begin{tikzpicture}
    \node[anchor=north west,inner sep=0pt] at (0,0){\includegraphics[width=0.22\textwidth]{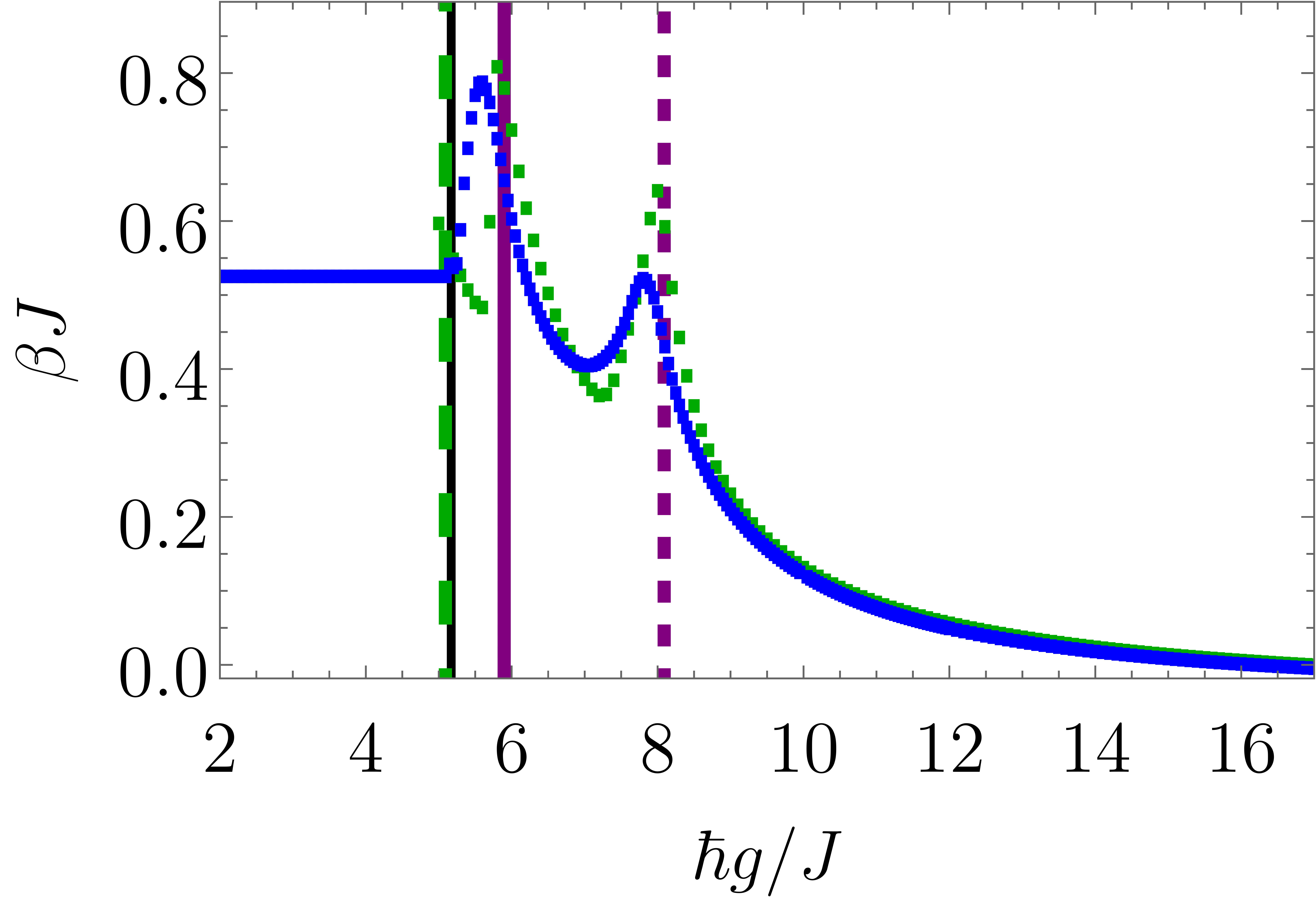}};
    \node[font=\normalfont] at (1ex,-2ex) {(b)};
    \end{tikzpicture}
    \begin{tikzpicture}
    \node[anchor=north west,inner sep=0pt] at (0,0){\includegraphics[width=0.22\textwidth]{Fig5c}};
    \node[font=\normalfont] at (1ex,-2ex) {(c)};
    \end{tikzpicture}
    \begin{tikzpicture}
    \node[anchor=north west,inner sep=0pt] at (0,0){\includegraphics[width=0.22\textwidth]{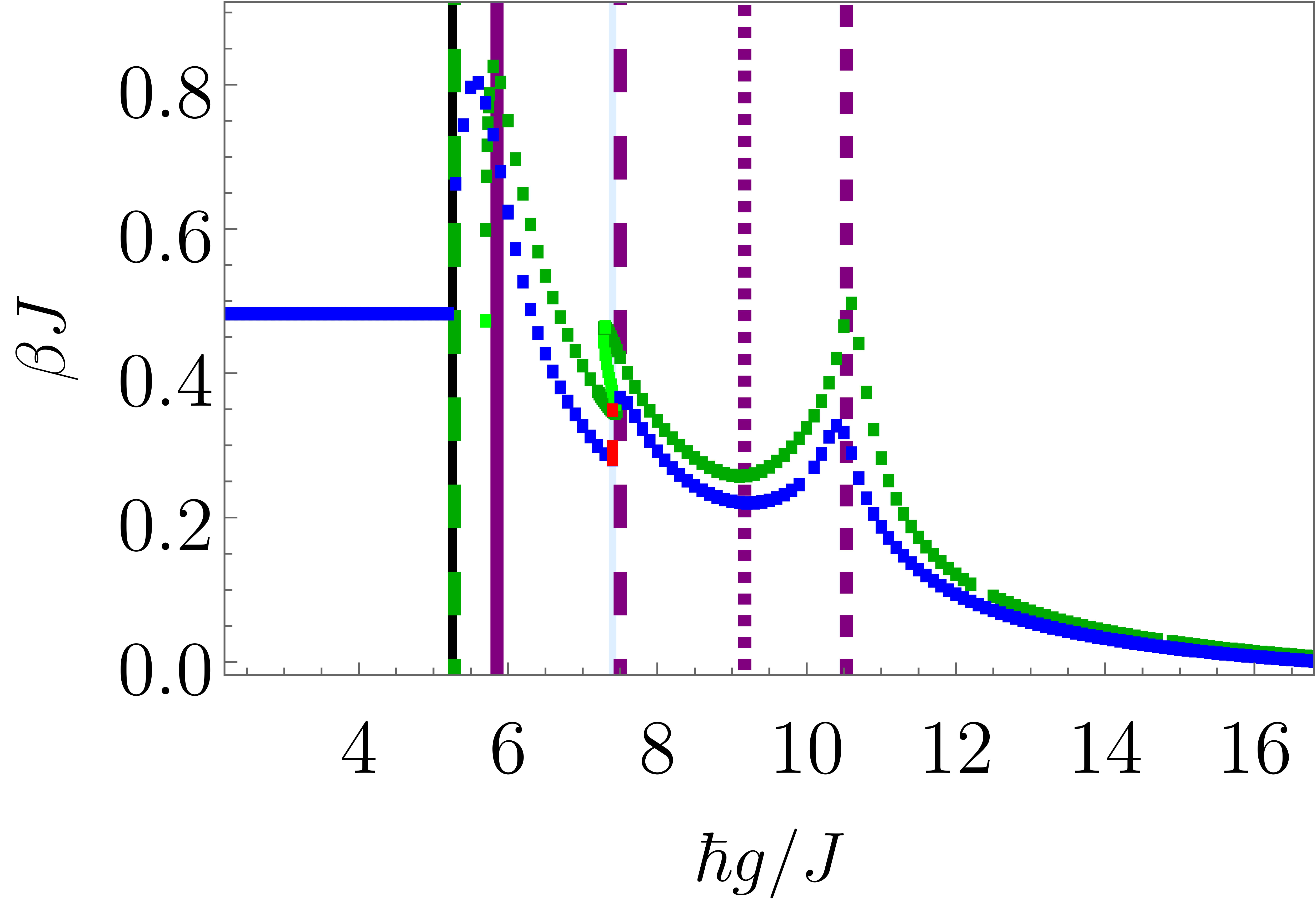}};
    \node[font=\normalfont] at (1ex,-2ex) {(d)};
    \end{tikzpicture}
    \includegraphics[width=0.48\textwidth]{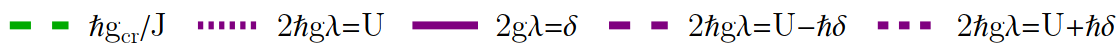}
    \vspace{-20pt}
    \caption[]{Cuts of the (a),~(c) sublattice imbalance $\langle\hat{\Delta}\rangle/L^d$ and (b),~(d) inverse temperature $\beta J$ as a function of atoms-cavity coupling $\hbar g/J$ of a finite size system $L\!=\!8$ for stable (blue) and unstable (red) results. The $T\!=\!0$ MF solution is shown as black solid line. We also show stable (dark green) and unstable (light green) results in the weak tunneling perturbation in the thermodynamic limit at quarter filling. The parameters used are $\hbar\delta/J\!=\!8$ and $\hbar\varGamma/J\!=\!1$ and in (a),~(b) at $U/J\!=\!8$ and in (c),~(d) at $U/J\!=\!20$. Vertical lines denote the critical coupling $g_\text{cr}/J$ and resonances as explained in the legend. The orange dashed line in (a) is the approximate scaling for $U/J\!\to\!\infty$ [Eqs.~(\ref{eq:imbalance_universal_scaling_Uinf})].
    }
    \label{fig:cuts_U8_gvar_delta8_Gamma1_and_U20_gvar_delta8_Gamma1}
\end{figure}

\begin{figure}[!hbtp]
\begin{flushleft}
    \begin{tikzpicture}
    \node[anchor=north west,inner sep=0pt] at (0,0){\includegraphics[height=0.18\textheight]{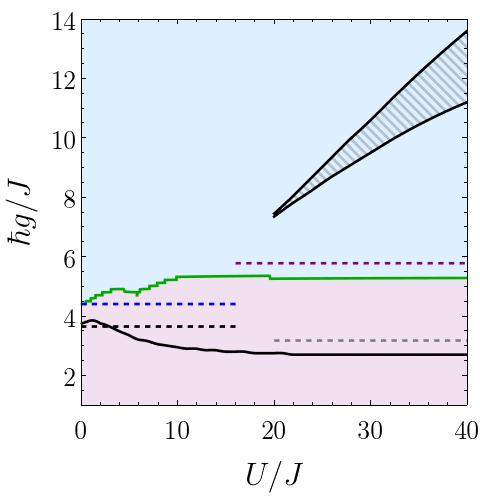}};
    \node[font=\normalfont] at (1ex,-2ex) {(a)};
    \end{tikzpicture}
    \begin{tikzpicture}
    \node[anchor=north west,inner sep=0pt] at (0,0){ \includegraphics[height=0.18\textheight]{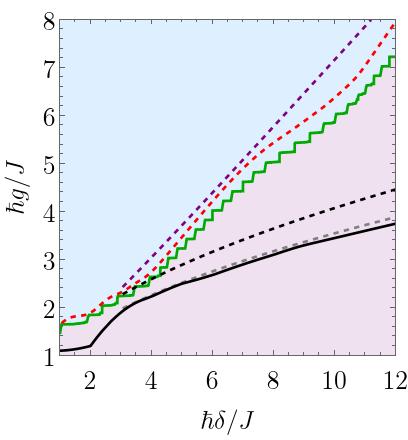}};
    \node[font=\normalfont] at (1ex,-2ex) {(b)};
    \end{tikzpicture}
    \begin{tikzpicture}
    \node[anchor=north west,inner sep=0pt] at (0,0){\includegraphics[height=0.18\textheight]{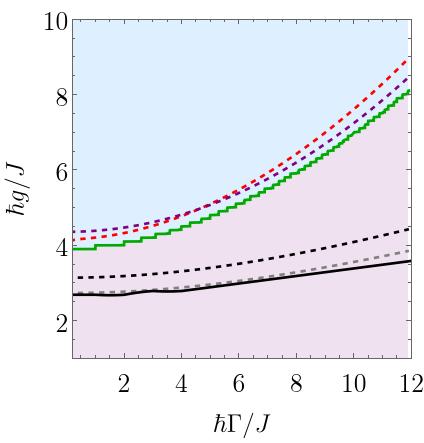} };
    \node[font=\normalfont] at (1ex,-2ex) {(c)};
    \end{tikzpicture}
    \end{flushleft}
    \vspace{-20pt}
    \caption[]{Simplified phase diagrams varying the atoms-cavity coupling $\hbar g/J$ and (a) the on-site interaction $U/J$ at $\hbar\delta/J\!=\!8$, $\hbar\varGamma/J\!=\!1$; (b) the pump-cavity detuning $\hbar\delta/J$ at $U/J\!=\!8$, $\hbar\varGamma/J\!=\!1$; (c) the dissipation rate $\hbar\varGamma/J$ at $\hbar\delta/J\!=\!6$, $U/J\!=\!8$, for a finite size system $L\!=\!8$ at quarter filling. The green line marks the transition between normal (purple region) and self-organized phase (blue region) at $\hbar g_\text{cr}/J$ determined by solving Eqs.~(\ref{eq:cav_MF})~\&~(\ref{eq:energy_transfer}) with ED. The red dashed line in (b),~(c) is calculated from Eq.~(\ref{eq:gcrMF}) with Eq.~(\ref{eq:F_AE_discrete}) using the self-consistently determined $\beta$ in the limit $U/J\!\to\!\infty$. For the purple dashed lines the large temperature limit Eq.~(\ref{eq:gcr_approx_highT}) is used. The blue dashed line in (a) is the equivalent result for $U/J\!=\!0$. The solid black line marks the critical coupling $\hbar g^\text{MF}_\text{cr}$ obtained for the $T\!=\!0$ MF approach solving Eq.~(\ref{eq:cav_MF}) via ED. We compare to the black (gray) dashed line, the approximate analytical result for $g_\text{cr}^\text{MF}$ [Eq.~(\ref{eq:gcrMF}) with Eq.~(\ref{eq:F_MF_discrete})] for $U/J\!=\!0$ ($U/J\!\to\!\infty$). The small hatched triangular region on the center right marks the fluctuation-induced bistability region within the self-organized phase.
 }
    \label{fig:gcr_plots_gvdelta_gvGamma}
\end{figure}

We begin by analyzing the self-organization transition of the atoms-cavity system. 
In Fig.~\ref{fig:cuts_U8_gvar_delta8_Gamma1_and_U20_gvar_delta8_Gamma1} we show both the $T\!=\!0$ MF results (black line) and also the results obtained taking the fluctuations into account (many-body adiabatic elimination method derived in Sec.~\ref{sec:mbae}) for two different interaction strength $U/J\!=8~\text{and}~20$, respectively. The blue/red symbols are the results for exact diagonalization and green symbols are from the weak tunneling perturbation in the thermodynamic limit (Sec.~\ref{sec:pertubation_J}). 
At low atoms-cavity coupling strength $g$, the system is in a normal state characterized by a vanishing photon number and vanishing atomic sublattice density imbalance [Fig.~\ref{fig:cuts_U8_gvar_delta8_Gamma1_and_U20_gvar_delta8_Gamma1}~(a),(c)]. As the atoms-cavity coupling $g$ is increased above $g_\text{cr}$ the system enters into the self-ordered phase, signaled by a simultaneous increase of the cavity field (not shown) and the sublattice imbalance, see for example the behavior of $\langle\hat{\Delta}\rangle/L^d$ around $\hbar g_\text{cr}/J\!\approx\!3$ for the $T\!=\!0$ MF result and around $\hbar g_\text{cr}/J\!\approx\!5.3$ taking the fluctuations into account for the parameters considered in Fig.~\ref{fig:cuts_U8_gvar_delta8_Gamma1_and_U20_gvar_delta8_Gamma1}~(a).
In both approaches, the self-ordering transition is present. However, the exact location and the nature of the transition can be very distinct as we discuss in more detail below.

The differences between the two approaches arise due to the influence of fluctuations in the atoms-cavity coupling. On the $T\!=\!0$ MF level the steady state is a pure state and the self-organization transition is a pure state transition. In contrast, the atomic state in the many-body adiabatic elimination that considers fluctuations is given by a thermal state. 
Even in the normal phase where the cavity field vanishes, the atoms still couple to the cavity field fluctuations, leading to an effective non-zero temperature. Let us note that in the normal phase regime the effective atomic Hamiltonian, Eq.~(\ref{eq:Heff_atoms}), reduces in one dimension to the integrable Fermi-Hubbard model, where the thermal state Ansatz, Eq.~(\ref{eq:rho_T}), has to be taken with care. However, we believe that the overall behavior is not changed and focus mainly on the self-organized phase where the Ansatz is justified. 

For the parameter regimes explored in this work, the temperature $k_BT\!=\!1/\beta$ has quite large values compared to the excitation energies between the lowest eigenstates of the effective model. For example, in Fig.~\ref{fig:cuts_U8_gvar_delta8_Gamma1_and_U20_gvar_delta8_Gamma1}~(b),(d) we observe that the effective temperature has values larger than $k_BT/J\!\approx\!1.25$ for both interactions strengths close to the transition. Thus, typically excited states contribute significantly to the steady state. This has as consequence that, even though the transition to the self-organized states is also captured by the $T\!=\!0$ MF approach \cite{RitschEsslinger2013, MivehvarRitsch2021}, the critical coupling $g^\text{MF}_\text{cr}$ is considerably lower compared to the approach including fluctuations for all parameters we consider. We also underline the fundamentally different nature of the transition. In the $T\!=\!0$ MF method the atoms are described by a pure state, whereas in the approach described in Sec.~\ref{sec:mbae} the atomic state is a mixed state with a self-consistent effective finite temperature determined by the fluctuations. These results are consistent with a similar comparison performed for interacting bosonic atoms coupled to a dissipative cavity in Ref.~\cite{BezvershenkoRosch2021}.

The effective inverse temperature $\beta$ is constant but finite in the normal phase and exhibits a steep increase above the transition followed by a maximum and subsequent decrease around the purple vertical line that marks the   atomic resonance $2g\lambda\!=\!\delta$ shown in Fig.~\ref{fig:cuts_U8_gvar_delta8_Gamma1_and_U20_gvar_delta8_Gamma1}~(b),(d). The resonance typically occurs close to the maximum in the density imbalance. We analyze the origin of the maxima in the effective temperature in Sec.~\ref{sec:cavity_cooling} and focus here to the behavior close to the transition. We observe that beside the different threshold as we increase the coupling strength above the transition, the increase of the sublattice imbalance is much more abrupt for the self-consistent finite temperature approach. The thermal fluctuations push the coupling strength needed for reaching the self-ordered phase to higher values. 

We depict this behavior in Fig.~\ref{fig:gcr_plots_gvdelta_gvGamma}, where we show a simplified state diagram. The emphasis in this diagram is laid on the transition values $g_\text{cr}$ for the finite temperature approach see green solid curve vs the $T\!=\!0$ MF results shown as black solid curve.
We extracted $g_\text{cr}$ from our numerical results by determining the $g$ at which the cavity order parameter becomes finite, i.e. more than the threshold $\lambda\!=\!10^{-6}$. 
The normal phase obtained by the finite temperature results is shown as a purple region in Fig.~\ref{fig:gcr_plots_gvdelta_gvGamma} and characterized by a vanishing photon number and vanishing atomic sublattice density imbalance. The corresponding self-organized phase is marked with blue. 

The different behavior of the approach with and without the fluctuations is clear in the dependence of the self-organization threshold with the on-site interactions in Fig.~\ref{fig:gcr_plots_gvdelta_gvGamma}~(a). Interestingly, in the $T\!=\!0$ MF approach the threshold decreases to lower coupling strengths with increasing $U/J$. In contrast for the finite temperature approach, we find that at small values of the interaction strength $U/J$ the critical coupling moves slightly to larger values $g_\text{cr}$. However, when the atomic interactions are large, here $U/J\!\gtrsim\!10$, in both approaches the critical coupling strength $g_\text{cr}$ depends only very weakly on $U/J$ [see green line in Fig.~\ref{fig:gcr_plots_gvdelta_gvGamma}~(a)].
This can be understood, since at low filling and high interaction strength the steady state and the lowest excited states have mainly singly occupied sites [see sketch Fig.~\ref{fig:Espectrum_IH_L8N4_U20_etavar}~(a) for $U/J\!>\!20$]. Thus, the behavior around the transition is dominated by the ground state and low lying excitations in which doubly occupied sites are rare.

To understand better the observed behavior, we derive analytical expressions for the critical coupling in the solvable non-interacting limit ($U/J\!=\!0$).
Similar approaches at the $T\!=\!0$ MF level were derived for non-interacting semiclassical particles without external optical lattice \cite{AsbothVukics2005, NiedenzuRitsch2011} as well as bosons \cite{KirtonDallaTorre2019, NagyDomokos2008}. 
Assuming that $\lambda$ is continuous at the transition, one can compute an approximate expression for the critical coupling $g_\text{cr}$, by searching for which value of the coupling the implicit equation can be solved to obtain a small non-zero value of $\lambda$ while using expansions for small $\lambda$. The assumption of a continuous $\lambda$ is valid for incommensurate values of the filling. 
The derivation of the critical coupling $g_\text{cr}$ is given in Appendix~\ref{app:gcr_approximations} for a one-dimensional system. The obtained expression is given by
\begin{equation}
\label{eq:gcrMF}
    g_\text{cr}=\sqrt{\frac{J\pi\big[\delta^2+(\varGamma/2)^2\big]}{\hbar\delta\text{F}}}.
\end{equation}
The equivalent expressions for higher dimensions can be obtained in an analogous way.
The function $\text{F}$ depends on the approximation used. We first describe in Eq.~(\ref{eq:F_MF_Linf}) and  Eq.~(\ref{eq:F_MF_discrete}) the results obtained for the $T\!=\!0$ MF approach and in Eq.~(\ref{eq:F_AE_discrete}) the results obtained with the approach including fluctuations around the mean-field coupling. For the $T\!=\!0$ MF approach the function $\text{F}$ only depends on the filling of the atomic system. In the thermodynamic limit $F\!=\!\text{F}^{\text{MF},L\to\infty}\!\left(\left.n \pi/2\right|1\right)$ where
\begin{subequations}
\begin{equation}
    \label{eq:F_MF_Linf}
    \text{F}^{\text{MF},L\to\infty}(\phi|m)\!=\!\int_0^{\phi}[1\!-\!m\sin^2(\theta )]^{-1/2}d\theta
\end{equation}
is the elliptic integral of the first kind. For a finite size non-interacting system we find
\begin{equation}
    \label{eq:F_MF_discrete}
    \text{F}^{\text{MF},L}\!=\!\frac{J\pi}{L}\sum_{k}\!\frac{1}{E(k,\lambda\!=\!0)},
\end{equation}
where the dispersion defined in Eq.~(\ref{eq:Ek}) is given by
\begin{equation*}
    \label{eq:Ek_selfconsistent}
    E(k,\lambda)=\sqrt{\epsilon(k)^2\!+\!(2\hbar g\lambda)^2}.
\end{equation*}
For the finite system the momentum sum has to be taken over all occupied momenta [black dashed line in Fig.~\ref{fig:gcr_plots_gvdelta_gvGamma}~(a)]. This approximation agrees very well with the numerical results obtained at low interaction strength using the $T\!=\!0$ MF approximation (black line in Fig.~\ref{fig:gcr_plots_gvdelta_gvGamma}).

We can make use of the derived expression to also obtain an approximation at $U/J\!\to\!\infty$. Here we assume that due to the infinite repulsion the ground state is approximated by non-interacting, but spinless, fermions with double the filling. 
For example at quarter filling, the spinless fermions would completely fill the lower band [gray dashed line  in Fig.~\ref{fig:gcr_plots_gvdelta_gvGamma}~(a)]. The numerical results at large interaction strength only slightly deviate from the results obtained within this approximation. 

Close to the transition we further obtain the scaling of $\lambda$ (see Appendix~\ref{app:gcr_approximations}) given by
\begin{equation}
    \lambda \propto J \big(g^\text{MF}_\text{cr}\big)^{3/2}\sqrt{g-g^\text{MF}_\text{cr}}.
\end{equation}
This scaling agrees with the one derived for the Dicke model $\alpha\!\propto\!(g'-g'_\text{cr})^{1/2}$ \cite{KirtonDallaTorre2019}  

with $g'$ the respective coupling strength and $\alpha\!=\!\langle\hat{a}\rangle$ the cavity order parameter. 
The mean-field Dicke scaling exponent $1/2$ is the same with and without photon losses.
The self-ordering transition of an ensemble of non-interacting semiclassical particles \cite{AsbothVukics2005, NiedenzuRitsch2011} shows the same scaling in a mean-field approach. However, in Ginzburg-Landau type theories a change in the scaling to $1/4$ at a tricritical point is derived for low pump-cavity detuning \cite{KeelingSimons2014}.

For the finite temperature approach the function $F$ becomes more involved and is given by
\begin{align}
\label{eq:F_AE_discrete}
\text{F}^{\text{fluc},L}\!&=\! \frac{J\pi}{L} \sum_{k}\!\frac{1}{E(k,\lambda\!=\!0)}\\&\times\Big[\frac{1}{e^{\beta(E(k,\lambda\!=\!0)-\mu)}\!+\!1}\!+\!\frac{1}{e^{\beta(-E(k,\lambda\!=\!0)-\mu)}\!+\!1}\Big],\nonumber
\end{align}
\end{subequations}
which depends on the self-consistently evaluated $\beta$ (see Appendix~\ref{app:gcr_approximations}). Therefore, the function $\text{F}$ in this case cannot be easily approximated by an analytical expression. Its evaluation is shown as the red dashed line in Fig.~\ref{fig:gcr_plots_gvdelta_gvGamma}~(b)-(c) for the infinitely strongly interacting gas.

A much simpler analytical expression can be found in the limit of large dissipation and high temperature, where 
\begin{equation}
    \label{eq:gcr_approx_highT}
   g_\text{cr}(\beta\!\ll\!1)\propto\frac{\delta^2\!+\!(\varGamma/2)^2}{\delta}.
\end{equation}
The obtained critical coupling is shown as blue dashed line in Fig.~\ref{fig:gcr_plots_gvdelta_gvGamma}~(a) for the non-interacting gas and as the purple dashed line in Fig.~\ref{fig:gcr_plots_gvdelta_gvGamma}~(a)-(c) for the infinitely strongly interacting gas, using the spinless fermion approximation described above. 

Thus, including the self-consistent effective temperature leads to a different critical coupling, the scaling of which we will discuss below. A similar result was obtained for bosonic atoms \cite{BezvershenkoRosch2021}. 

We now compare our analytical findings to what we observe in the phase diagram in more detail varying either the pump-cavity detuning $\delta$ or the dissipation rate $\varGamma$ in Fig.~\ref{fig:gcr_plots_gvdelta_gvGamma}~(b),(c). Even though the expressions for the critical coupling are only valid in the non-interacting limit and approximately for the limit $U/J\!\to\! \infty$, we find a good agreement of the overall dependence on $\delta$ and $\varGamma$ at $U/J\!=\!8$. In particular, the drastically different increase with both $\delta$ and $\varGamma$ is well approximated by the analytical expressions.
The scaling in the $T\!=\!0$ MF result is for $\delta\!\gg\!\varGamma$ simply $\hbar g_\text{cr}^\text{MF}/J\!\sim\!\sqrt{\hbar\delta/J}$, while in the presence of fluctuations it becomes linear in this regime, i.e.~$g_\text{cr}\!\sim\!\delta$ [Fig.~\ref{fig:gcr_plots_gvdelta_gvGamma}~(b)]. 

We attribute the increase of the critical couplings with the detuning $\delta$ to the effect that the effective long-range cavity-mediated coupling decreases as the pump gets further detuned. Thus, to reach the same effective atoms-cavity coupling, a larger amplitude of the pump field is required. Simultaneously for large detuning increasing $\delta$ also leads to an increased effective temperature around the self-ordering transition, which further contributes to the shift of the transition towards higher couplings explaining the much faster increase of the critical coupling if the fluctuations are taken into account.

When increasing the dissipation strength $\hbar \varGamma/J$, the critical pump strength $\hbar g_\text{cr}^\text{MF}/J$ increases approximately linearly for the $T\!=\!0$ MF approach [Eq.~(\ref{eq:gcrMF}) with Eq.~(\ref{eq:F_MF_discrete})] and approximately quadratically for the finite temperature approach [Fig.~\ref{fig:gcr_plots_gvdelta_gvGamma}~(c)]. Increasing $\varGamma$ also leads to an increased effective temperature, explaining the much faster growth of $g_\text{cr}$. Thus, we have shown that the dependence of the critical coupling marking the self-ordering transition changes drastically if fluctuations are taken into account.

\subsubsection{Fluctuation-effects in the ordered phase}
\label{sec:cavity_cooling}

In this section, we discuss the behavior of the steady states in the self-organized phase, focusing on the effects stemming from taking into account the fluctuations in the atoms-cavity coupling. 

\begin{figure}[!hbtp]
    \begin{tikzpicture}
	\node[anchor=north west,inner sep=0pt] at (0,0){\includegraphics[width=0.23\textwidth]{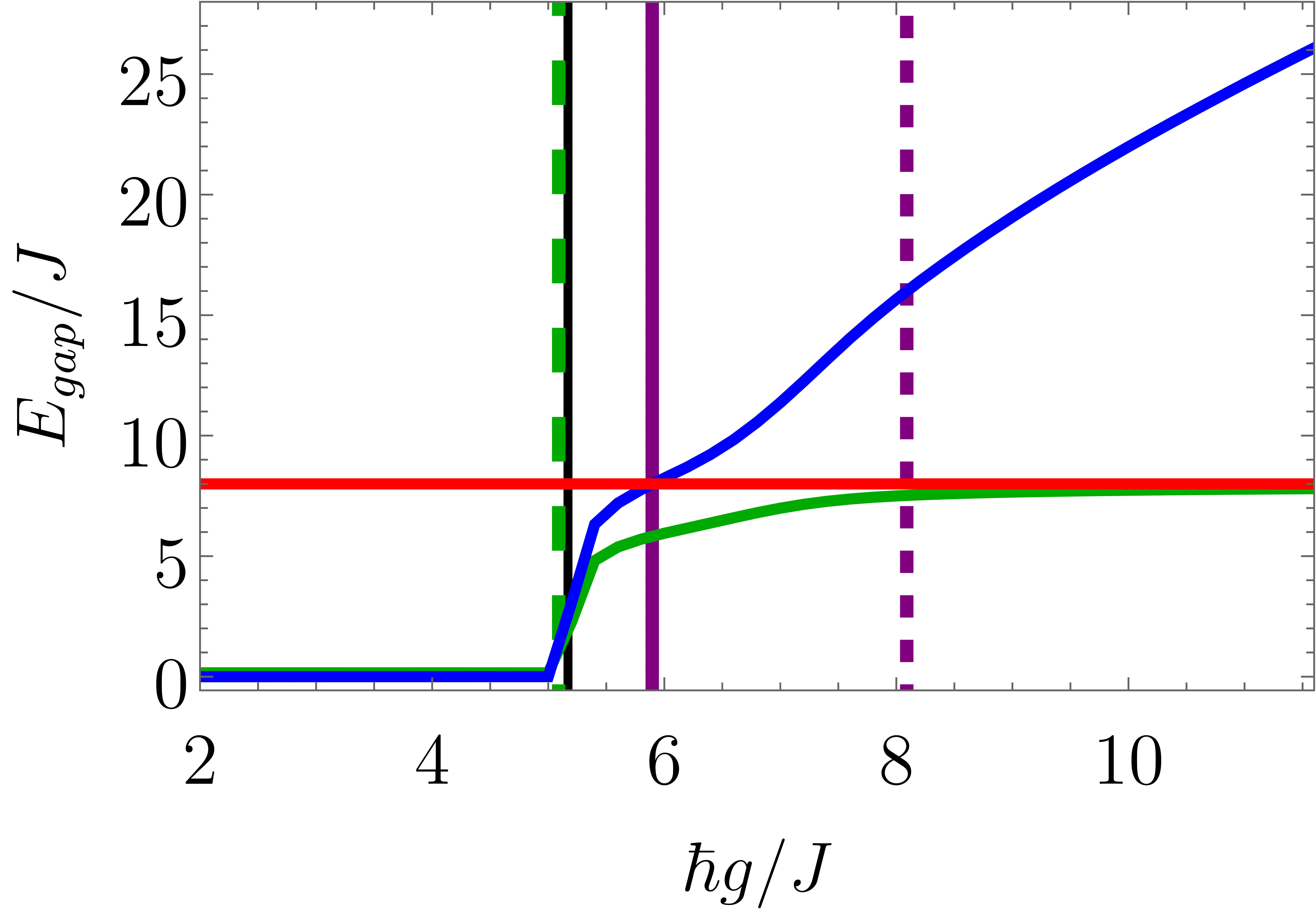}};
	\node[font=\normalfont] at (1.5ex,0ex) {(a)};
    \end{tikzpicture}
    \begin{tikzpicture}
	\node[anchor=north west,inner sep=0pt] at (0,0){\includegraphics[width=0.23\textwidth]{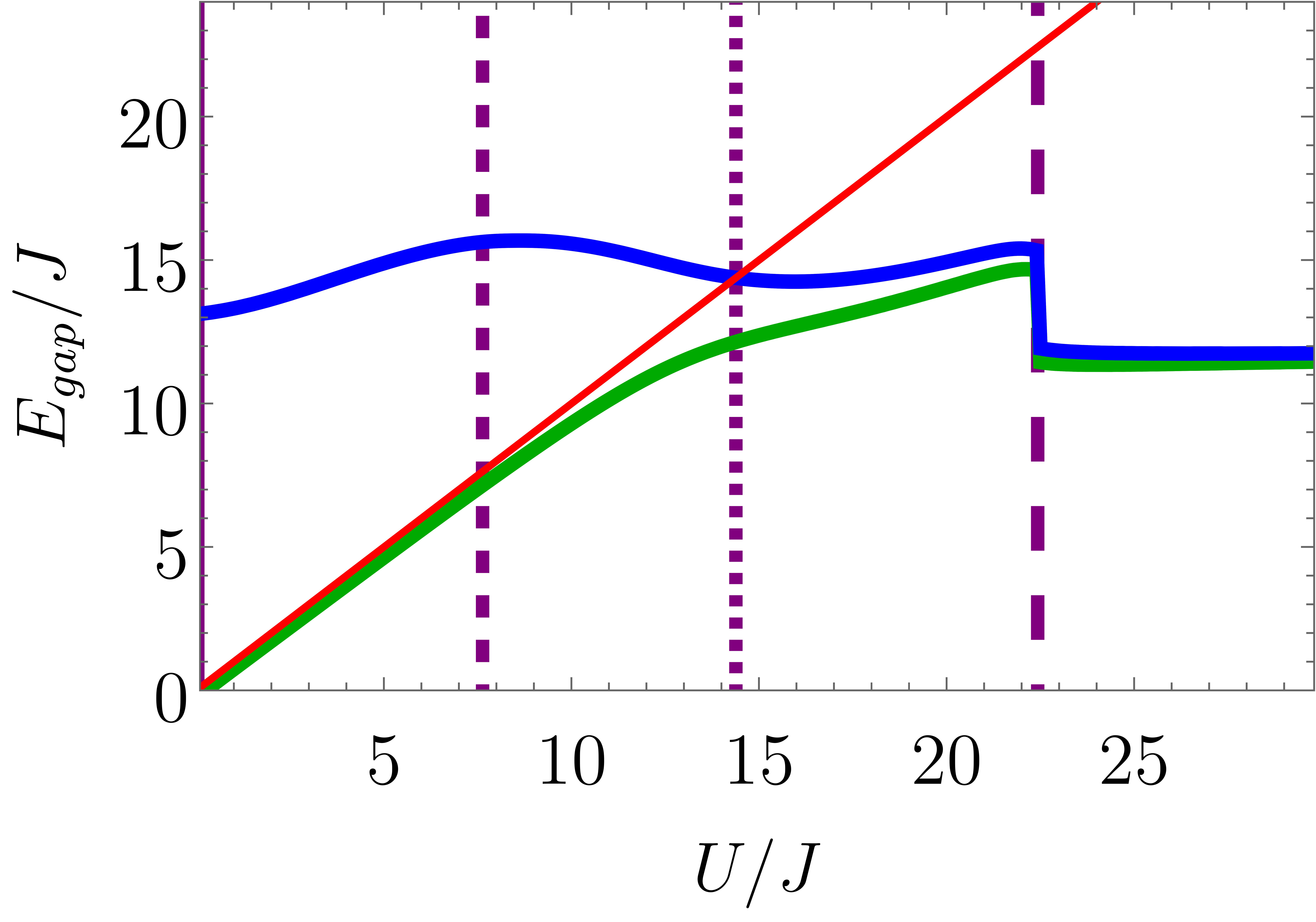}};
	\node[font=\normalfont] at (1.5ex,0ex) {(b)};
    \end{tikzpicture}
    \includegraphics[width=0.48\textwidth]{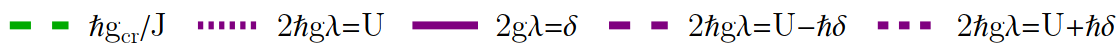}
    \vspace{-10pt}
    \caption[]{Energy excitation gap (green solid curve) between eigenstates defined by $\langle\hat{\Delta}\rangle\!=\!N$ and $\sum_j\langle\hat{n}_{j\uparrow}\hat{n}_{j\downarrow}\rangle\!=\!0$ and the lowest energy states differing in either of the quantities 
    in the eigen-spectrum of the effective atomic model $\hat{H}_\text{eff}$. We show the excitation gap of a finite size system $L\!=\!8$ at quarter filling in (a) as a function of atoms-cavity coupling $\hbar g/J$ at $U/J\!=\!8$ and in (b) as a function of the on-site interaction strength $U/J$ at $\hbar g/J\!=\!8$. The other parameters used are $\hbar\delta/J\!=\!8$, $\hbar\varGamma/J\!=\!1$. The effective sublattice potential imbalance $2\hbar g\lambda/J$ is plotted in blue, the on-site interaction $U/J$ in red. Vertical lines are explained in the legend.}
    \label{fig:cut_Egap_U8_gvar_delta8_Gamma1}
\end{figure}

As in our results the excited states of the effective model $\hat{H}_\text{eff}$ are present in the steady state density matrix, it is important to understand the nature of the excited states and the behavior of the gaps.
We investigate the excitation gap from the ground state in the energy spectrum $E_\text{gap}$ that is obtained from the eigenspectrum of $\hat{H}_\text{eff}$. At low filling, in the atomic limit we obtain an approximate band structure where all eigenstates in the lowest energy band are defined by maximal sublattice density imbalance $\langle\hat{\Delta}\rangle\!=\!N$ and exclusively singly occupied sites $\sum_j\langle\hat{n}_{j\uparrow}\hat{n}_{j\downarrow}\rangle\!=\!0$.
We calculate the excitation gap between the lowest energy band and the lowest energy eigenstate with differing $\langle\hat{\Delta}\rangle$ or $\sum_j\langle\hat{n}_{j\uparrow}\hat{n}_{j\downarrow}\rangle$.
In the parameter regime considered here the finite gap between states with equal density imbalance and number of double occupancies, occurring from the lifting of the degeneracy by the finite size effects and higher order hoppings, can be omitted. As shown in Fig.~\ref{fig:cut_Egap_U8_gvar_delta8_Gamma1}, the excitation gap (green curve) increases above the phase transition and approaches $U/J$ for $2\hbar g\lambda\!\gg\! U$. Thus, the nature of the excitation gap changes from band-insulator to Mott-insulator on the low-potential sublattice sites.

\begin{figure}[!hbtp]
\begin{flushleft}
\begin{tikzpicture}
    \node[anchor=north west,inner sep=0pt] at (0,0){\includegraphics[width=0.235\textwidth]{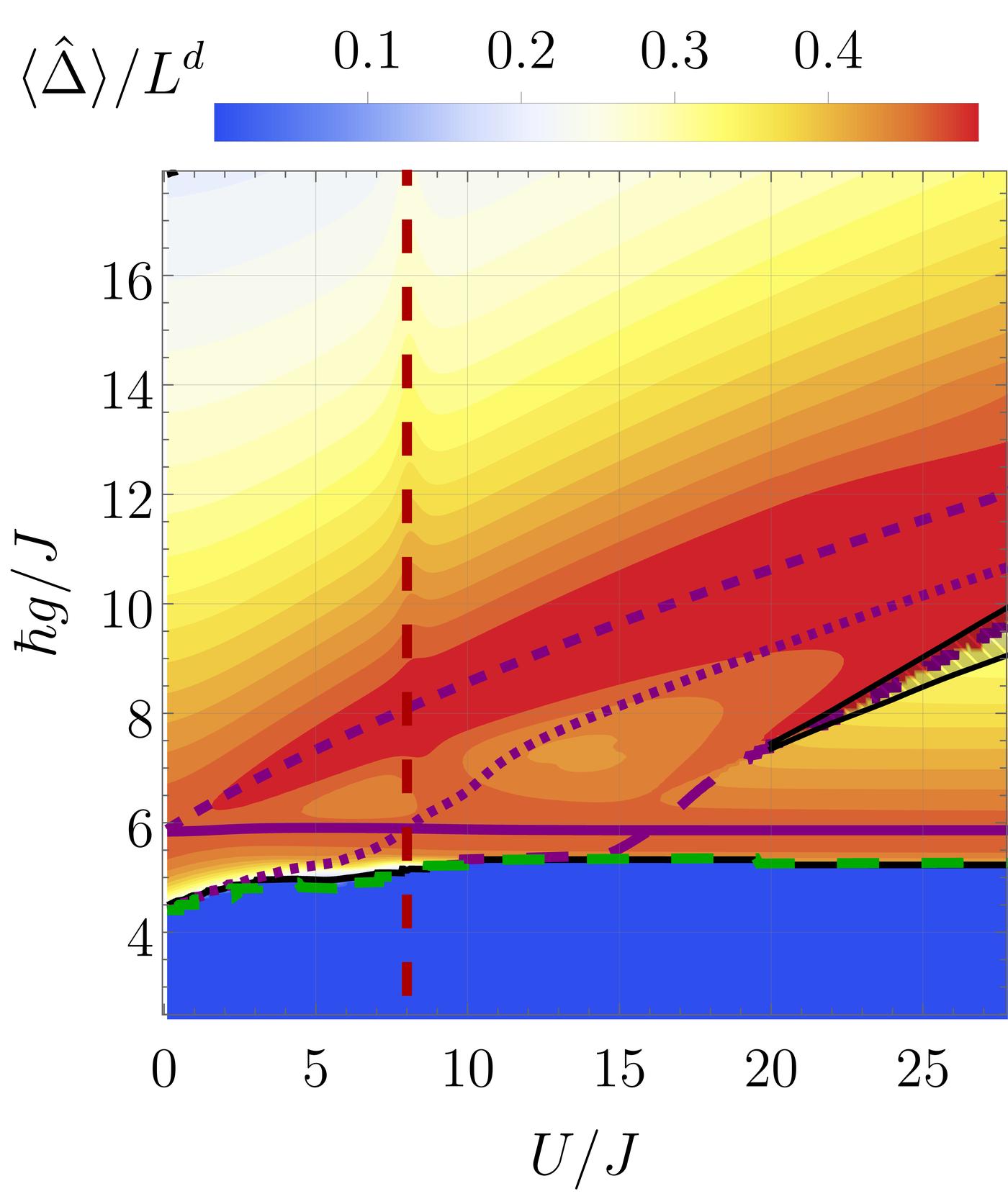}};
    \node[font=\normalfont] at (2ex,-6ex) {(a)};
    \end{tikzpicture}
    \begin{tikzpicture}
    \node[anchor=north west,inner sep=0pt] at (0,0){\includegraphics[width=0.235\textwidth]{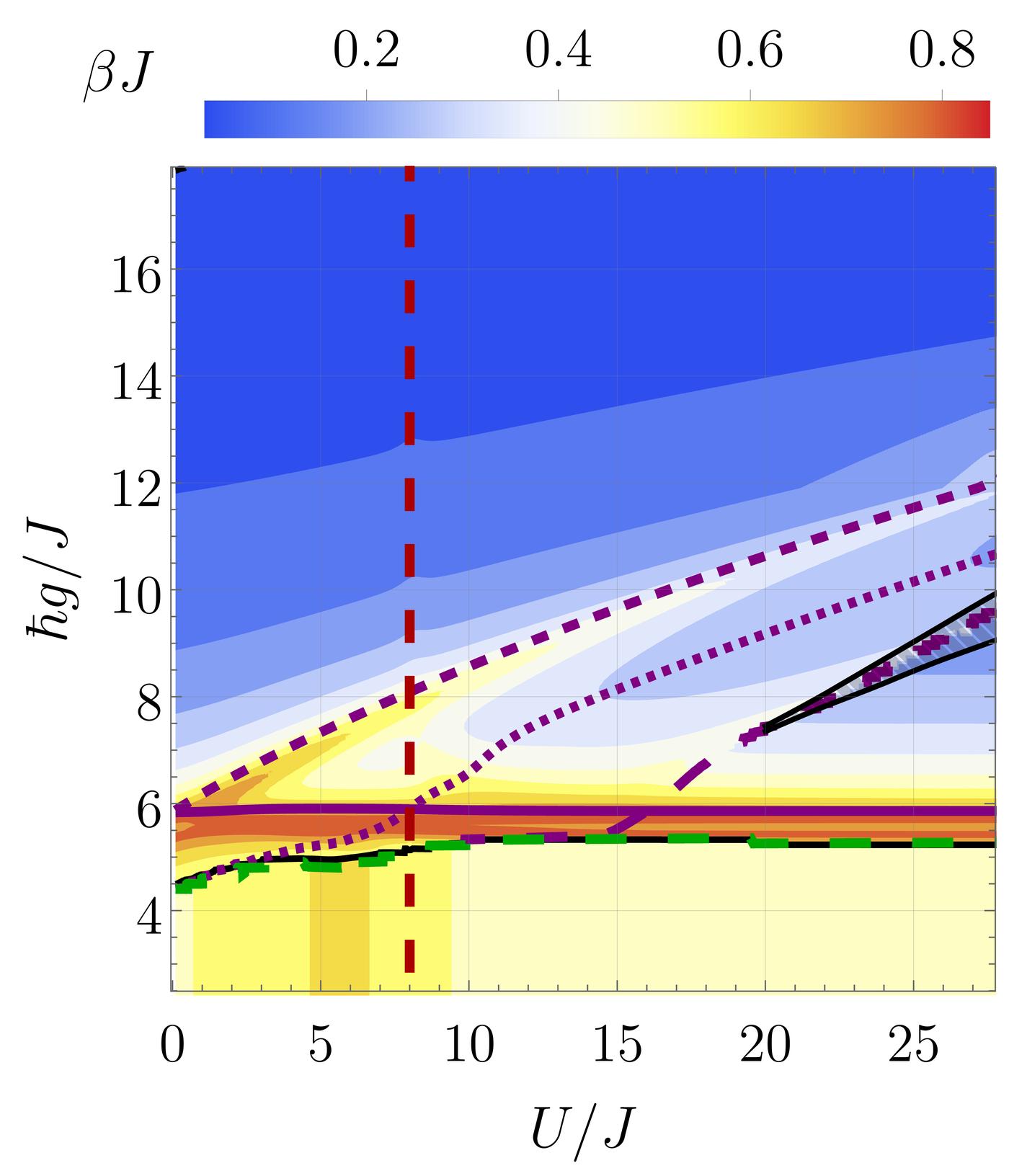}};
    \node[font=\normalfont] at (2ex,-6ex) {(b)};
    \end{tikzpicture}
    \includegraphics[width=0.32\textwidth]{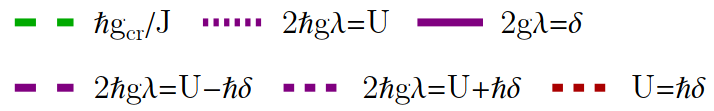}
    \vspace{-10pt}
    \caption[]{(a) Sublattice imbalance $\langle\hat{\Delta}\rangle/L^d$ and (b) inverse temperature $\beta J$ as a function of atoms-cavity coupling $\hbar g/J$ and on-site interaction $U/J$ of a finite size system $L\!=\!8$ at quarter filling. The parameters used are $\hbar\delta/J\!=\!8$, $\hbar\varGamma/J\!=\!1$. The small hatched triangular region on the center right marks the fluctuation-induced bistability region within the self-organized phase. Lines denote the critical coupling $\hbar g_\text{cr}/J$ and resonances as explained in the legend.
   }
    \label{fig:density_plots_gvU_delta8_Gamma1}
    \end{flushleft}
\end{figure}

\begin{figure}[!hbtp]
    \begin{tikzpicture}
    \node[anchor=north west,inner sep=0pt] at (0,0){\includegraphics[width=0.235\textwidth]{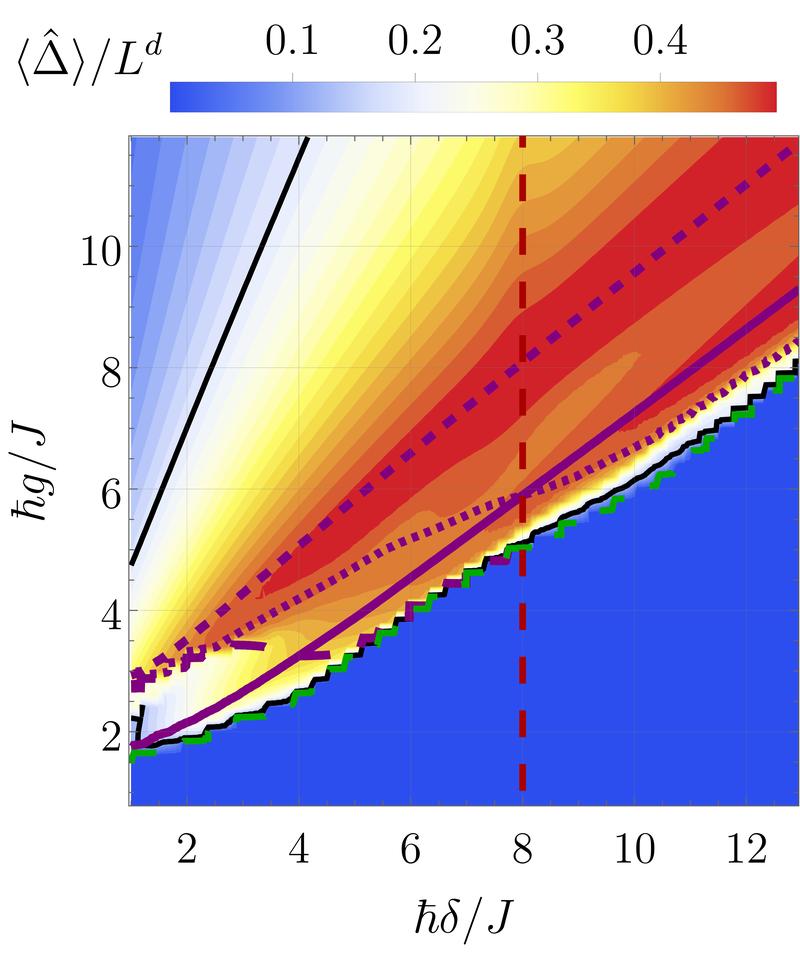}};
    \node[font=\normalfont] at (2ex,-6ex) {(a)};
    \end{tikzpicture}
    \begin{tikzpicture}
    \node[anchor=north west,inner sep=0pt] at (0,0){\includegraphics[width=0.235\textwidth]{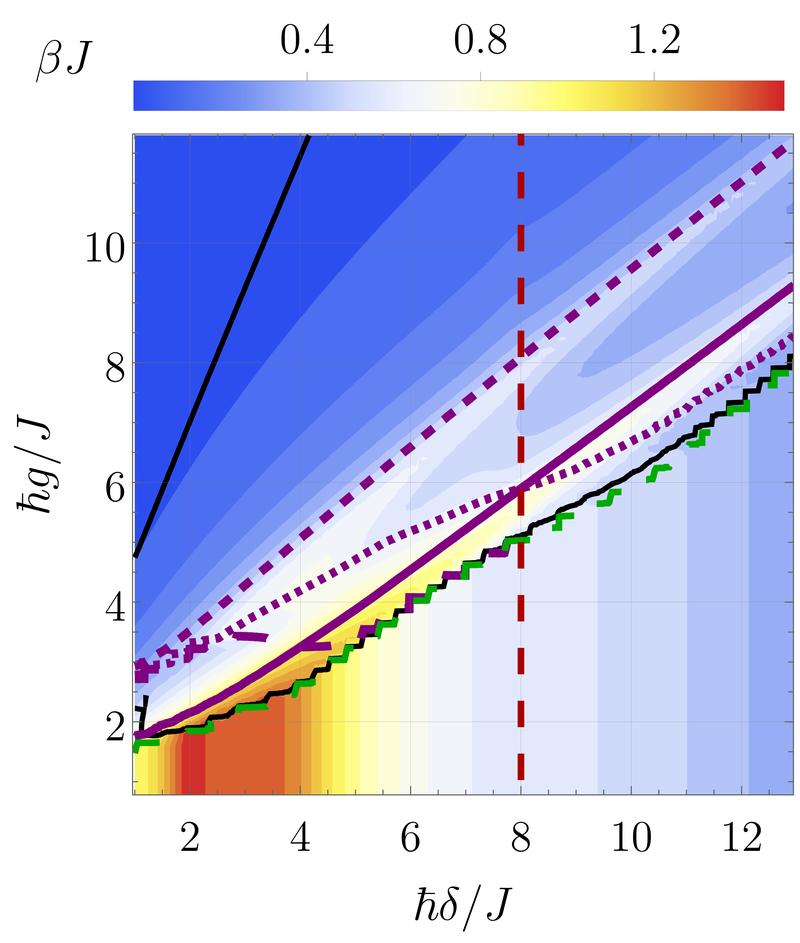}};
    \node[font=\normalfont] at (2ex,-6ex) {(b)};
    \end{tikzpicture}
    \includegraphics[width=0.38\textwidth]{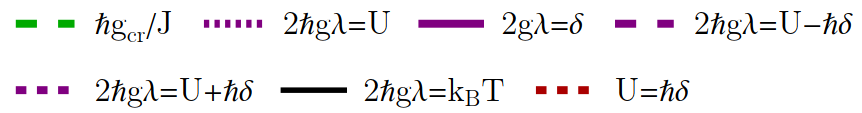}
    \vspace{-10pt}
    \caption[]{(a) Sublattice imbalance $\langle\hat{\Delta}\rangle/L$ and (b) inverse temperature $\beta J$ as a function of atoms-cavity coupling $\hbar g/J$ and detuning $\hbar\delta/J$ of a finite size system $L\!=\!8$ at quarter filling. The parameters used are $U/J\!=\!8$ and $\hbar\varGamma/J\!=\!1$. Lines denote the critical coupling $\hbar g_\text{cr}/J$ and resonances in the atomic limit.}
    \label{fig:density_plots_gvdelta_U8_G1}
\end{figure}

By increasing the coupling strength above the $g_\text{cr}$ we observe that the sublattice imbalance $\langle\hat{\Delta}\rangle$ and the effective temperature have a non-monotonic behavior. The imbalance shows an increase above $g_\text{cr}$ to a maximum between $\hbar g/J\!\approx\!5$ to $6$ in Fig.~\ref{fig:cuts_U8_gvar_delta8_Gamma1_and_U20_gvar_delta8_Gamma1}~(a),(c). At even higher coupling strength one or two further maxima are seen (which will be analyzed further in Sec.~\ref{sec:many_body_cooling}), followed by a decrease for larger couplings. Also the inverse temperature $\beta$ shows a strong increase to a maximum followed by one or two additional maxima [Fig.~\ref{fig:cuts_U8_gvar_delta8_Gamma1_and_U20_gvar_delta8_Gamma1}~(b),(d)] on an overall decrease. 
These maxima corresponds to cooling effects, and a local minimum in temperature on the overall heating. 
The atoms are cooled by the cavity in this regime. We find that the first minimum in temperature occurs approximately when the photon energy matches the height of the self-organized potential, $2g\lambda\!\approx\!\delta$ (purple line in Fig.~\ref{fig:cuts_U8_gvar_delta8_Gamma1_and_U20_gvar_delta8_Gamma1}). The position of the first maximum in the imbalance and the corresponding minimum in the effective temperature can be very well followed in Fig.~\ref{fig:density_plots_gvU_delta8_Gamma1} and Fig.~\ref{fig:density_plots_gvdelta_U8_G1}, where we show the dependence on the different system parameters. In both figures the first maximum/minimum for the imbalance/inverse temperature follows approximately the purple line indicating the condition $2g\lambda\!\approx\!\delta$.

\begin{figure}[!hbtp]
    \includegraphics[width=0.4\textwidth]{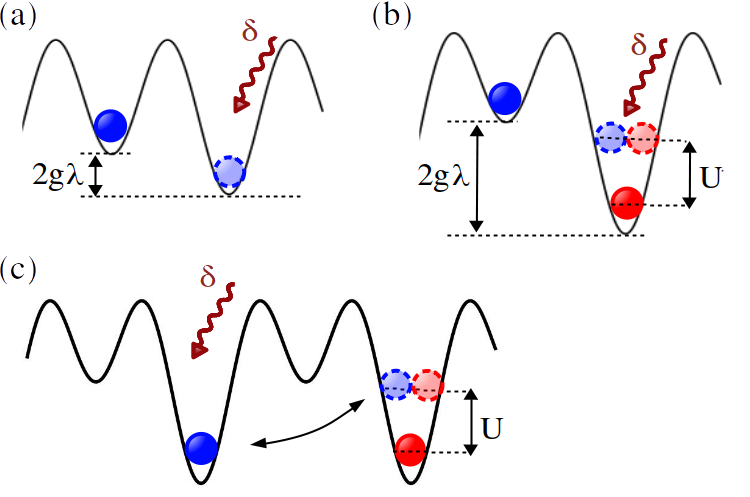} 
    \caption[]{Sketches of the energy transfer transitions captured by Eq.~(\ref{eq:EOM_1D}).~(a) First line in Eq.~(\ref{eq:EOM_1D}): hopping from a high- to an empty neighboring low potential sublattice site, adding a photon to/removing a photon from the cavity field $\Delta E_{nm}\!=\!\pm(2\hbar g\lambda\!\pm\!\hbar\delta)$.~(b) Second line in Eq.~(\ref{eq:EOM_1D}) with $p\!=\!1$: hopping from a high- to a neighboring low potential sublattice site, creating a double occupancy and adding/removing a photon to the cavity field $\Delta E_{nm}\!=\!\pm(2\hbar g\lambda\!-\!U\!\pm\!\hbar\delta)$.~(c) Next neighbor hopping process resonant at $\Delta E_{nm}\!=\!\pm(U\!-\!\hbar\delta)$}
    \label{fig:sketch_transition_EOM}
\end{figure}

\begin{figure}[!hbtp]
    \begin{tikzpicture}
    \node[anchor=north west,inner sep=0pt] at (0,0){\includegraphics[width=0.225\textwidth]{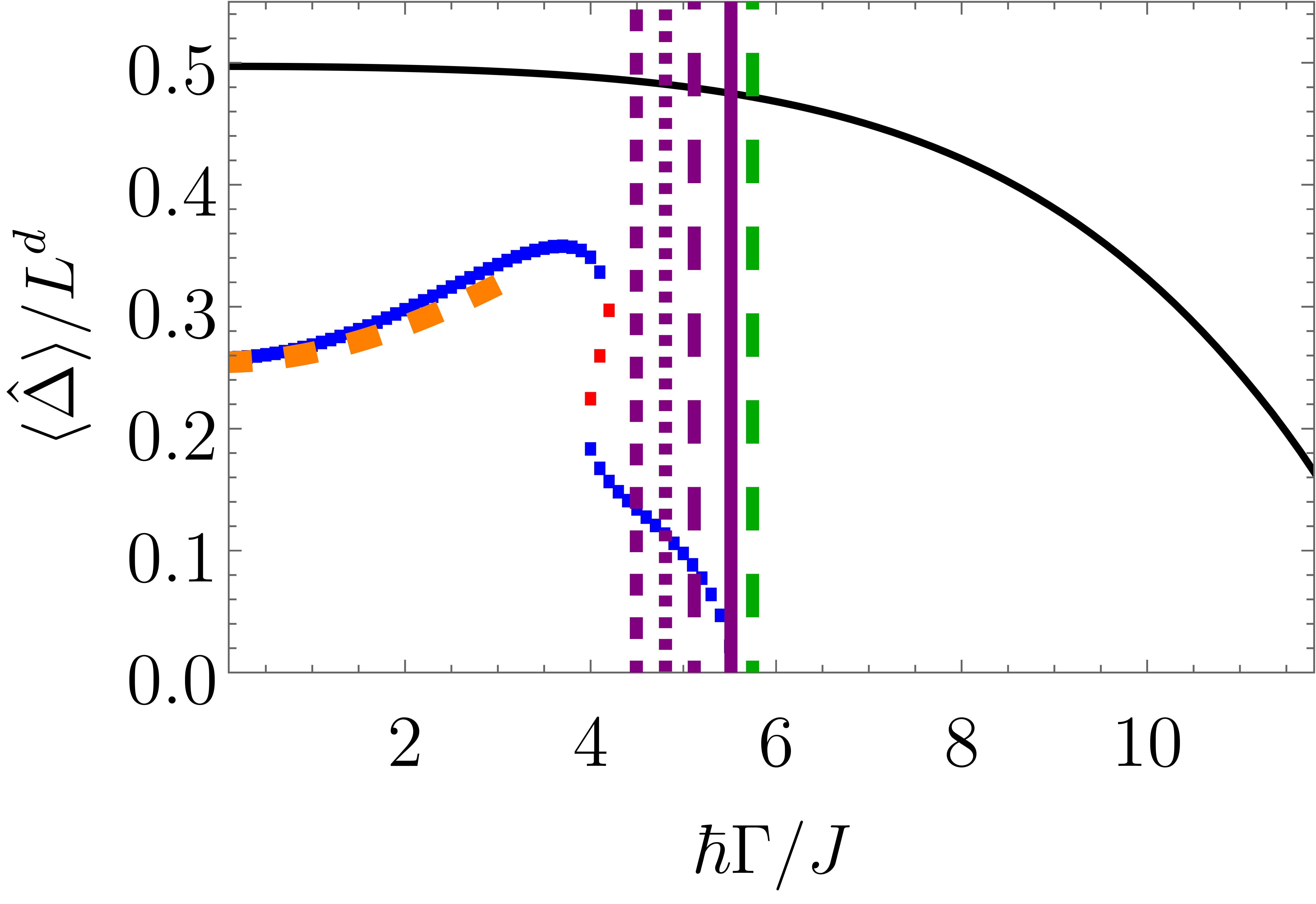}};
    \node[font=\normalfont] at (1.5ex,-1ex) {(a)};
    \end{tikzpicture}
    \begin{tikzpicture}
    \node[anchor=north west,inner sep=0pt] at (0,0){\includegraphics[width=0.225\textwidth]{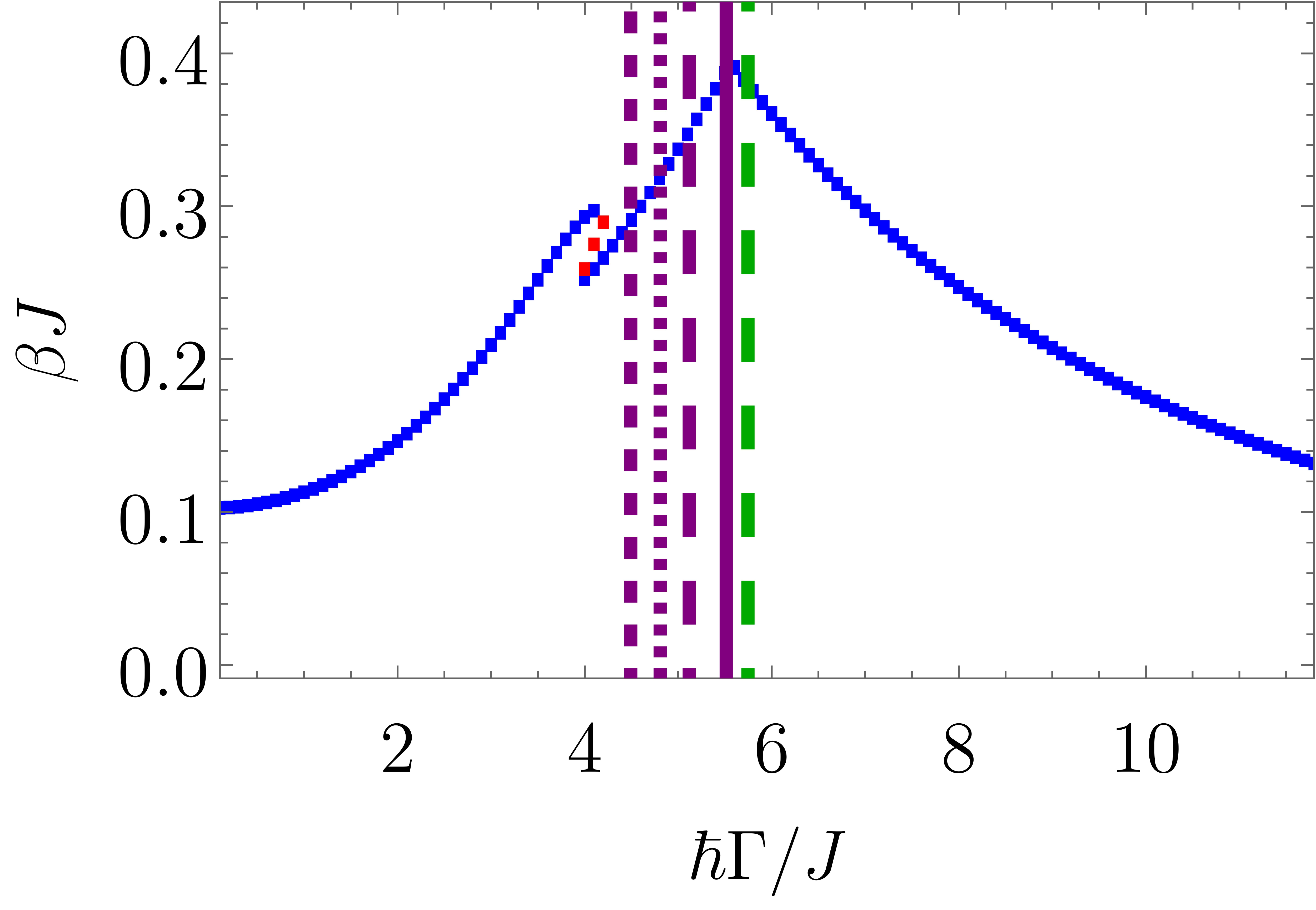}};
    \node[font=\normalfont] at (1ex,-1ex) {(b)};
    \end{tikzpicture}
    \includegraphics[width=0.48\textwidth]{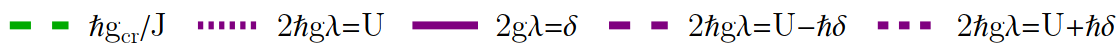}
    \vspace{-10pt}
    \caption[]{The (a)~sublattice imbalance $\langle\hat{\Delta}\rangle/L^d$ and (b)~inverse temperature $\beta J$ as a function of $\hbar\varGamma/J$ of a finite size system $L\!=\!8$ at quarter filling for stable (blue) and unstable (red) results. The $T\!=\!0$ MF solution is shown as black solid line. The parameters are $\hbar g/J\!=\!5$, $U/J\!=\!8$ and $\hbar\delta/J\!=\!2$. Vertical lines denote the critical coupling $\hbar g_\text{cr}/J$ and resonances. The orange dashed line is the approximate scaling for $U/J\!\to\!\infty$ [Eqs.~(\ref{eq:imbalance_universal_scaling_Uinf})].}
    \label{fig:cuts_U8_g8_delta2_Gammavar_and_U8_gvar_delta2_Gamma1}
\end{figure}

\begin{figure}[!hbtp]
\begin{tikzpicture}
    \node[anchor=north west,inner sep=0pt] at (0,0){\includegraphics[width=0.235\textwidth]{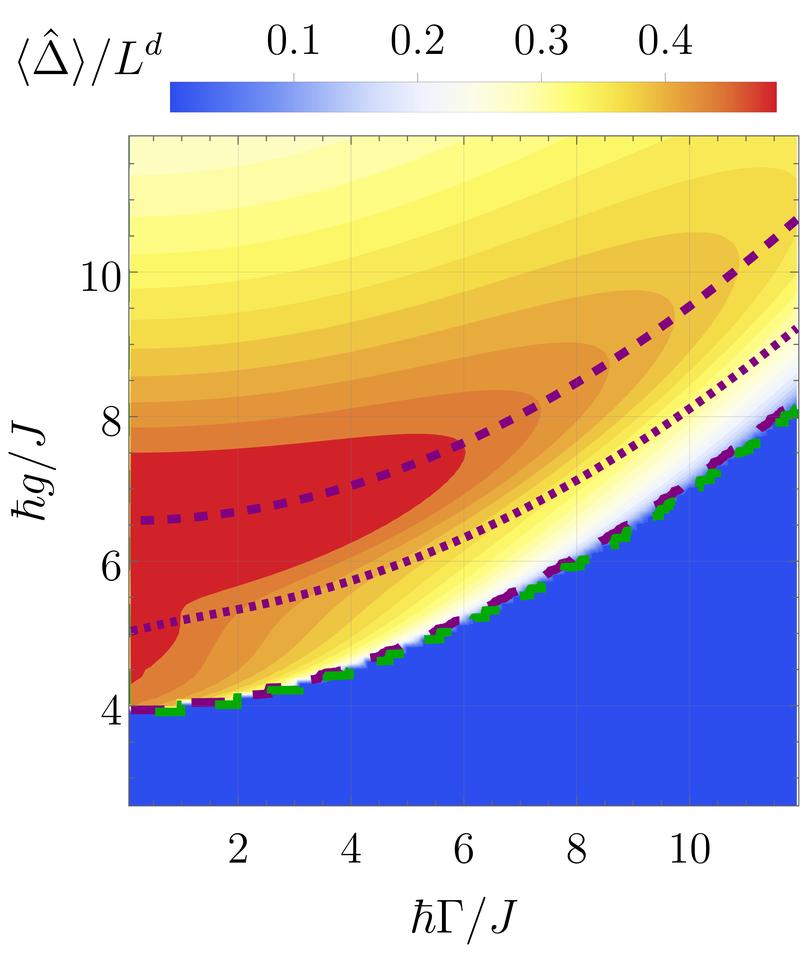}};
    \node[font=\normalfont] at (2ex,-6ex) {(a)};
    \end{tikzpicture}
    \begin{tikzpicture}
    \node[anchor=north west,inner sep=0pt] at (0,0){\includegraphics[width=0.235\textwidth]{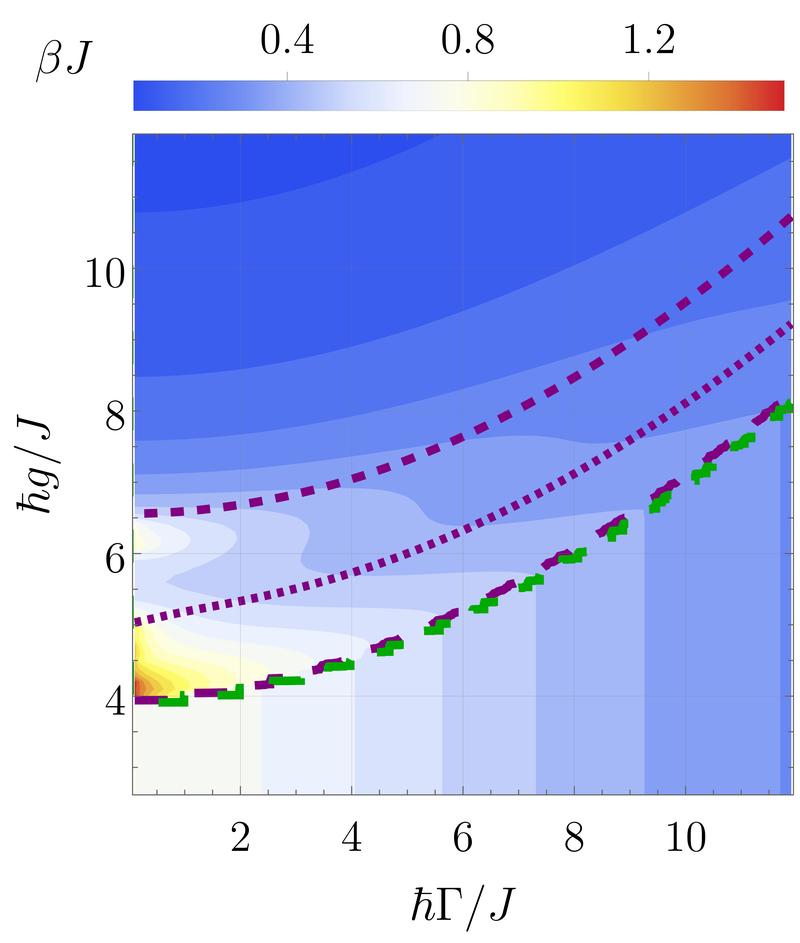}};
    \node[font=\normalfont] at (2ex,-6ex) {(b)};
    \end{tikzpicture}
    \includegraphics[width=0.42\textwidth]{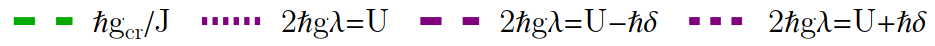}
    \vspace{-10pt}
    \caption[]{(a) Sublattice imbalance $\langle\hat{\Delta}\rangle/L^d$ and (b) inverse temperature $\beta J$ as a function of atoms-cavity coupling $\hbar g/J$ and dissipation $\hbar\varGamma/J$ of a finite size system $L\!=\!8$ at quarter filling. The parameters used are $U/J\!=\!8$, $\hbar\delta/J\!=\!6$. Lines denote the critical coupling $\hbar g_\text{cr}/J$ and resonances in the atomic limit.}
    \label{fig:density_plots_gvGamma_U8_delta6}
\end{figure}

\begin{figure}[!hbtp]
\begin{flushleft}
    \begin{tikzpicture}
    \node[anchor=north west,inner sep=0pt] at (0,0){\includegraphics[width=0.23\textwidth]{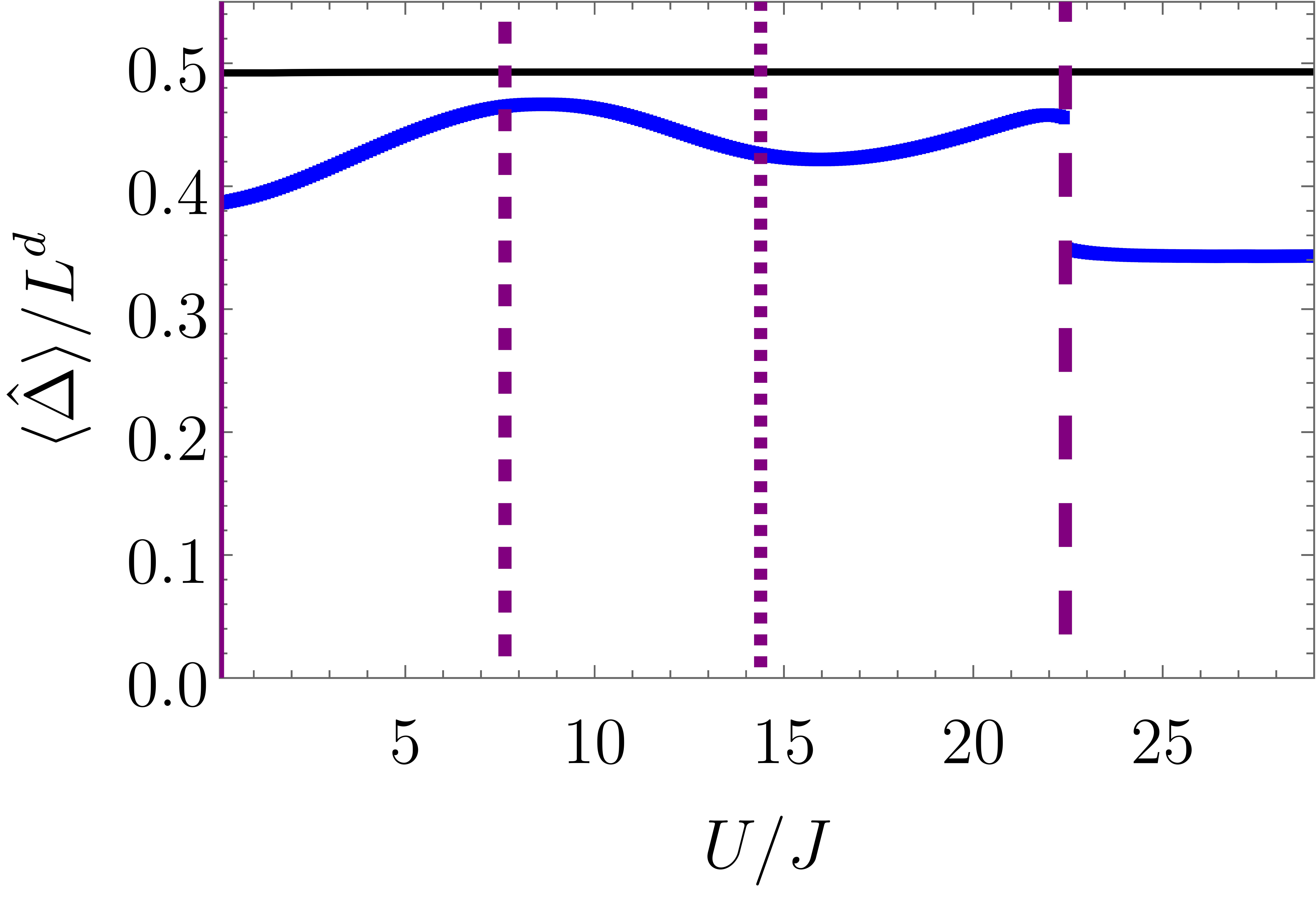}};
    \node[font=\normalfont] at (1.5ex,-1ex) {(a)};
    \end{tikzpicture}
    \begin{tikzpicture}
    \node[anchor=north west,inner sep=0pt] at (0,0){\includegraphics[width=0.23\textwidth]{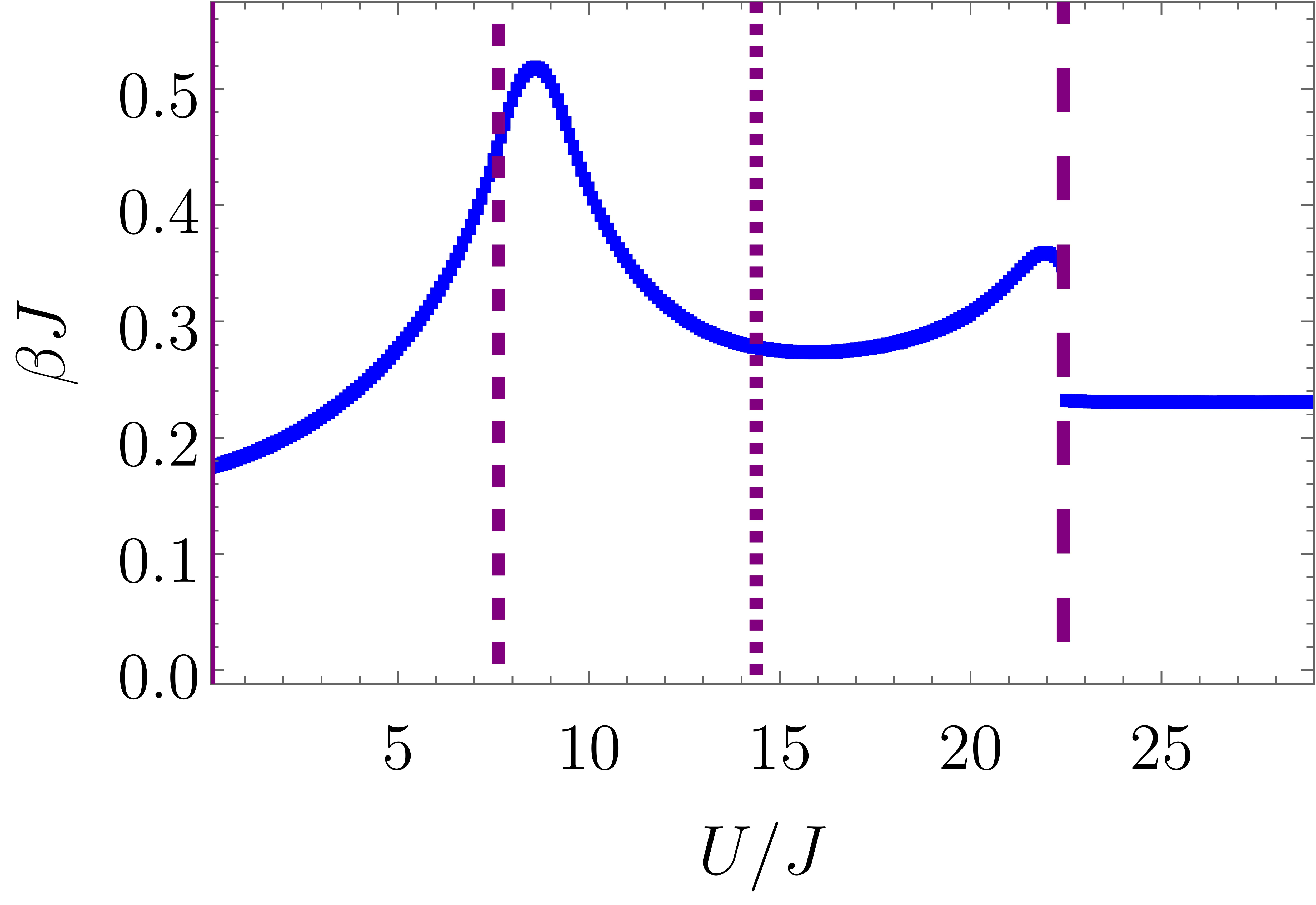}};
    \node[font=\normalfont] at (1.5ex,-1ex) {(b)};
    \end{tikzpicture}
    \includegraphics[width=0.31\textwidth]{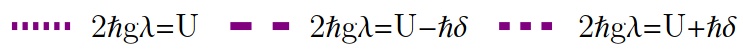}
    \vspace{-10pt}
    \end{flushleft}
    \caption[]{The (a)~sublattice imbalance $\langle\hat{\Delta}\rangle/L^d$ and (b)~inverse temperature $\beta J$ as a function of $U/J$ of a finite size system $L\!=\!8$ at quarter filling. The $T\!=\!0$ MF solution is shown as black solid line in (a). The fixed parameters used are $\hbar g/J\!=\!8$, $\hbar\delta/J\!=\!8$ and $\hbar\varGamma/J\!=\!1$. Vertical lines denote resonances, in the atomic limit.}
    \label{fig:cuts_Uvar_g8_delta8_Gamma1}
\end{figure}

\begin{figure}[!hbtp]
    \begin{tikzpicture}
    \node[anchor=north west,inner sep=0pt] at (0,0){\includegraphics[width=0.23\textwidth]{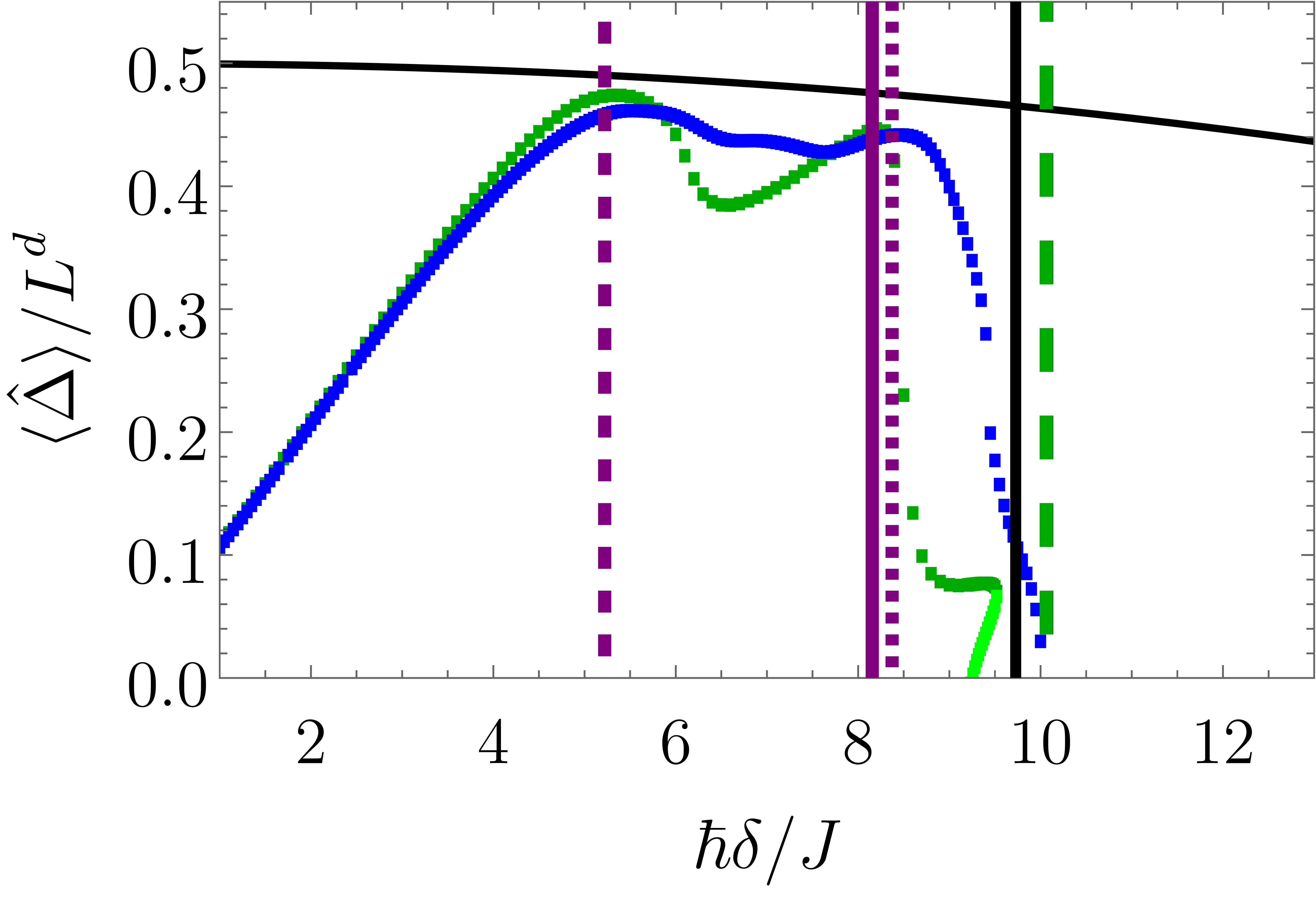}};
    \node[font=\normalfont] at (1.5ex,-1ex) {(a)};
    \end{tikzpicture}
    \begin{tikzpicture}
    \node[anchor=north west,inner sep=0pt] at (0,0){\includegraphics[width=0.23\textwidth]{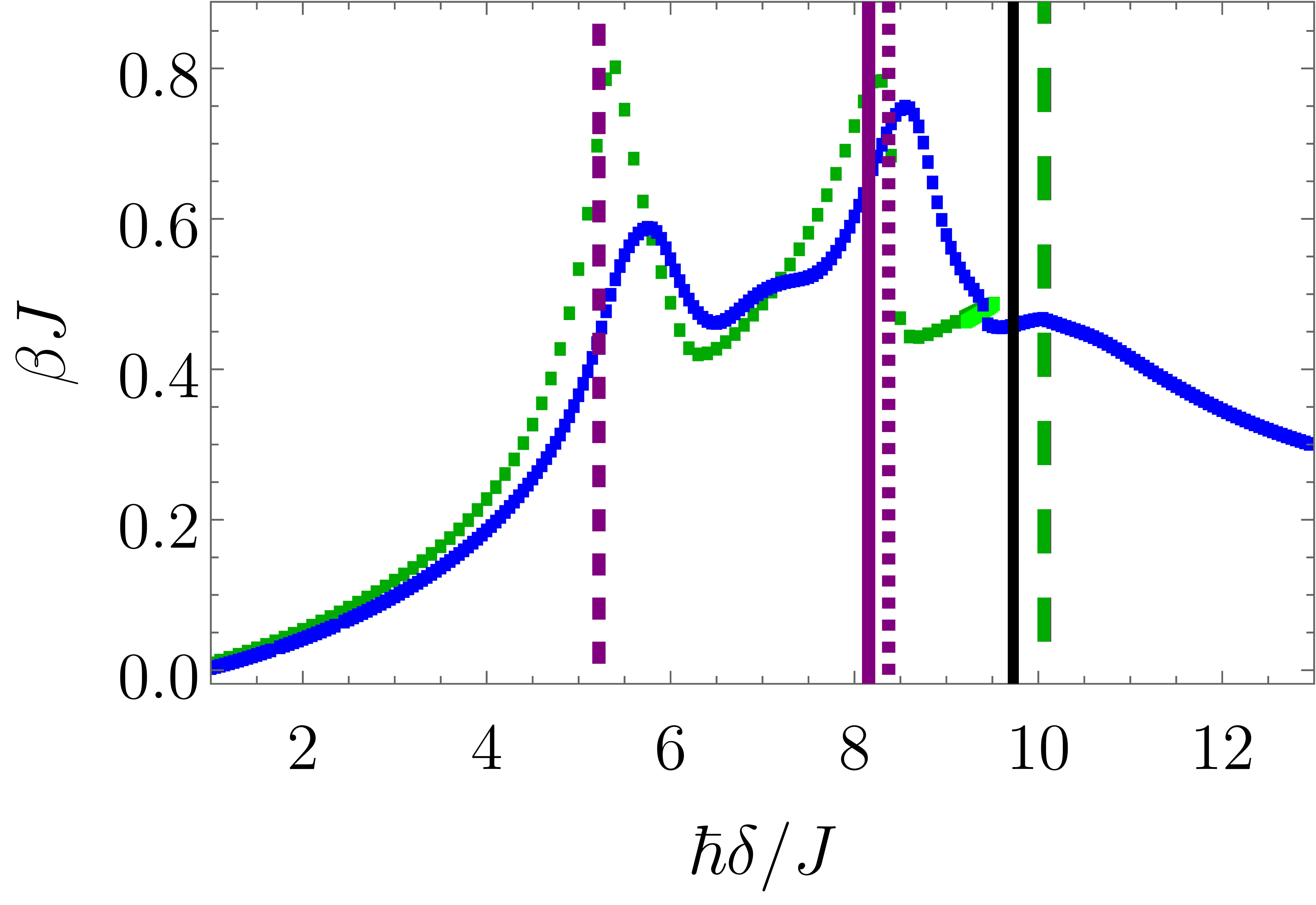}};
    \node[font=\normalfont] at (1.5ex,-1ex) {(b)};
    \end{tikzpicture}
    \includegraphics[width=0.48\textwidth]{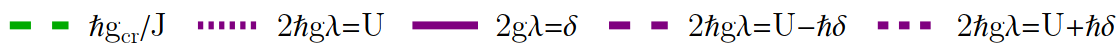}
    \vspace{-10pt}
    \caption[]{The~(a)~sublattice imbalance $\langle\hat{\Delta}\rangle/L^d$ and~(b)~inverse temperature $\beta J$ as a function of $\hbar\delta/J$; of a finite size system $L\!=\!8$ stable (blue); in the weak tunneling perturbation in the thermodynamic limit [Eqs.~(\ref{eq:system_of_equations_1D})] stable (dark green) and unstable (light green) at quarter filling. The $T\!=\!0$ MF solution is shown as black solid line. The parameters are $\hbar g/J\!=\!6$, $U/J\!=\!8$ and $\hbar\varGamma/J\!=\!1$. Vertical lines denote the critical coupling $\hbar g_\text{cr}/J$ and resonances.}
    \label{fig:cuts_U8_g6_deltavar_Gamma1}
\end{figure}

The physical interpretation of the condition $2g\lambda\!=\!\delta$ is a resonance between an atom hopping from a high potential sublattice site to a neighboring low potential site with an approximate energy gain $2g\lambda$ and a photon creation in the cavity with approximate energy cost $\delta$ as sketched in Fig.~\ref{fig:sketch_transition_EOM}~(a).
By this process, close to the resonance an efficient energy transfer from the atomic system to the cavity field is enabled. We find that this leads to `cooling', i.e.~a reduced effective atomic temperature in the steady state. The reduced temperature causes an increase in the sublattice density imbalance and photon number. 
This process corresponds to the \emph{cavity-cooling} phenomena discussed for many different implementations of atoms-cavity coupled systems \cite{RitschEsslinger2013}. 
Whereas the $T\!=\!0$ MF approach cannot cover this effect, the formalism we employ (as described in Sec.~\ref{sec:mbae}) can capture the cooling process. 
The key ingredient stems from the fluctuations in the atoms-cavity coupling which lead to a finite effective temperature.
Furthermore, in Sec.~\ref{sec:many_body_cooling} we show that a similar cooling process also occurs at many-body resonances, which explains the observed minima in the temperature.

At very large coupling strength one would expect from the $T\!=\!0$ MF approach that the density imbalance $\langle\hat{\Delta}\rangle$ saturates to its maximal value $N$ [see black curve in Fig.~\ref{fig:cuts_U8_gvar_delta8_Gamma1_and_U20_gvar_delta8_Gamma1}~(a),(c)]. At the same time the photon number increases quadratically with the coupling as obtained from the steady state condition for $\langle\hat{a}\rangle$ [see Eq.~(\ref{eq:time_dependence_a})] with maximal density imbalance $\langle N_\text{pho}\rangle\!\propto\frac{g^2}{\delta^2\!+\!(\varGamma/2)^2}n^2$. 
However, this behavior is in stark contrast with the results obtained with a self-consistent temperature. 
Within this approach a decrease of the imbalance at intermediate to larger coupling strengths is found for the considered parameter regimes [see blue curves in 
Fig.~\ref{fig:cuts_U8_gvar_delta8_Gamma1_and_U20_gvar_delta8_Gamma1}~(a),(c)]. 
We attribute the origin of this decrease to the rapidly increasing effective temperature of the atomic state [see blue curves in 
Fig.~\ref{fig:cuts_U8_gvar_delta8_Gamma1_and_U20_gvar_delta8_Gamma1}~(b),(d)]. When the effective temperature surpasses the energy scale given by the height of the self-organized potential $2\hbar g\lambda$ the atoms are thermally excited to the lattice sites corresponding to a higher potential leading to a decrease of the sublattice imbalance. 

For the non-interacting case ($U/J\!=\!0$) in the strong coupling limit, we can obtain analytic expressions for the inverse temperature, imbalance and photon number (see Appendix~\ref{app:universal_scaled_photon_number})
\begin{align}
\label{eq:imbalance_universal_scaling_U0}
    \beta J&\approx\frac{(\varGamma/2)^2+\delta^2}{g^2n(2-n)}, \\
     \frac{\langle\hat{\Delta}\rangle}{L}&=\frac{\sqrt{(2-n)n((\varGamma/2)^2+\delta ^2)}}{2g}\nonumber, \\
    \frac{\langle\hat{a}^\dagger\hat{a}\rangle}{L}&\approx\frac{n}{4}(2-n)\nonumber.
\end{align}
While for $U/J\!\to\!\infty$ in the strong coupling regime we have
\begin{align}
\label{eq:imbalance_universal_scaling_Uinf}
    \beta&\approx\frac{(\varGamma/2)^2+\delta^2}{2\hbar\delta g^2n(1-n)}, \\
     \frac{\langle\hat{\Delta}\rangle}{L}&=\frac{\sqrt{(1-n)n((\varGamma/2)^2+\delta^2)}}{\sqrt{2} g}\nonumber \\
    \frac{\langle\hat{a}^\dagger\hat{a}\rangle}{L}&\approx\frac{n}{2}(1-n) \nonumber
\end{align}
Thus, we obtain that at large coupling strengths the imbalance decreases as $\langle\hat{\Delta}\rangle\!\propto\!g^{-1}$ and the photon number saturates at a value independent of the atoms-cavity coupling.

We note that the result for spinful fermions obtained here differs from the one obtained for a Bose-Hubbard system in Ref.~\cite{BezvershenkoRosch2021}, where the photon number scales as $n(n\!+\!1)$ with the particle density.
We observe in our results both the scaling of the temperature, $\beta\!\propto\!g^{-2}$, and the sublattice imbalance, $\langle\hat{\Delta}\rangle\! \propto\! g^{-1}$, as we increase the coupling strength (see orange dashed line in 
Fig.~\ref{fig:cuts_U8_gvar_delta8_Gamma1_and_U20_gvar_delta8_Gamma1}). The obtained dependence on $\varGamma$ also matches the results (see orange dashed line in 
Fig.~\ref{fig:cuts_U8_g8_delta2_Gammavar_and_U8_gvar_delta2_Gamma1}). 
In Fig.~\ref{fig:cuts_U8_g8_delta2_Gammavar_and_U8_gvar_delta2_Gamma1} and Fig.~\ref{fig:density_plots_gvGamma_U8_delta6}
one also sees clearly that for large values of the dissipation heating is induced in the atomic system. With increasing dissipation the effective temperature rises approximately as $k_B T\sim\varGamma^2$ as predicted in Eq.~(\ref{eq:beta_approx_large_dissipation}). Only the resonances described in Sec.~\ref{sec:many_body_cooling} lead to a strongly non-monotonic behavior. 

The expressions in Eq.~(\ref{eq:imbalance_universal_scaling_U0}) and Eq.~(\ref{eq:imbalance_universal_scaling_Uinf}) are in stark contrast with the expectations from the $T\!=\!0$ MF approach.

\begin{figure}[!hbtp]
\begin{tikzpicture}
	\node[anchor=north west,inner sep=0pt] at (0,0){\includegraphics[width=0.23\textwidth]{Fig15a}};
	\node[font=\normalfont] at (1.5ex,0ex) {(a)};
\end{tikzpicture}
    \begin{tikzpicture}
	\node[anchor=north west,inner sep=0pt] at (0,0){\includegraphics[width=0.23\textwidth]{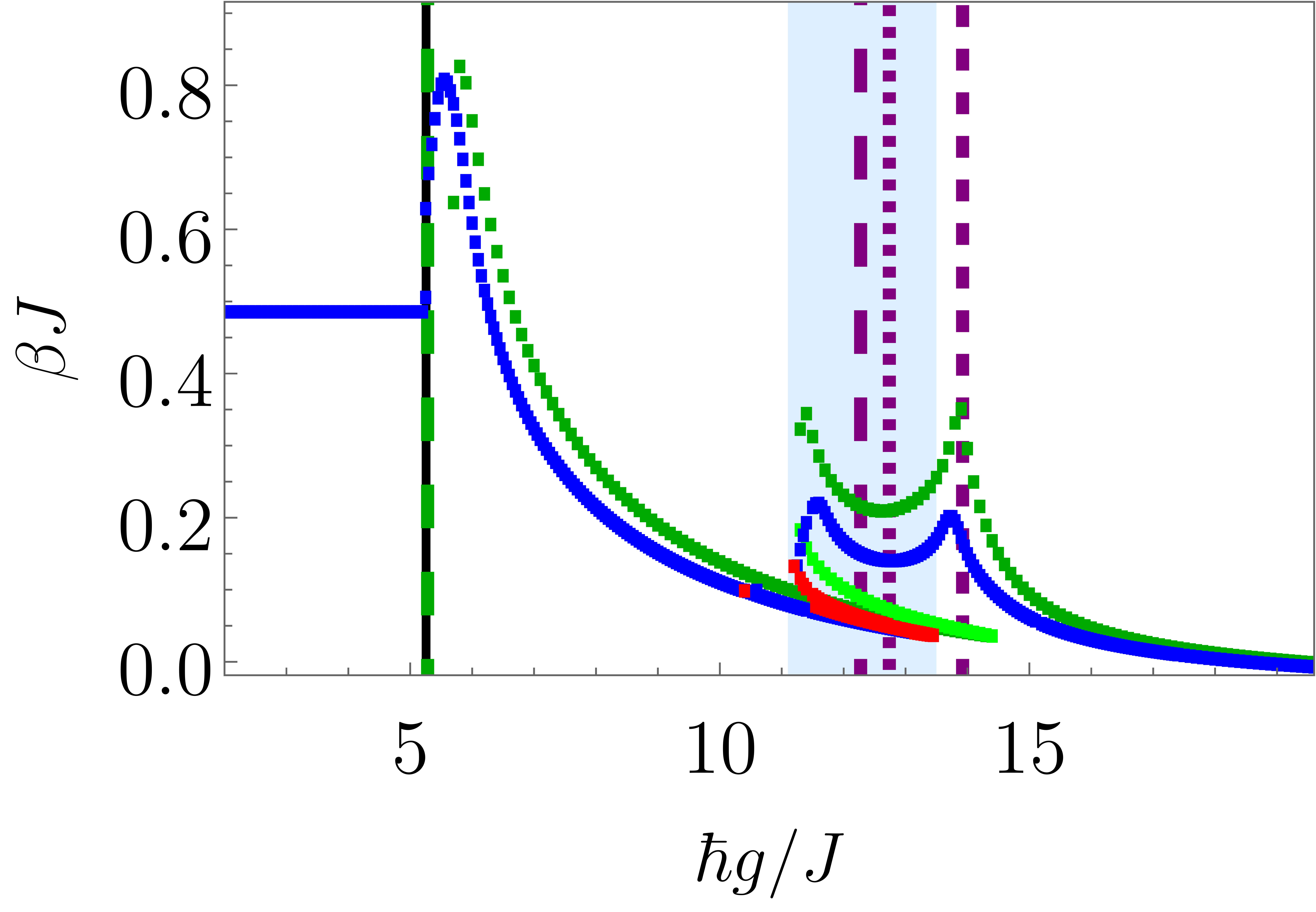}};
	\node[font=\normalfont] at (1.5ex,0ex) {(b)};
\end{tikzpicture}
    \begin{tikzpicture}
	\node[anchor=north west,inner sep=0pt] at (0,0){\includegraphics[width=0.23\textwidth]{Fig15c}};
	\node[font=\normalfont] at (1.5ex,0ex) {(c)};
\end{tikzpicture}
\begin{tikzpicture}
	\node[anchor=north west,inner sep=0pt] at (0,0){\includegraphics[width=0.23\textwidth]{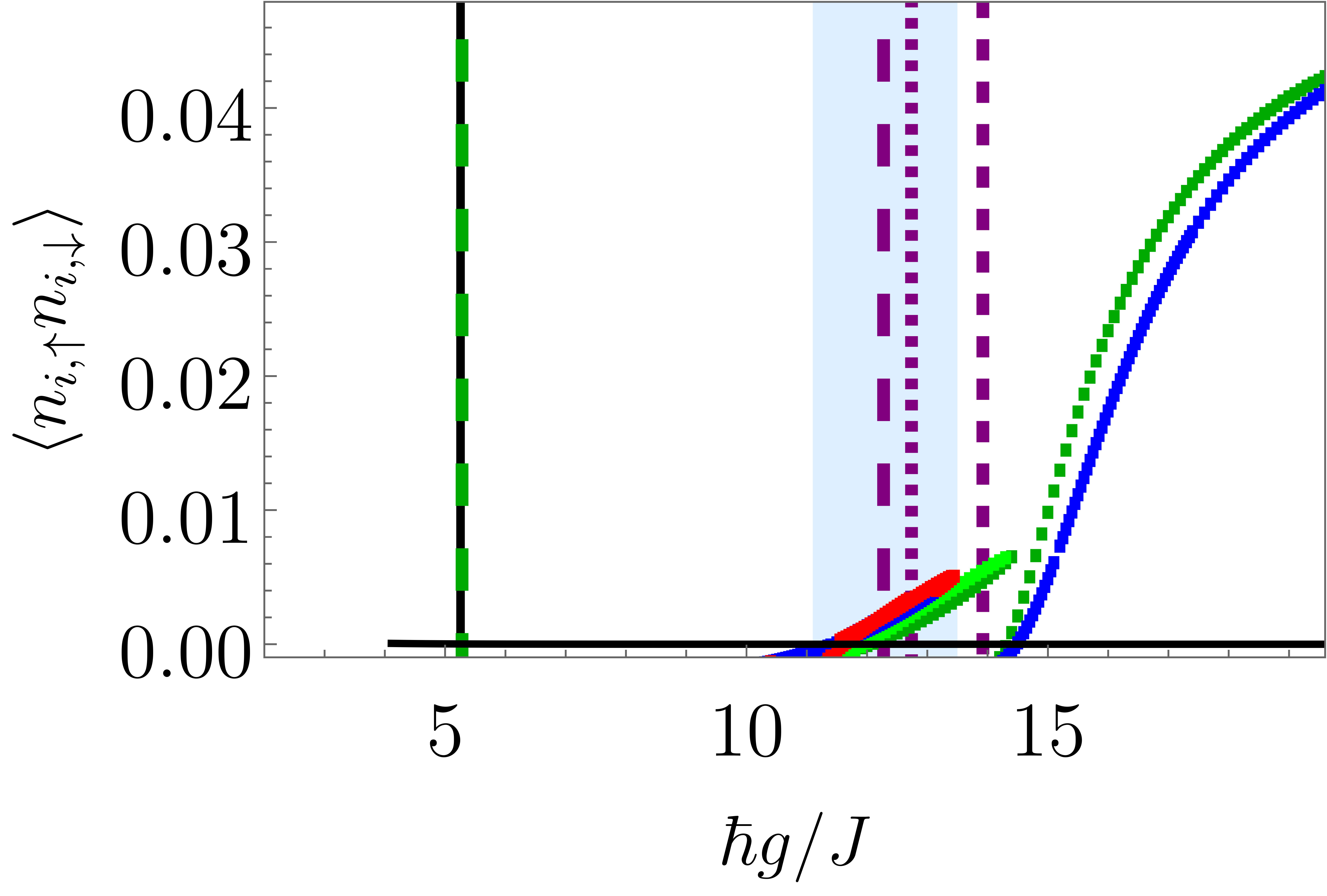} };
	\node[font=\normalfont] at (1.5ex,0ex) {(d)};
\end{tikzpicture}
    \includegraphics[width=0.42\textwidth]{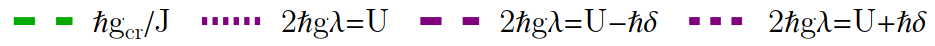}
    \vspace{-10pt}
    \caption[]{Cuts of the (a) scaled photon number, (b) inverse temperature $\beta J$, (c) sublattice imbalance $\langle\hat{\Delta}\rangle/L^d$ and (d) average double-occupancy as a function of atoms-cavity coupling $\hbar g/J$ at $U/J\!=\!40$; of a finite size system $L\!=\!8$ stable (blue) and unstable (red). The light blue interval marks the bistability region; in the weak tunneling perturbation in the thermodynamic limit [Eqs.~(\ref{eq:system_of_equations_1D})] stable (dark green) and unstable (light green) at quarter filling. The $T\!=\!0$ MF solution is shown as black solid line. The other parameters used are $\hbar\delta/J\!=\!8$, $\hbar\varGamma/J\!=\!1$. Vertical lines denote the critical coupling $\hbar g_\text{cr}/J$ and resonances. The orange dashed lines are the scalings of the analytical approximations for $U/J\!\to\!\infty$ [Eqs.~(\ref{eq:imbalance_universal_scaling_Uinf})].}
    \label{fig:cuts_U40_gvar_delta8_Gamma1}
\end{figure}

\subsubsection{Many-body cooling processes \label{sec:many_body_cooling}}

In this section, we highlight that the phenomena of cavity-cooling, here represented by a minimum in the steady state temperature, also occurs due to many-body resonances. 
We show that when the cavity detuning approximately matches gaps in the spectrum of the many-body effective Hamiltonian a cooling mechanism can emerge leading to lower self-consistently determined effective temperatures of the atomic states. 
The presence of these resonances explains the non-monotonic structure in the inverse temperature occurring at intermediate coupling strength in the self-organized phase, e.g.~seen in Fig.~\ref{fig:cuts_U8_gvar_delta8_Gamma1_and_U20_gvar_delta8_Gamma1}~(b),(d) or Fig.~\ref{fig:density_plots_gvU_delta8_Gamma1} and Fig.~\ref{fig:density_plots_gvGamma_U8_delta6}~(b). The effect can also be observed varying $U/J$ or $\hbar\delta/J$ at fixed intermediate coupling strength in Fig.~\ref{fig:cuts_Uvar_g8_delta8_Gamma1} and Fig.~\ref{fig:cuts_U8_g6_deltavar_Gamma1}~(b).

It is important to note that the cooling effects can only arise due to the fact that in our method the temperature is self-consistently determined from the perturbative effect of the cavity fluctuations (see Sec.~\ref{sec:mbae}).% Such phenomena would not be captured by imposing a fixed temperature for the atomic state.

In order to understand the non-monotonic behavior of the steady states of the coupled atoms-cavity system, it is useful to briefly revisit the energy gaps of the effective Hamiltonian at quarter filling (discussed in Sec.~\ref{sec:ionic_hubbard}). The weight of the terms that capture the transition between eigenstates $\ket{n}$ and $\ket{m}$ in Eq.~(\ref{eq:energy_transfer_spectral}) is strongly influenced by resonances $\hbar\delta\approx E_n\!-\!E_m$. The sign of the terms in Eq.~(\ref{eq:energy_transfer_spectral}) implies either the cooling or heating nature of the processes in the self-consistent dynamics towards the steady state solution where the processes are balanced \cite{BezvershenkoRosch2021}.

To identify the physical processes associated with the dominant heating or cooling processes, we look at the equation derived in the weak tunneling limit, Eq.~(\ref{eq:EOM_1D}). This allows us to connect the atomic resonances to certain atomic processes which take place during the exchange of energy between the atoms and cavity, as we discuss in the following. Note that the resonance conditions in the weak tunneling perturbation are given for the limit of vanishing tunneling. Thus, small deviations are to be expected for the finite $J$ results.
 
In the isolated ionic Hubbard model, excitations from the ground state, as sketched in Fig.~\ref{fig:sketch_transition_EOM}~(b), are suppressed by the conservation of energy. In contrast, for the Hubbard model coupled to the cavity, the cavity field is able to facilitate such transitions by taking or providing the energy cost or gain. 

One can interpret the contributions in the upper line of Eq.~(\ref{eq:EOM_1D}) as originating from eigenstates of $\hat{H}_\text{eff}$, which differ by a single particle being on the lower potential or upper potential sub-lattice [sketched in Fig.~\ref{fig:sketch_transition_EOM}~(a)]. The energy difference~$E_n\!-\!E_m\!\approx\! 2\hbar g\lambda$, corresponds to transitions between the ground state and an excited state of the effective ionic Hubbard model (see Fig.~\ref{fig:Espectrum_IH_L8N4_U20_etavar}).
We found a minimum of the effective temperature close to the resonance given by this energy difference, $2g\lambda\approx \delta$, as we discussed in Sec.~\ref{sec:cavity_cooling}. 
In plots the resonance is marked by a purple line, e.g.~in Fig.~\ref{fig:density_plots_gvU_delta8_Gamma1} and Fig.~\ref{fig:density_plots_gvdelta_U8_G1}, or as vertical line in the cuts in Fig.~\ref{fig:cuts_U8_gvar_delta8_Gamma1_and_U20_gvar_delta8_Gamma1}, Fig.~\ref{fig:cuts_Uvar_g8_delta8_Gamma1} and Fig.~\ref{fig:cuts_U8_g6_deltavar_Gamma1}.

Interestingly, similar non-trivial effects appear also when the cavity is resonant to transitions between two excited states involving the interaction energy, i.e.~where $2\hbar g\lambda\!=\!U\!\pm\!\hbar\delta$ is fulfilled.
This is expressed in the lower line of Eq.~(\ref{eq:EOM_1D}) stemming from transitions that simultaneously change the sublattice density imbalance and number of double occupancies, i.e.~$E_n\!-\!E_m\!\approx\!2\hbar g\lambda\mp U$. This process is sketched in Fig.~\ref{fig:sketch_transition_EOM}~(b).
The cooling mechanism leads to an efficient energy transfer from the atomic subsystem to the cavity field at these resonances and the resulting steady state effective temperature shows a minimum.
Here, the specific resonant processes are transitions between states with two neighboring singly occupied sites or a double occupancies on the low potential site while adding/removing a photon from the cavity field [sketched in Fig.~\ref{fig:sketch_transition_EOM}~(b)]. 
This drives the system towards a steady state with lower temperature and increased density sublattice imbalance. 
The resonance conditions 
$2g\lambda_2\!=\!U/\hbar\!\pm\!\delta$ are marked by purple, long(-)/short(+)-dashed lines, e.g.~in Fig.~\ref{fig:density_plots_gvU_delta8_Gamma1}, Fig.~\ref{fig:density_plots_gvdelta_U8_G1} and Fig.~\ref{fig:density_plots_gvGamma_U8_delta6}, or as vertical lines in the cuts in Fig.~\ref{fig:cuts_U8_gvar_delta8_Gamma1_and_U20_gvar_delta8_Gamma1} and Fig.~\ref{fig:cuts_Uvar_g8_delta8_Gamma1}.
The resonances involving doubly occupied sites are also important within the emerging fluctuation-induced bistable regime that we discuss in the next section Sec.~\ref{sec:fluctuation_induced_bistability}.

The cooling effect emerging at the resonance condition can also be observed around $\hbar\delta\!\sim\!U$. Here transitions resembling the ones sketched in Fig.~\ref{fig:sketch_transition_EOM}~(c) become resonant.
We observe in our results that when the condition $U\!\approx\!\hbar\delta$ is satisfied (dark red dashed lines in Fig.~\ref{fig:density_plots_gvU_delta8_Gamma1} and Fig.~\ref{fig:density_plots_gvdelta_U8_G1}) the value of the effective temperature is slightly reduced and the sublattice imbalance exhibits a small maximum, by varying both the $U/J$ (Fig.~\ref{fig:density_plots_gvU_delta8_Gamma1}) or $\hbar\delta/J$ (Fig.~\ref{fig:density_plots_gvdelta_U8_G1}).
We note that a similar resonant cooling effect was also identified for interacting bosonic atoms in Ref.~\cite{BezvershenkoRosch2021}.

We can gain additional intuition for the impact of the $U\!\approx\!\hbar\delta$ resonance as in the following, the imaginary part of the retarded susceptibility in Eq.~(\ref{eq:energy_transfer_susceptibility}) is peaked around $\hbar\omega\!=\!\pm2g\lambda$ with additional peaks at $\hbar\omega\!=\!\pm U$, suppressed by a factor of $J^2/\hbar g\lambda$ since they require second order hopping processes. Processes without energy transfer are additionally suppressed since their contribution in Eq.~(\ref{eq:energy_transfer_susceptibility}) is proportional to $\Delta E_{nm}$. 
These terms only become important if a resonance $U/J\!\sim\!\hbar\delta/J$ is present.
The dominant cooling terms driving the system to the steady state thus stem from $\hbar\omega\!\sim\!-U$, so the processes creating a double occupancy while adding a photon to the cavity field. Contrarily, the heating is dominated by the peaks at $\hbar\omega\!\sim\!U$.

We can view the $U\!\approx\!\hbar\delta$ resonance more generally as an instance in which the energy of the cavity field matches the gap between the ground state and the first excited state of the effective Hamiltonian $E_\text{gap}\!=\!\hbar\delta$. We can exemplify this in Fig.~\ref{fig:density_plots_gvU_delta8_Gamma1} and Fig.~\ref{fig:density_plots_gvdelta_U8_G1}, where the $E_\text{gap}\!=\!\hbar\delta$ resonance matches $U\!\approx\!\hbar\delta$ for the $\hbar g/J\!\gtrsim\!9$ regime and goes towards the condition $2g\lambda\!=\!\delta$, marked by the purple solid curve.
This further allows us to understand maxima in the sublattice density imbalance away from regimes where the nature of the eigenstates of $\hat{H}_\text{eff}$ can be easily identified, as for example around $U/J\!\approx\!9.8$, $\hbar g/J\!\approx\!6.8$ in Fig.~\ref{fig:density_plots_gvU_delta8_Gamma1}.

\subsubsection{Fluctuation-induced bistability}

\label{sec:fluctuation_induced_bistability}

\begin{figure}[!hbtp]
\begin{flushleft}
    \begin{tikzpicture}
    \node[anchor=north west,inner sep=0pt] at (0,0){\includegraphics[height=0.18\textheight]{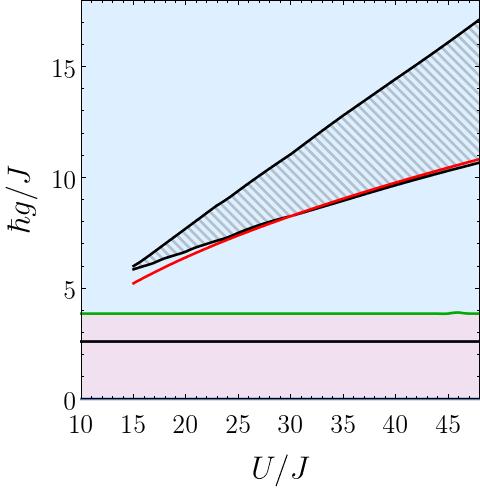}};
    \node[font=\normalfont] at (2ex,-2ex) {(a)};
    \end{tikzpicture}
    \begin{tikzpicture}
    \node[anchor=north west,inner sep=0pt] at (0,0){ \includegraphics[height=0.18\textheight]{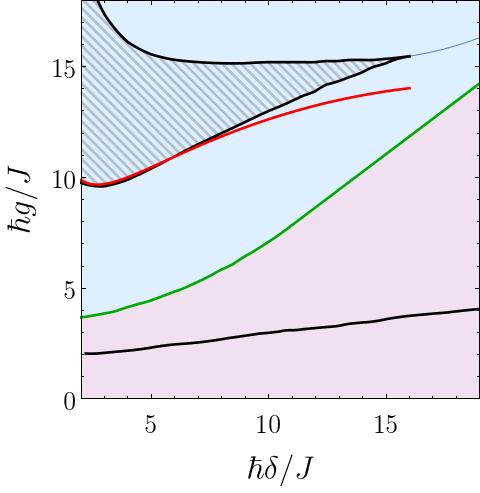}};
    \node[font=\normalfont] at (2ex,-2ex) {(b)};
    \end{tikzpicture}
    \begin{tikzpicture}
    \node[anchor=north west,inner sep=0pt] at (0,0){\includegraphics[height=0.18\textheight]{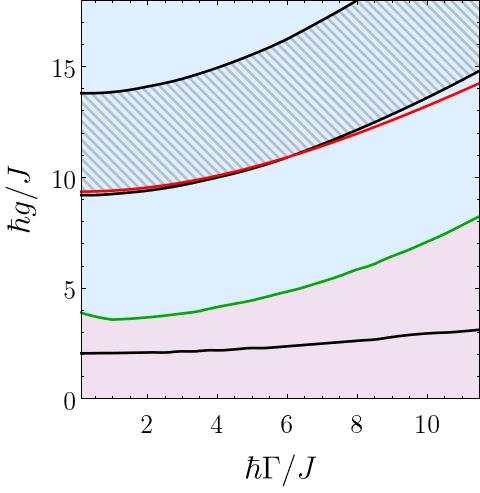}};
    \node[font=\normalfont] at (2ex,-2ex) {(c)};
    \end{tikzpicture}
    \end{flushleft}
    \vspace{-20pt}
    \caption[]{Simplified phase diagrams varying the atoms-cavity coupling $\hbar g/J$ and (a) the on-site interaction $U/J$ at $\hbar\delta/J\!=\!5$, $\hbar\varGamma/J\!=\!3$; (b) the pump-cavity detuning $\hbar\delta/J$ at $U/J\!=\!40$, $\hbar\varGamma/J\!=\!5$; (c) the dissipation rate $\hbar\varGamma/J$ at $\hbar\delta/J\!=\!5$, $U/J\!=\!40$ in the weak tunneling perturbation in the thermodynamic limit $L\!\to\!\infty$ at quarter filling. 
    The green line marks the transition between normal (purple region) and self-organized phase (blue region) at $\hbar g_\text{cr}/J$ determined by solving Eqs.~(\ref{eq:cav_MF})~\&~(\ref{eq:energy_transfer}), the solid black line in the purple region marks the critical coupling $\hbar g^\text{MF}_\text{cr}$ obtained for the $T\!=\!0$ MF approach solving Eq.~(\ref{eq:cav_MF}) via ED. The black lines bordering the hatched region mark the couplings $\hbar g_\text{bi,1(2)}$ at finite temperature. The red line marks the onset of the bistability approximated by Eq.~(\ref{eq:gbi1_approx}).}
    \label{fig:bistable_region_plots_quarter_filling}
\end{figure}

As we go to large atoms-cavity couplings in the regime of strong atomic interactions we observe the emergence of multiple stable solutions (blue shaded regions in Fig.~\ref{fig:cuts_U40_gvar_delta8_Gamma1} and hatched regions in Fig.~\ref{fig:bistable_region_plots_quarter_filling}). 
This is surprising as in $T\!=\!0$ MF approach only a single solution is found in this regime (black curve in Fig.~\ref{fig:cuts_U40_gvar_delta8_Gamma1}). 
We first observed and discussed this effect in Ref.~\cite{TolleHalati2024} and named it \emph{fluctuation-induced bi-(multi-)stability}. The discussion of this regime will revisit the results of Ref.~\cite{TolleHalati2024} and complement them by additional details and interpretation.
In the thermodynamic limit, we identify a regime with two stable solutions, $\lambda_{1 (2)}$ with lower (higher) cavity field (darker green points in Fig.~\ref{fig:cuts_U40_gvar_delta8_Gamma1}), respectively.
These two solutions are connected by a third unstable branch (light green points in Fig.~\ref{fig:cuts_U40_gvar_delta8_Gamma1}). The stability is determined by the stability condition derived in Eq.~(\ref{eq:condition}). For small system sizes finite size effects can lead to additional solutions within the bistability regions [blue (stable) and red (unstable) points in Fig.~\ref{fig:cuts_U40_gvar_delta8_Gamma1}].

\begin{figure}[!hbtp]
    \includegraphics[width=0.4\textwidth,valign=t]{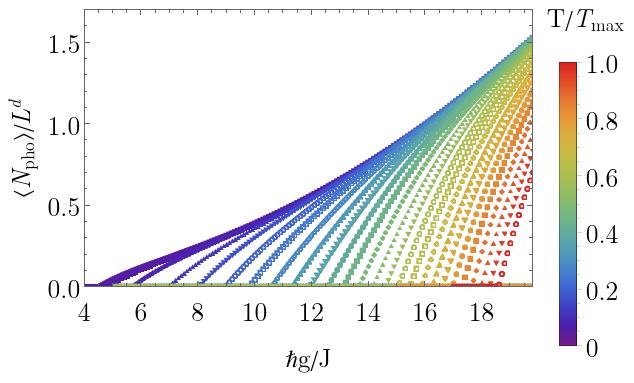} 
    \caption[]{Solution of the system of equations Eqs.~(\ref{eq:system_of_equations_1D}) with color-coded fixed values of $k_BT=1/\beta$ varying $\hbar g/J$ at $U/J\!=\!40$, $\hbar\delta/J\!=\!8$, $\hbar\varGamma/J\!=\!1$, $k_BT_\text{max}/J\!=\!25$ in the weak tunneling perturbation in the thermodynamic limit at quarter filling. (adapted from Ref.~\cite{TolleHalati2024})}
    \label{fig:Npho_beta_fixed_U40_delta8_Gamma1}
\end{figure}

\begin{figure}[!hbtp]
    \includegraphics[width=0.28\textwidth,valign=t]{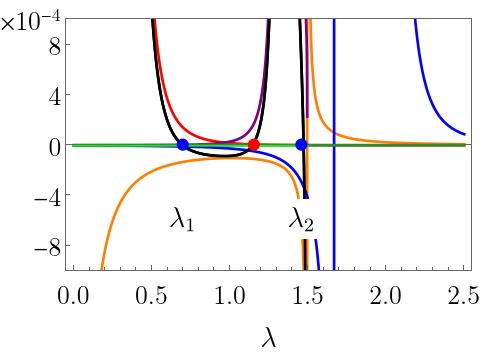} \includegraphics[width=0.48\textwidth,valign=t]{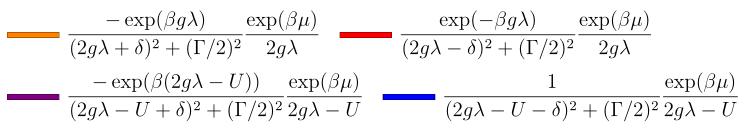}
    \vspace{-10pt}
    \caption[]{Individual terms of Eq.~(\ref{eq:EOM_1D}) and full equation (black line) at $\hbar g/J\!=\!12$, $U/J\!=\!40$, $\hbar\delta/J\!=\!8$, $\hbar\varGamma/J\!=\!1$ in the weak tunneling perturbation in the thermodynamic limit at quarter filling. The system of equations has two stable (blue points, $\lambda_{1(2)}$) and one unstable (red point) solutions at the zero crossings of the black line. Contributions from second line terms in Eq.~(\ref{eq:EOM_1D}) with $p\!=\!-1$ are shown as green lines (close to 0).}
    \label{fig:EOM_terms_U40_g12_delta8_Gamma1}
\end{figure}

Importantly, the bistability does not appear in the $T\!=\!0$ MF approach (black line, Fig.~\ref{fig:cuts_U40_gvar_delta8_Gamma1}). 
The crucial ingredient which causes this effect are the quantum fluctuations in the coupling term between the atoms and the cavity field which act on top of the mean-field decoupling. The fluctuations imprint the dissipative nature of the cavity field onto the atoms, resulting in a heating of the atoms to a finite temperature mixed state. 
The effective temperature to which the system stabilizes is determined by the complex interplay of the cavity energy scales and the spectrum of the effective atomic Hamiltonian.
Whereas at low interaction strength we obtain that the effective temperature is uniquely determined for each coupling strength, in the strongly interacting regime there exists regions where multiple solutions appear. 
The effective temperature and the strength of the cavity field in these stable solutions differ considerably.
We find that the second solution arises with the lower effective temperature. We verified that if the temperature is fixed externally, and not determined by the energy flow between the atoms and the cavity, the bistability is not present. We see in Fig.~\ref{fig:Npho_beta_fixed_U40_delta8_Gamma1} that regardless of the value of the externally fixed temperature a single stable solution is obtained. 
Thus, the origin of the fluctuation-induced bistability is fundamentally different from the ones observed in $T\!=\!0$ MF methods, which are typically caused by steady states of different nature that are coexisting \cite{RitschEsslinger2013}, e.g.~around a first-order dissipative phase transition \cite{LarsonLewenstein2008, Minganti_Ciuti_2018, HelsonBrantut2022,FerriEsslinger2021}, or due to spontaneously breaking of a weak symmetry \cite{BaumannEsslinger2010, Keeling_Simons_2010, Mivehvar_2024}.
We observe a mean-field bistability for the considered model at half filling ($n\!=\!1$), which we discuss in Sec.~\ref{sec:half_filling}.

For the fluctuation-induced bistability, the emerging cooling mechanisms discussed in Sec.~\ref{sec:many_body_cooling} lead to multiple stable states, with different contributions from the eigenstates of the effective Hamiltonian, which determine the physical properties.
The competition of on-site atom-atom interactions and the global range interaction mediated by the cavity field is crucial for the existence of the bistability.
For a better understanding of this competition, we explain the fluctuation-induced bistability within the perturbative approach derived for values of large $U/J$ and the thermodynamic limit, as described in Sec.~\ref{sec:pertubation_J}. 
This approach is valid in the regime where the fluctuation-induced bistability appears and for arbitrary lattice dimension.
Thus, all results discussed in this section are valid also for higher dimensional atomic lattices.
As other models would lead to similar equations of motion within this perturbation theory, e.g.~interacting bosonic atoms coupled to the cavity field, it implies that the described bistability is more general effect and not particular to our model.

The solution of the equations of motion, Eq.~(\ref{eq:system_of_equations_1D}), depends on balancing the transition rates between eigenstates $\ket{n},\ket{m}$ of $\hat{H}_\text{eff}$. 
For solutions with lower cavity field the important states are those with single site occupations on the lower or upper potential sub-lattice in the self-consistent dynamics towards the steady state. Double occupancies are high energy excitations in this regime and can effectively be eliminated by reducing the description to the low-energy sector which does not contain two particles on the same site. 
These processes between states with only single site occupations have energy changes in the atomic subsystem of $E_n\!-\!E_m\!\approx\! 2\hbar g\lambda_1$, as captured by the first line terms in Eq.~(\ref{eq:EOM_1D}). 
We plot the LHS of the energy transfer equation as well as the individual terms as a function of $\lambda$ for fixed values of the parameters in Fig.~\ref{fig:EOM_terms_U40_g12_delta8_Gamma1}. In the limit $U\!\gg\!J$ the solution of the particle number equation Eq.~(\ref{eq:pc_1D}) shows to be almost independent of $\lambda$ around the onset of the bistability (further details are given in the Supplemental Material of Ref.~\cite{TolleHalati2024}). 
We derive the approximation $\mu\!\approx\!\log(1/2)/\beta$ and use Eq.~(\ref{eq:beta_approx_large_glambda}). Inserting these in the system of equations Eq.~(\ref{eq:system_of_equations_1D}), the system is solved by finding values of $\lambda$ for which the steady state condition is fulfilled (zero crossings of black line). 
For example, in Fig.~\ref{fig:EOM_terms_U40_g12_delta8_Gamma1} we identify the two stable solutions $\lambda_{1(2)}$ (blue points) and one unstable solution (red point).
This allows us to get an intuition which terms in the energy transfer equation dominantly contribute and are the most important in determining the solutions. 

In the considered parameter ranges the low $\lambda_1$ solution balances the cooling $\propto\!1/[(2g\lambda\!+\!\delta)^2\!+\!(\varGamma/2)^2]$ (light orange line in Fig.~\ref{fig:EOM_terms_U40_g12_delta8_Gamma1}) and heating $\propto\!1/[(2g\lambda\!-\!\delta)^2\!+\!(\varGamma/2)^2]$ (dark orange line) at relatively high temperature $k_BT_1\!>\!8 zJ$, where $z$ is the lattice coordination number that adapts the scaling to arbitrary dimensions. 
This higher temperature can be seen in Fig.~\ref{fig:cuts_U40_gvar_delta8_Gamma1}~(b) where $J/k_BT$ quickly decreases. All other terms are vanishing around $\lambda_1$.
We note that the photon number corresponding to the $\lambda_1$ solution quickly saturates as the coupling strength is increased and agrees with the limit of the universal scaled photon number derived in Appendix~\ref{app:universal_scaled_photon_number}, resulting in Eq.~(\ref{eq:imbalance_universal_scaling_Uinf})
\begin{align}
\frac{\langle\hat{a}^\dagger\hat{a}\rangle}{L}&\approx\frac{n}{2}(1-n) \nonumber.
\end{align}
With larger values of the coupling strength, the temperature also increases. These effects compensate each other such that the photon number saturates and imbalance decreases as $\langle\hat{\Delta}\rangle\!\propto\!g^{-1}$, see Eq.~(\ref{eq:imbalance_universal_scaling_Uinf}), shown by the orange dashed line in Fig.~\ref{fig:cuts_U40_gvar_delta8_Gamma1}~(a),(c), as discussed in Sec.~\ref{sec:cavity_cooling}.

In contrast, the self-consistently determined effective temperature for the $\lambda_2$ solution with higher scaled photon number is considerably lower.
More specifically, the cooling mechanism that we described in Sec.~\ref{sec:many_body_cooling}, originating from the efficient transfer of energy from the atoms to the cavity mode due to resonances between excited states and the photonic energy, drives the atoms to a steady state with a much lower temperature. 
In the bistability region these resonances typically appear in the $\lambda_2$-solution.
The atoms consequently show a higher degree of cavity-induced density order and the effective sublattice potential becomes large enough to be comparable with $U$. 
In this regime, we have the resonant energy transfer between the atoms and cavity which includes the on-site interaction, i.e.~$E_n\!-\!E_m\!\approx\!2\hbar g\lambda_2\!-\!U$.
One can therefore no longer omit transitions between states with either two neighboring singly occupied sites or a double occupancies described in the second line of Eq.~(\ref{eq:EOM_1D}) with $p\!=\!1$ (purple and blue lines in Fig.~\ref{fig:EOM_terms_U40_g12_delta8_Gamma1}).
Contributions from transitions between high energy states i.e.~$E_n\!-\!E_m\!\approx\!2\hbar g\lambda_2\!+\!U$, captured by the second line of Eq.~(\ref{eq:EOM_1D}) with $p\!=\!-1$ (green lines in Fig.~\ref{fig:EOM_terms_U40_g12_delta8_Gamma1}) remain vanishingly small compared to the other processes throughout the region where our system has a solution.

Increasing the pump strength further to $\hbar g_{\text{bi},2}/J\!=\!13.5$ the bistability disappears as the low $\lambda_1$ solution becomes unstable. The unstable solution is often characterized by a decreasing photon number with increasing the coupling. 
However, this behavior is not generally the hallmark of unstable solutions, as the decrease can also come from the temperature effects as seen for even large couplings, but one needs to employ the criterion given in Eq.~(\ref{eq:condition}).

For higher global coupling, the temperature starts to rise drastically [see Fig.~\ref{fig:cuts_U40_gvar_delta8_Gamma1}~(b)], which leads to a strong decrease of the scaled photon number and the atomic density imbalance between the sublattices [Fig.~\ref{fig:cuts_U40_gvar_delta8_Gamma1}~(a),(c)], but also an increased occupation of the excited states for the $\lambda_2$ solution, leading to a growing average double occupancy [Fig.~\ref{fig:cuts_U40_gvar_delta8_Gamma1}~(d)].

The width of the bistability increases approximately linearly with $U/J$ as can be seen in Fig.~\ref{fig:bistable_region_plots_quarter_filling}~(a). At fixed $\hbar\delta/J$ the interaction strength controls the position of the resonances between the excited states from which the bistability originates. 

At large $U/J$ by analyzing Eq.~(\ref{eq:system_of_equations_1D}) we derive an approximate expression for lower onset of the bistability $g_{\text{bi},1}$ (see Supplemental Material of Ref.~\cite{TolleHalati2024}). In the limit $\beta U\!\gg\!1$, we find from Eq.~(\ref{eq:pc_1D}) that fixing the particle number the chemical potential scales approximately linearly with the temperature, i.e. $\mu\!\approx\! -\log (2) k_B T$. Using this expression for the chemical potential, the self-consistency equation Eq.~(\ref{eq:sc_1D}) can be simplified and solved for the temperature. 

Both expressions can be inserted into Eq.~(\ref{eq:EOM_1D}). Analyzing this equation further by keeping only the terms which are important around the lower onset of the bistability and using Padé-approximations around the point $\lambda\!=\! \delta g\tanh (2)/(\varGamma^2/4\!+\!\delta^2)$  \cite{Baker_Graves-Morris_1996} we find that the onset follows approximately (see Supplemental Material of Ref.~\cite{TolleHalati2024})
\begin{equation}
    \label{eq:gbi1_approx}
    g_{\text{bi},1}\!=\!\sqrt{(U/\hbar-\delta)\left[(\varGamma/2)^2+\delta^2\right]/(2\delta)}.
\end{equation}
This is in stark contrast to the almost $U/J$-independent critical coupling $g_\text{cr}(\beta\!\ll\!1)$ which can be approximated as in Eq.~(\ref{eq:gcr_approx_highT}) for the considered parameters and emphasizes the importance of the on-site interaction and the cavity energy scales for the appearance bistability region
(see red line in Fig.~\ref{fig:bistable_region_plots_quarter_filling}). 

A direct comparison of the analytic approximation (red line in Fig.~\ref{fig:bistable_region_plots_quarter_filling}) and the numerical results (lower black line in Fig.~\ref{fig:bistable_region_plots_quarter_filling}) shows a very good agreement of both results. The increase of $g_{\text{bi},1}$ with the interaction strength is well reproduced in Fig.~\ref{fig:bistable_region_plots_quarter_filling}~(a) and only small deviations at lower value of $U/J$ are found. Also the increase of $g_{\text{bi},1}$ with the detuning  in Fig.~\ref{fig:bistable_region_plots_quarter_filling}~(b) is well reproduced by the approximation, and only deviations occur for large $\delta$. Additionally, the dependence on the dissipation strength is very well covered by the approximation and almost no deviations are seen. 

The upper boundary $g_{\text{bi},2}$ seems to evolve almost linearly with $U/J$, such that a wider bistable region is present at larger interactions. In contrast, increasing $\delta$ causes the bistability region to become smaller and to disappear beyond a certain large value of $\delta$.
Whereas the lower boundary of the bistability region shows a similar rise with $\delta$ as the self-organization transition, the upper boundary becomes almost independent of the detuning above $\delta\!\sim\!\varGamma$.
Varying the dissipation in Fig.~\ref{fig:density_plots_gvGamma_U8_delta6} at sufficiently large on-site interactions, a narrow bistability region appears, but disappears at larger values of $\hbar\varGamma/J$. 
As with increasing dissipation the effective temperature of the system also grows drastically, most features are washed out due to the loss of coherence. Therefore, thermal fluctuations dominate and no self-organization is found. 
At higher on-site interactions $U\!\gg\!\hbar\varGamma$, the width of the bistability region is almost independent of $\varGamma$, whereas the coupling strengths marking its edges increase simultaneously consistent to a quadratic scaling [hatched region in Fig.~\ref{fig:bistable_region_plots_quarter_filling}~(c)]

In this section, we discussed the self-ordering transition at low filling and identified the cooling mechanisms related to atomic resonances as well as the fluctuations-induced bistability within the self-ordered phase. Since our analytical results are derived for general low, incommensurate filling, the effects prevail for different fillings $n\!\leq\!1/2$. Numerical results at $n\!\leq\!1/3$ confirmed that the observations discussed stay qualitatively the same.
In the following, we investigate if these phenomena prevail and how they are different at commensurate filling.

\subsection{Half filling ($n\!=\!1$) \label{sec:half_filling}}

In this section, we discuss the case of half filling ($n\!=\!1$) for the atoms in the Hubbard model. At this commensurate filling, the Hubbard model is well known to exhibit additionally a Mott-insulating phase. Thus, we can expect a different behavior upon the coupling to the cavity for this filling.

In the absence of the pump beam realizing the coupling to the cavity, the Fermi Hubbard system at half filling can be in a liquid or a Mott-insulating phase depending on the interaction strength. 
Switching on the coupling, a finite density imbalance can only be achieved with an increased density of double occupancies in the semiclassical limit on the low potential sites (see Sec.~\ref{sec:ionic_hubbard} for details).
We expect that across the self-ordering transition an initial Mott-insulating state is destroyed and the character of the fermionic state changes to a state with finite density imbalance in the steady state. 
Taking fluctuations in the coupling perturbatively into account we still find a bistability, i.e.~a region of pumping strengths for which two or more steady states with a distinct finite photon number exist. 
However, the nature of this bistability at half filling is fundamentally different from the bistability we introduced in Sec.~\ref{sec:fluctuation_induced_bistability}. In the case of half filling the nature of the ground state of the effective model, changes its character due to a closing of the gap to the lowest excitations from a Mott-insulator to a charge density wave. 
Such bistabilities are typically associated with a dissipative first-order transition \cite{LarsonLewenstein2008, LarsonLewenstein2008b, Minganti_Ciuti_2018}, therefore, they also present at the $T\!=\!0$ MF level. However, to our knowledge this is the first time the bistability occurring at the self-organization transition of the half-filled Fermi-Hubbard model coupled to the cavity has been described.
Furthermore, we show that by varying the cavity parameters, the detuning $\delta$ and the dissipation strength $\varGamma$, we obtain important deviations between the self-consistent thermal state determined by the fluctuations and the $T\!=\!0$ MF approach.

\subsubsection{Self-ordering transition and bistable solutions}
\label{sec:self_organization_transition_half_filling}

\begin{figure}[h]
    \begin{tikzpicture}
    \node[anchor=north west,inner sep=0pt] at (0,0){\includegraphics[width=0.225\textwidth]{Fig19a}};
    \node[font=\normalfont] at (1ex,-2ex) {(a)};
    \end{tikzpicture}
    \begin{tikzpicture}
    \node[anchor=north west,inner sep=0pt] at (0,0){\includegraphics[width=0.225\textwidth]{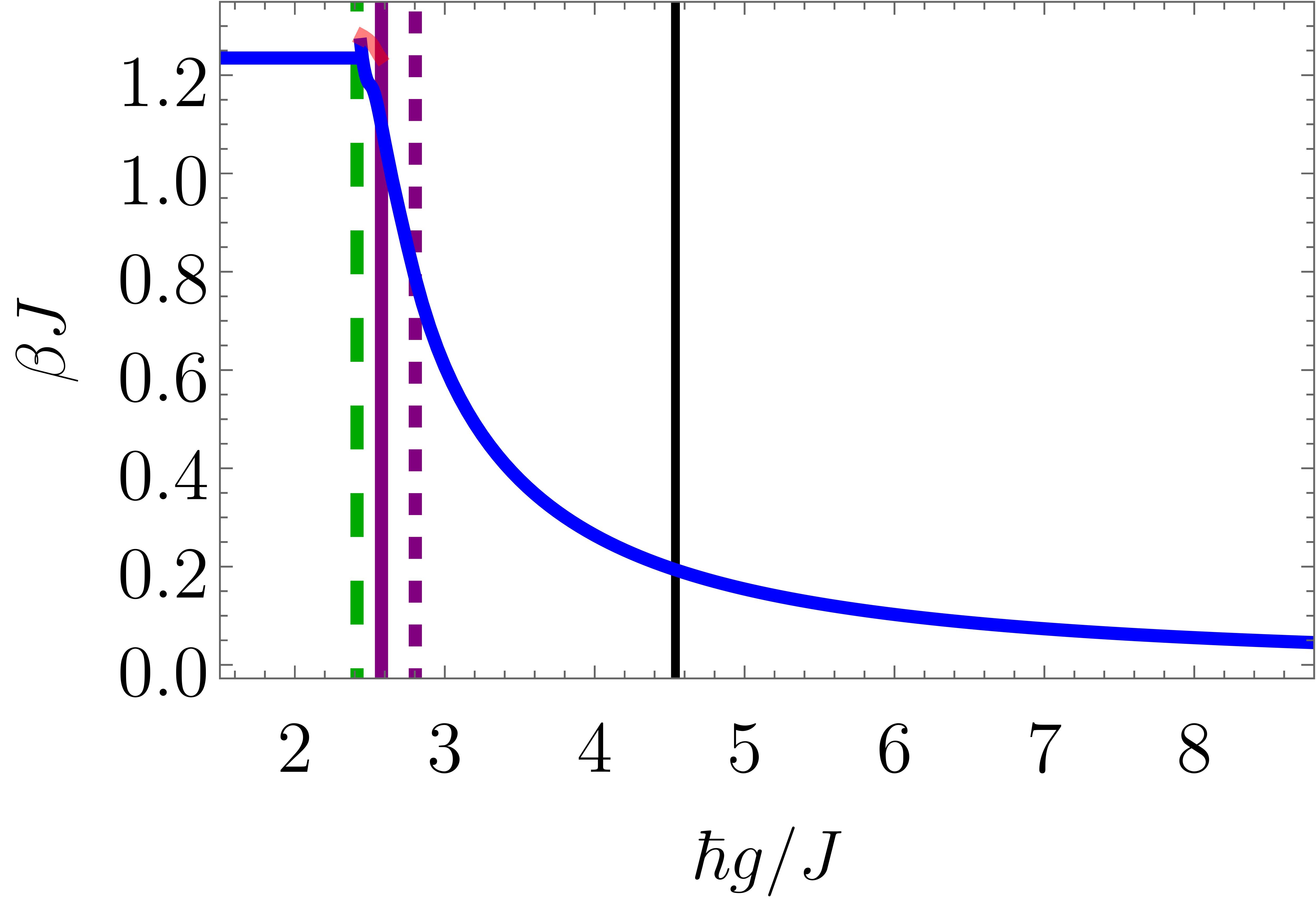}};
    \node[font=\normalfont] at (1ex,-2ex) {(b)};
    \end{tikzpicture}
    \begin{tikzpicture}
    \node[anchor=north west,inner sep=0pt] at (0,0){\includegraphics[width=0.225\textwidth]{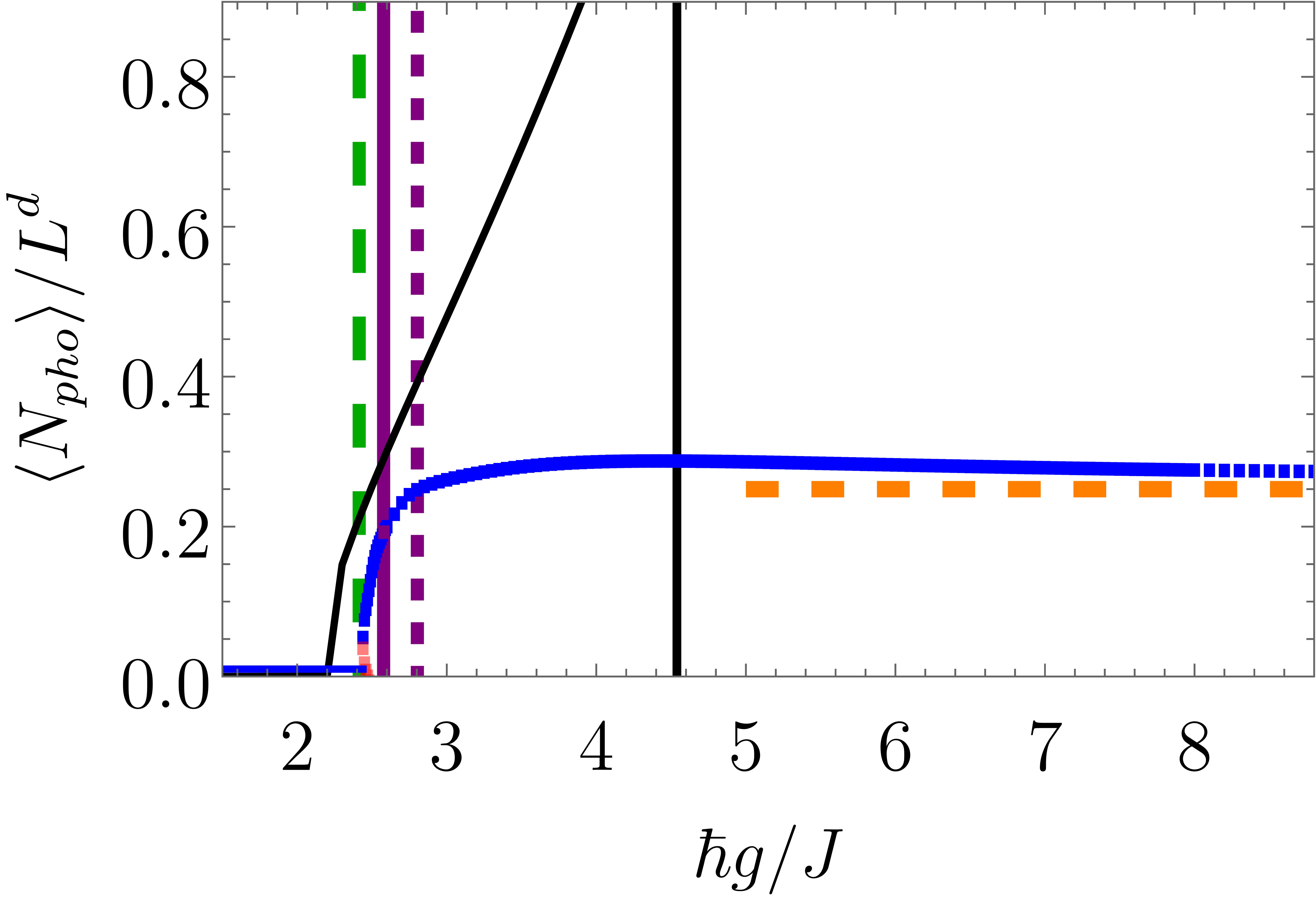}};
    \node[font=\normalfont] at (1ex,-2ex) {(c)};
    \end{tikzpicture}
    \begin{tikzpicture}
    \node[anchor=north west,inner sep=0pt] at (0,0){\includegraphics[width=0.225\textwidth]{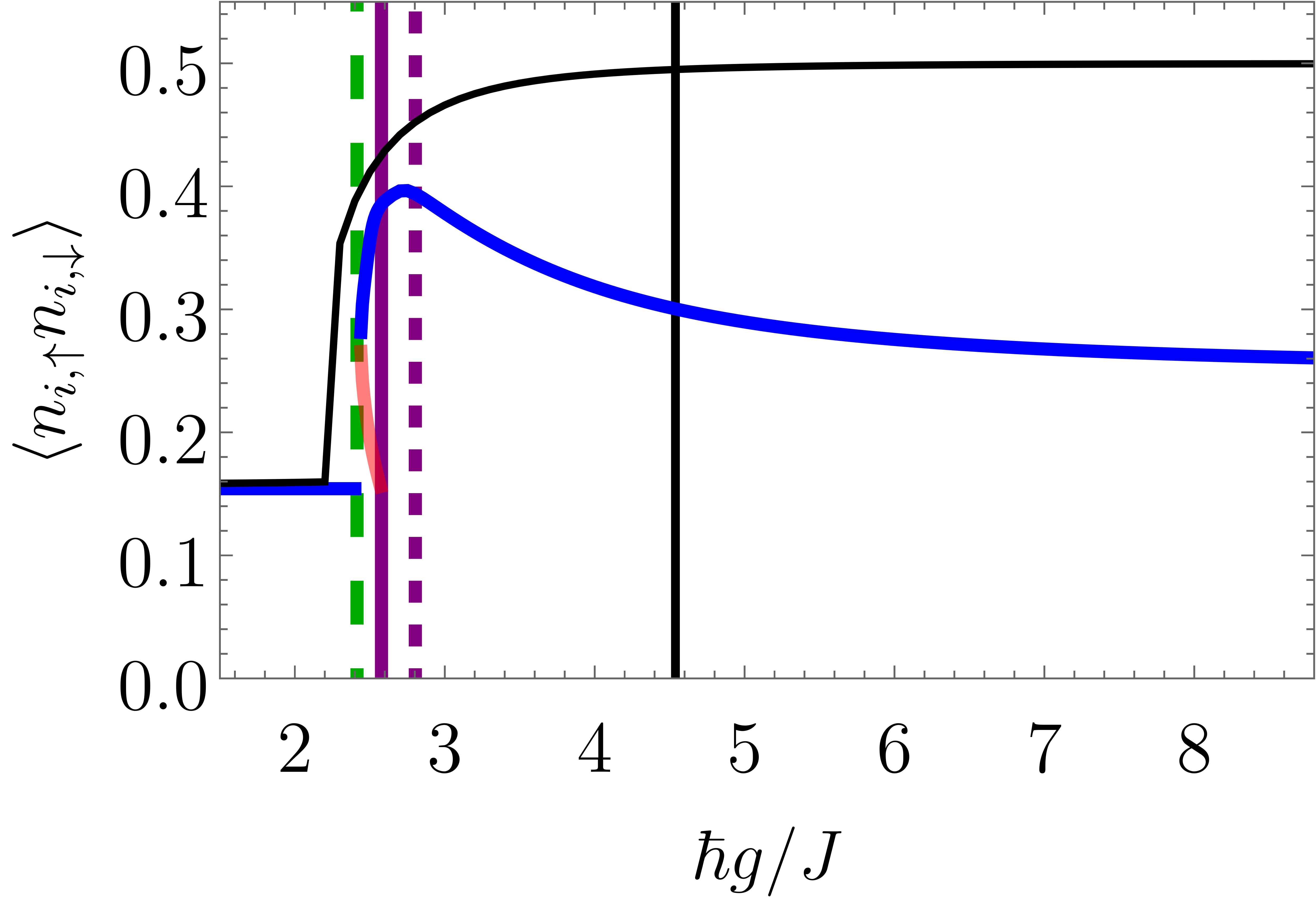}};
    \node[font=\normalfont] at (1ex,-2ex) {(d)};
    \end{tikzpicture}
    \includegraphics[width=0.4\textwidth]{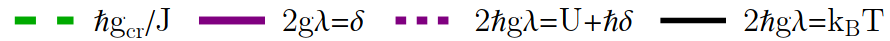}
    \vspace{-10pt}
    \caption[]{(a) Sublattice imbalance $\langle\hat{\Delta}\rangle/L^d$, (b) inverse temperature $\beta J$, (c) scaled photon number and (d) average number of double occupancies as a function of atoms-cavity coupling $\hbar g/J$ at $\hbar\delta/J\!=\!4$; of a finite size system $L\!=\!6$ at half filling for stable (blue) and unstable (red) results. The $T\!=\!0$ MF solution is shown as black solid curve. The parameters used are $U/J\!=\!2$, $\hbar\varGamma/J\!=\!1$. Vertical lines denote the critical coupling $\hbar g_\text{cr}/J$ and resonances in the atomic limit. The orange dashed lines are the approximate scalings for $U/J\!=\!0$ [Eqs.~(\ref{eq:imbalance_universal_scaling_U0})].}
    \label{fig:cuts_U2_gvar_delta4_Gamma1_half_filling}
\end{figure}

\begin{figure}[h]
    \begin{tikzpicture}
    \node[anchor=north west,inner sep=0pt] at (0,0){\includegraphics[width=0.225\textwidth]{Fig20a}};
    \node[font=\normalfont] at (1ex,-2ex) {(a)};
    \end{tikzpicture}
    \begin{tikzpicture}
    \node[anchor=north west,inner sep=0pt] at (0,0){\includegraphics[width=0.225\textwidth]{Fig20b}};
    \node[font=\normalfont] at (1ex,-2ex) {(b)};
    \end{tikzpicture}
    \begin{tikzpicture}
    \node[anchor=north west,inner sep=0pt] at (0,0){\includegraphics[width=0.225\textwidth]{Fig20c}};
    \node[font=\normalfont] at (1ex,-2ex) {(c)};
    \end{tikzpicture}
    \begin{tikzpicture}
    \node[anchor=north west,inner sep=0pt] at (0,0){\includegraphics[width=0.225\textwidth]{Fig20d}};
    \node[font=\normalfont] at (1ex,-2ex) {(d)};
    \end{tikzpicture}
    \vspace{-10pt}
    \caption[]{(a) Sublattice imbalance $\langle\hat{\Delta}\rangle/L^d$ and (b) inverse temperature $\beta J$, (c) scaled photon number and (d) average number of double occupancies as a function of atoms-cavity coupling $\hbar g/J$ at $\hbar\delta/J\!=\!4$; of a finite size system $L\!=\!6$ for stable (blue) and unstable (red) results; in the weak tunneling perturbation in the thermodynamic limit [Eqs.~(\ref{eq:system_of_equations_1D})] stable (dark green) at half filling. The $T\!=\!0$ MF solution is shown as a black (stable) or gray (unstable) solid curve. The parameters used are $U/J\!=\!8$ and $\hbar\varGamma/J\!=\!1$. Vertical lines denote the critical coupling $\hbar g_\text{cr}/J$ (green dashed). The orange dashed line is the approximate scaling for $U/J\!=\!0$ [Eqs.~(\ref{eq:imbalance_universal_scaling_U0})].}
    \label{fig:cuts_U8_gvar_delta4_Gamma1_half_filling}
\end{figure}

To start we discuss the results of the $T\!=\!0$ MF results. 
We show an example of the self-organization transition for intermediate interaction strength $U/J\!=\!2$ in Fig.~\ref{fig:cuts_U2_gvar_delta4_Gamma1_half_filling} and for larger interaction strength $U/J\!=\!8$ in Fig.~\ref{fig:cuts_U8_gvar_delta4_Gamma1_half_filling}, where we show with the black line the result of the $T\!=\!0$ MF method.
Within the $T\!=\!0$ MF method, the photon number, the density imbalance and the average value of the double occupancies exhibit a large value in the self-organized solution at the transition for both shown values of $U/J$. 
This stems from the property that at commensurate filling a finite density imbalance coincides with the creation of double-occupancies on the low-potential sublattice. At larger interaction strength, $U/J\!=\!8$, the emergence of a bistable behavior around the self-ordering transition can be seen in Fig.~\ref{fig:cuts_U8_gvar_delta4_Gamma1_half_filling} for $T\!=\!0$ MF. The stable solution with homogeneous density distribution and zero photon field and the self-ordered steady state above a critical coupling $g^\text{MF}_\text{cr}$ (black points) coexist and are connected by an unstable branch (gray points). 
For the plot shown here, the gray line is partly covered by the red curve.
The sublattice imbalance and the double occupancy approach their maximal value and the photon number continues to rise very steeply with the pump strength.
At intermediate interaction strength $U/J\!=\!2$ a bistability also appears to occur, but the region of the instability seems to be much more narrow (Fig.~\ref{fig:cuts_U2_gvar_delta4_Gamma1_half_filling}). 
Due this very narrow bistable region it becomes numerically difficult to identify it unambiguously.

\begin{figure}[!hbtp]
\begin{flushleft}
    \begin{tikzpicture}
    \node[anchor=north west,inner sep=0pt] at (0,0){\includegraphics[height=0.18\textheight]{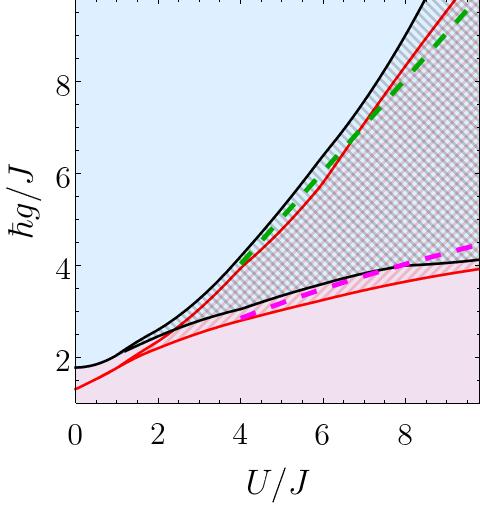}};
    \node[font=\normalfont] at (2ex,-2ex) {(a)};
    \end{tikzpicture}
    \begin{tikzpicture}
    \node[anchor=north west,inner sep=0pt] at (0,0){ \includegraphics[height=0.18\textheight]{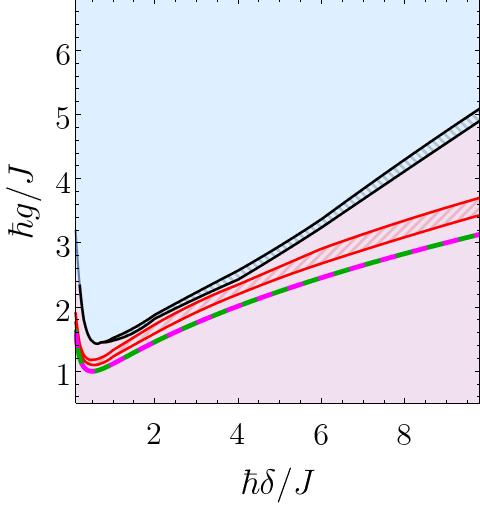}};
    \node[font=\normalfont] at (2ex,-2ex) {(b)};
    \end{tikzpicture}
    \begin{tikzpicture}
    \node[anchor=north west,inner sep=0pt] at (0,0){\includegraphics[height=0.18\textheight]{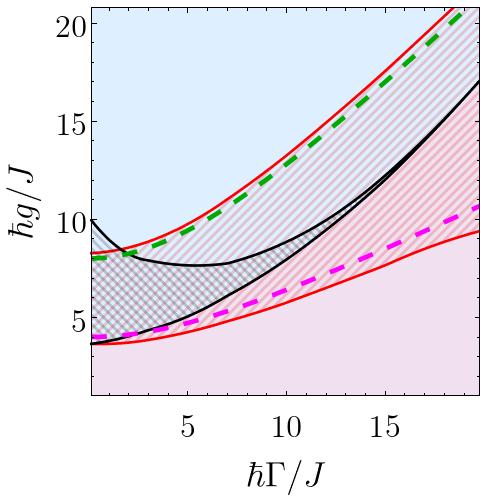}};
    \node[font=\normalfont] at (2ex,-2ex) {(c)};
    \end{tikzpicture}
    \end{flushleft}
    \vspace{-20pt}
    \caption[]{Simplified phase diagrams varying the atoms-cavity coupling $\hbar g/J$ and (a) the on-site interaction $U/J$ at $\hbar\delta/J=4$, $\hbar\varGamma/J\!=\!1$; (b) the pump-cavity detuning $\hbar\delta/J$ at $U/J\!=\!2$, $\hbar\varGamma/J\!=\!1$; (c) the dissipation rate $\hbar\varGamma/J$ at $\hbar\delta/J\!=\!4$, $U/J\!=\!8$ of a finite size system $L\!=\!6$ at half filling. The lower black line marks the transition between normal [purple region] and the high-$\lambda$ solution in the self-organized phase [blue region] determined by solving Eq.~(\ref{eq:cav_MF}) and Eq.~(\ref{eq:energy_transfer}) with ED, $\hbar g_\text{cr}$. At half filling it coincides with the onset of the bistability $\hbar g_\text{bi,1}$. The black hatched region marks the region with two stable solutions between $\hbar g_\text{bi,1}$ and $\hbar g_\text{bi,2}$. The lower red line marks the critical coupling $\hbar g^\text{MF}_\text{cr}$ obtained for the $T\!=\!0$ MF approach that coincides with  $\hbar g^\text{MF}_\text{bi,1}$. The red hatched region marks the bistability between $\hbar g^\text{MF}_\text{bi,1}$ and the upper red line $\hbar g^\text{MF}_\text{bi,2}$.
    The dashed lines are approximations of the critical couplings $g_\text{cr}^\text{MF}$ [Eq.~(\ref{eq:gcrMFhalffilling})] (magenta dashed) and $\hbar g^\text{MF}_\text{bi,2}$ (green dashed), determined by solving Eq.~(\ref{eq:cav_MF}) at maximal imbalance.}
    \label{fig:bistable_region_plots_half_filling}
\end{figure}

The narrowing at small interaction strength of the parameter region where two stable solutions are present can be seen varying the on-site interaction $U/J$ and atoms-cavity coupling $\hbar g/J$ in 
Fig.~\ref{fig:bistable_region_plots_half_filling}~(a), marked for the $T\!=\!0$ MF results by the red hatched region. At small interaction strength it is very hard to identify whether a bistability is present. In contrast, at large interaction strength a broad bistable region is present. We would like to emphasize that the nature of the bistabilty at half filling differs from the fluctuation-induced bistability at low filling. 
The onset of the bistability here coincides with the self-organization transition. We attribute this to the change of the nature of the ground state of the effective ionic Hubbard model from the Mott-insulating state to a charge density wave \cite{LarsonLewenstein2008b}. This transition we expect around $2\hbar g\lambda\! \approx \!U$ for large $U$ and large $g\lambda$ (as discussed in Sec.~\ref{sec:ionic_hubbard}), where $\lambda$ corresponds to the stable self-ordered solution. 
The self-consistency condition assuming maximal density imbalance is given by $\langle\hat{\Delta}\rangle\!=\!N$.
The resulting critical coupling scales approximately as 
\begin{equation}
\label{eq:gcrMFhalffilling}
    g_\text{cr}^{MF}\approx\sqrt{\frac{U}{2n}\frac{\delta^2+(\varGamma/2)^2}{\delta}}
\end{equation}
(magenta dashed line in Fig.~\ref{fig:bistable_region_plots_half_filling}).
Here, the dependence on the detuning and the dissipation enters as already seen at lower fillings. The approximate scaling with the $\sqrt{U}$ agrees roughly with the scaling obtained numerically in Fig.~\ref{fig:bistable_region_plots_half_filling}~(a). For the commensurate filling we cannot make the assumption of a small cavity field. Therefore, the considerations made for the critical coupling strength at lower fillings in Sec.~\ref{sec:self_organization_transition} do not apply here.

Another approach to determine the bistability edges involves the stability condition derived in Eq.~(\ref{eq:stability_condition_MF}). Assuming the solution connecting low-and high field stable solution to be unstable, the coupling for which the condition has a zero with multiplicity 2 is roughly at the two critical couplings. We explored this approach numerically and find reasonable agreement with our previously obtained results from solving Eq.~(\ref{eq:cav_MF}) and Eq.~(\ref{eq:energy_transfer}) via ED.

We study in the following what happens to the predicted behavior around the transition when adding the contributions from the fluctuations in the atoms-cavity coupling. Whereas the bistability survives, many other drastic differences occur. 
Including fluctuations, the behavior of the average photon number at low couplings looks similar to the $T\!=\!0$ MF one. In Fig.~\ref{fig:cuts_U2_gvar_delta4_Gamma1_half_filling} and Fig.~\ref{fig:cuts_U8_gvar_delta4_Gamma1_half_filling} we show the results for a numerical solution of the finite system for the self-ordered phase solution (blue points) which are connected by the unstable branch (red points) to the normal phase. 
Additionally, the results for the perturbative calculations in the self-ordered phase, that also hold for higher dimensions of the atomic sublattice (see Sec.~\ref{sec:pertubation_J}) are plotted as dark green points.
The self-organized solution is connected by the unstable solution to the normal phase, which we omit in the plots for better readability. A sudden rise of the photon occupation occurs at the critical coupling $g_\text{cr}$, both for $U/J\!=\!2$ (see Fig.~\ref{fig:cuts_U2_gvar_delta4_Gamma1_half_filling}) and at larger interactions, $U/J\!=\!8$, at the bistability (see Fig.~\ref{fig:cuts_U8_gvar_delta4_Gamma1_half_filling}). 
The emergence of a bistable behavior around the self-ordering transition can be seen clearly in Fig.~\ref{fig:cuts_U8_gvar_delta4_Gamma1_half_filling} around $\hbar g/J\!\approx\!3.8$.

Despite those similarities of the photon number to the $T\!=\!0$ MF study, important differences due to the presence of fluctuations in the atoms-cavity coupling are evident. Even below the self-ordering transition an effective finite temperature is induced by the coupling [see Fig.~\ref{fig:cuts_U2_gvar_delta4_Gamma1_half_filling} and Fig.~\ref{fig:cuts_U8_gvar_delta4_Gamma1_half_filling}~(b)]. 
The effective temperature is already sizable, around $k_B T\!\approx\!J$, causing a mixed steady state.
Comparing the transition point to the self-organized phase obtained in the presence of fluctuations, it appears at higher atoms-cavity coupling compared to the $T\!=\!0$ MF results. 
We find that a slight shift in the lower and the upper critical couplings to larger values of $\hbar g/J$ is seen. We interpret this that the finite temperature destroys small density orderings, preventing the system at lower coupling from entering the self-ordered phase.

\begin{figure}[h]
\begin{flushleft}
    \begin{tikzpicture}
    \node[anchor=north west,inner sep=0pt] at (0,0){\includegraphics[width=0.225\textwidth]{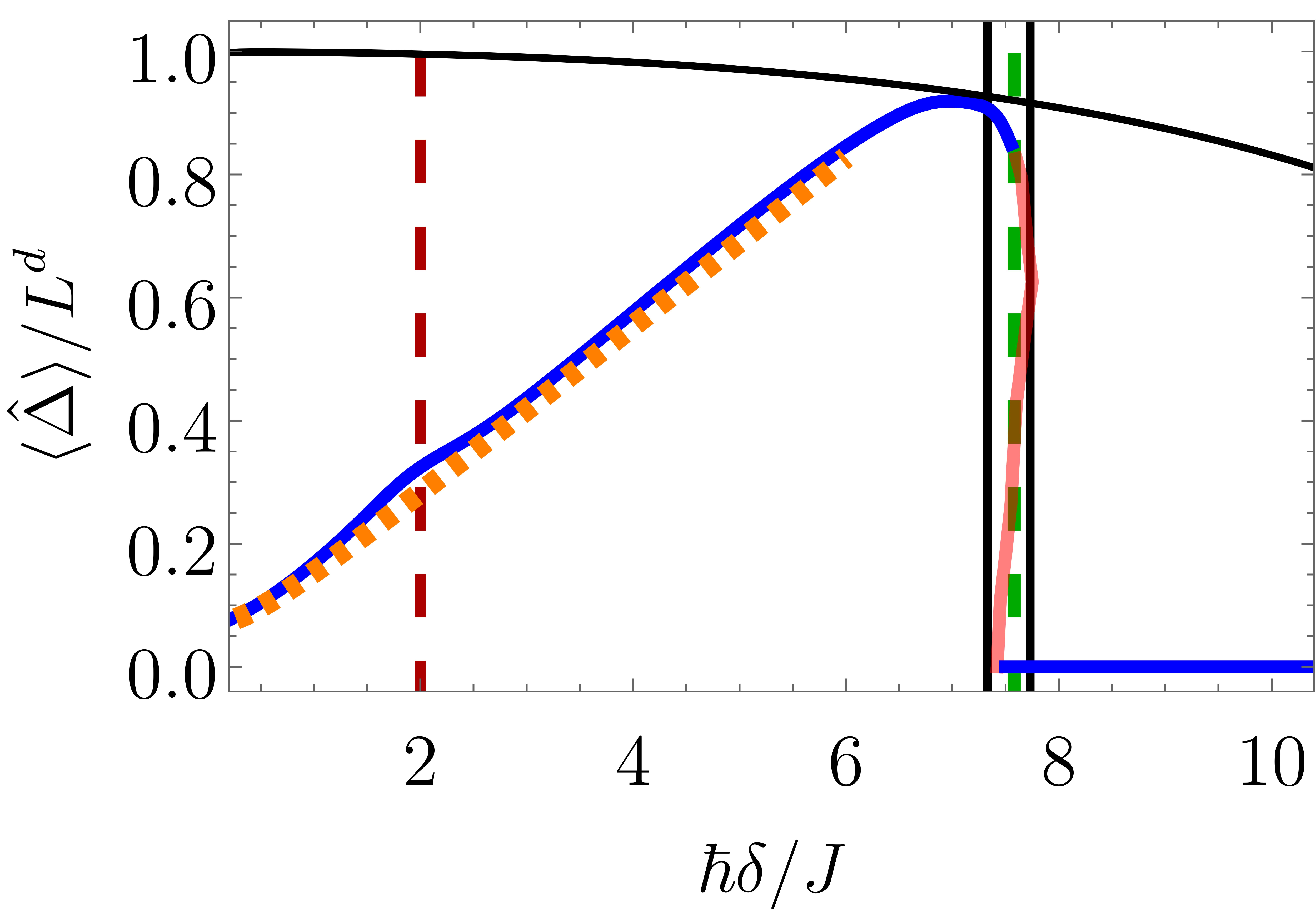}};
    \node[font=\normalfont] at (1ex,-2ex) {(a)};
    \end{tikzpicture}
    \begin{tikzpicture}
    \node[anchor=north west,inner sep=0pt] at (0,0){\includegraphics[width=0.225\textwidth]{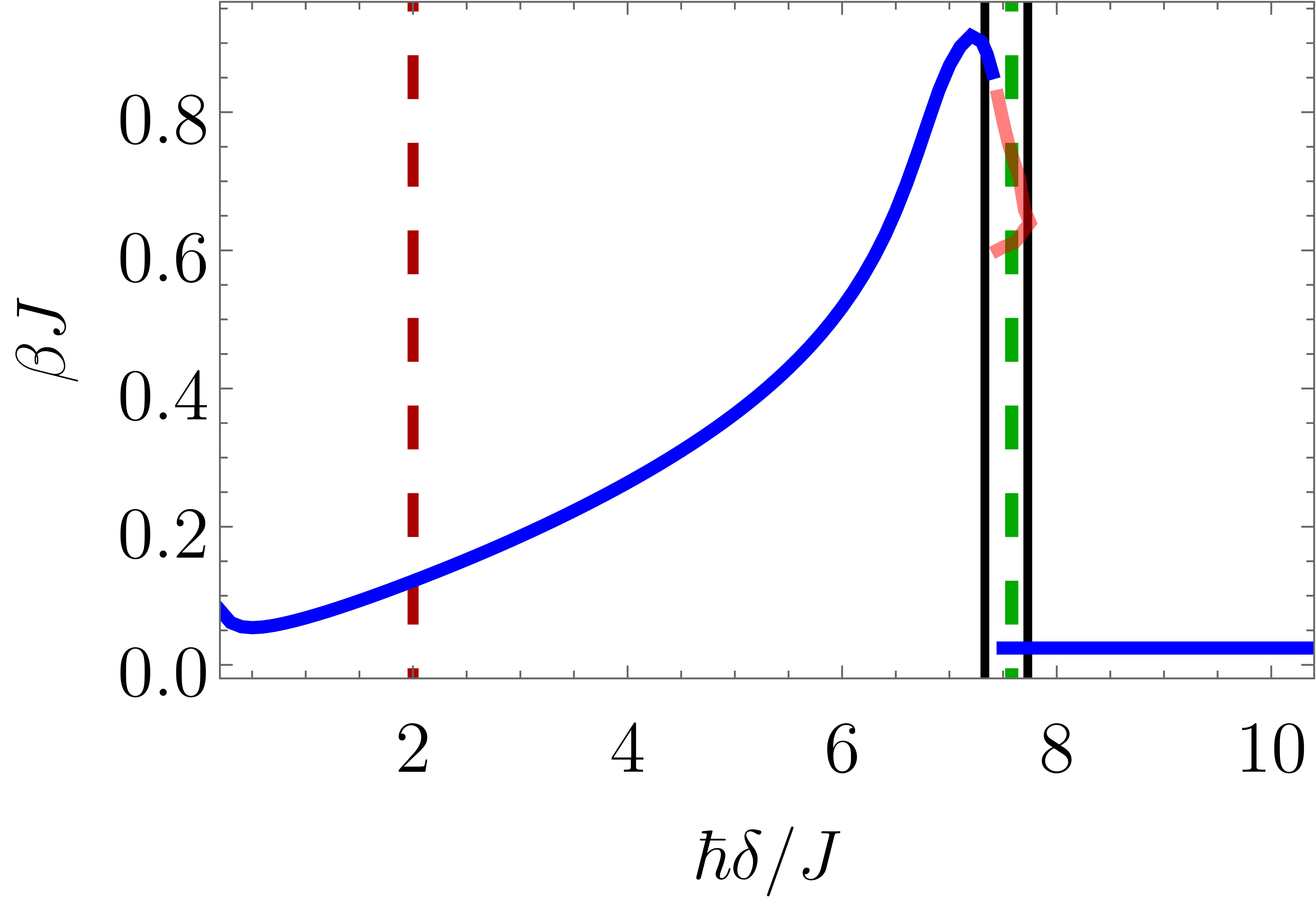}};
    \node[font=\normalfont] at (1ex,-2ex) {(b)};
    \end{tikzpicture}
    \begin{tikzpicture}
    \node[anchor=north west,inner sep=0pt] at (0,0){\includegraphics[width=0.225\textwidth]{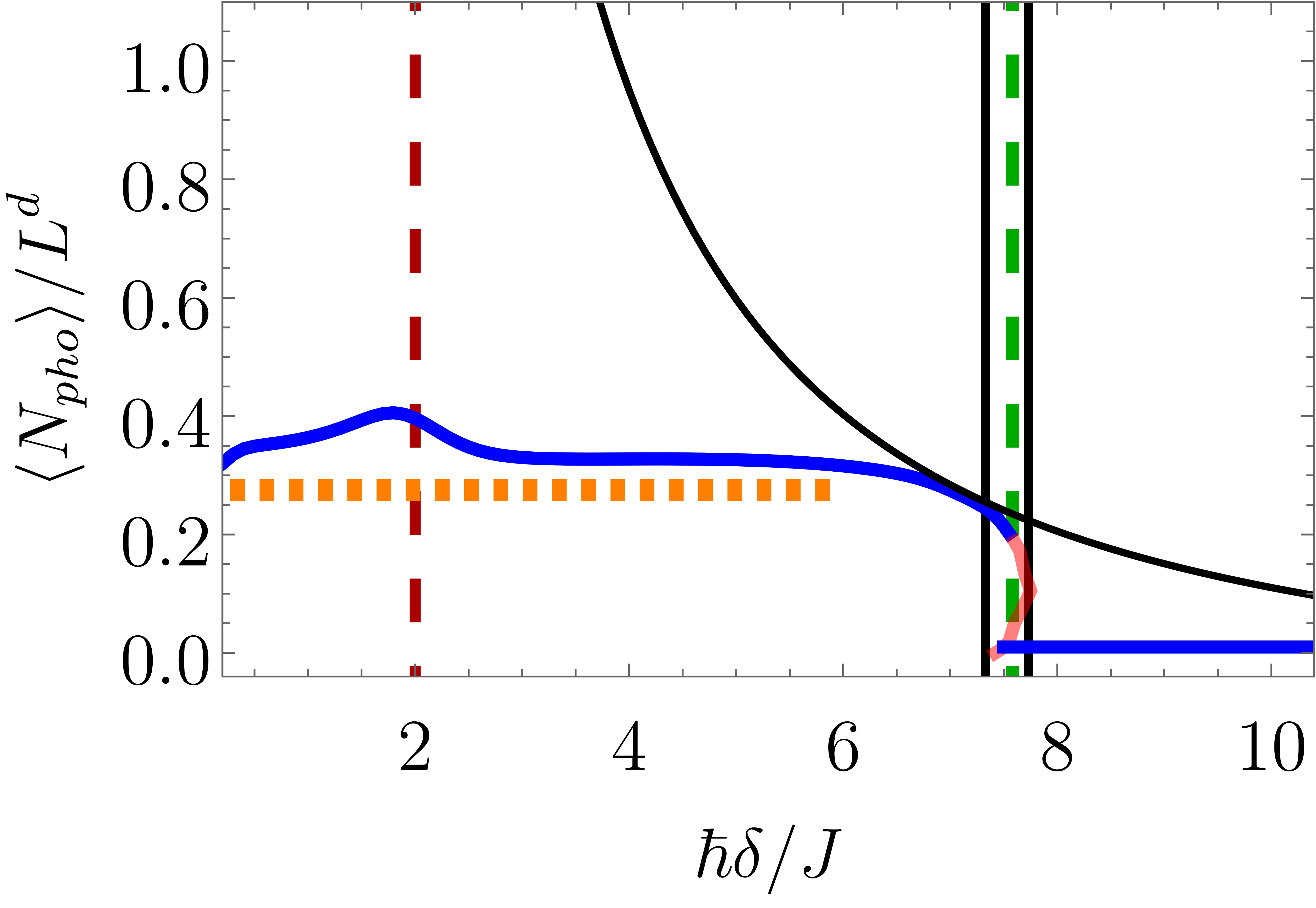}};
    \node[font=\normalfont] at (1ex,-2ex) {(c)};
    \end{tikzpicture}
    \begin{tikzpicture}
    \node[anchor=north west,inner sep=0pt] at (0,0){\includegraphics[width=0.225\textwidth]{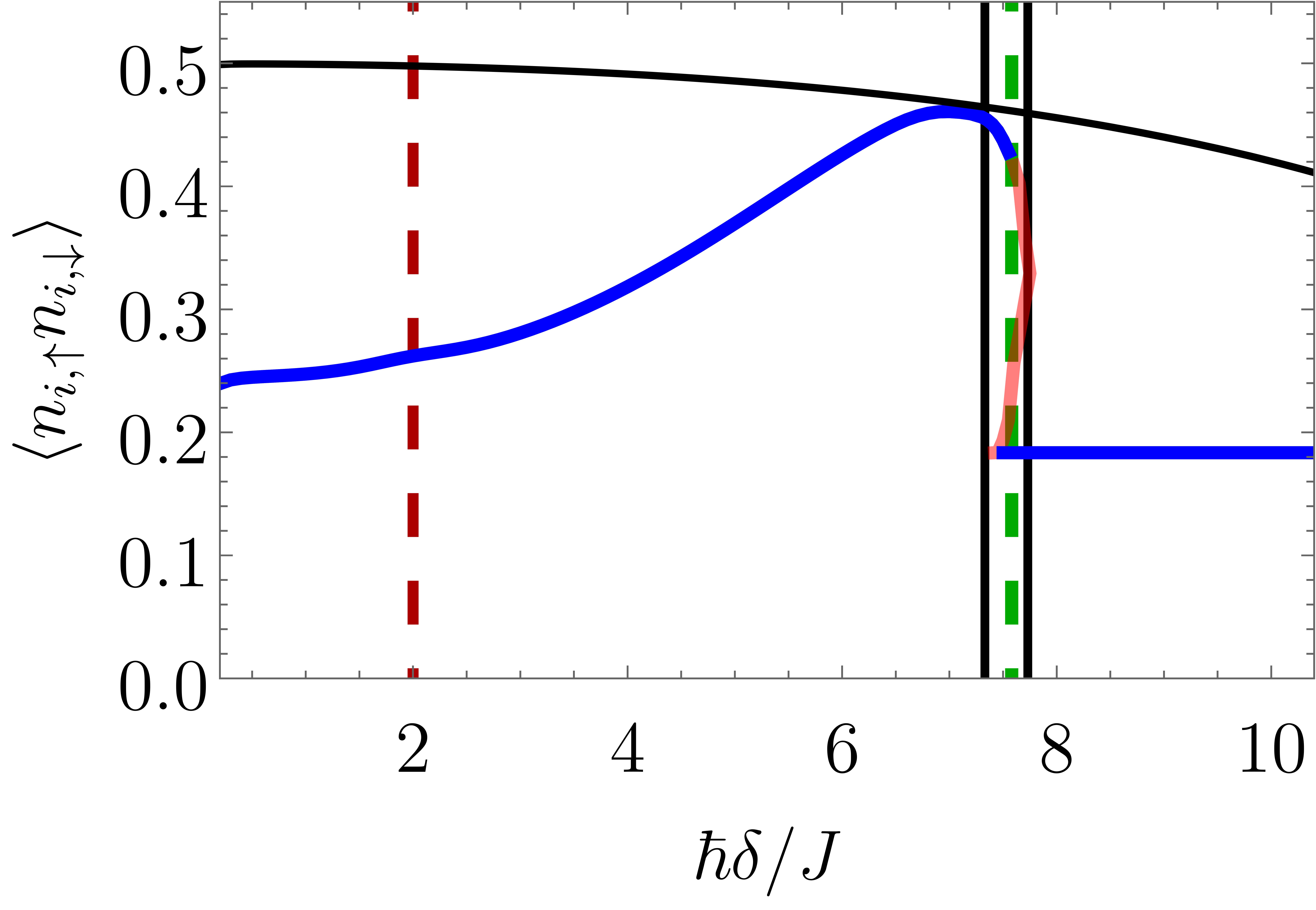}};
    \node[font=\normalfont] at (1ex,-2ex) {(d)};
    \end{tikzpicture}
    \includegraphics[width=0.3\textwidth]{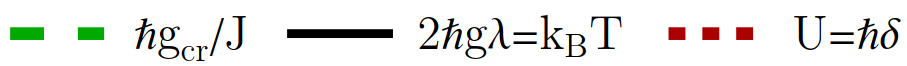}
    \end{flushleft}
    \vspace{-10pt}
    \caption[]{(a) Sublattice imbalance $\langle\hat{\Delta}\rangle/L^d$, (b) inverse temperature $\beta J$, (c) scaled photon number and (d) average number of double occupancies as a function of $\hbar\delta/J$ at $\hbar g/J\!=\!4$ of a finite size system $L\!=\!6$ at half filling for stable (blue) and unstable (red) results. The $T\!=\!0$ MF solution is shown as black solid curve. The parameters used are $U/J\!=\!2$, $\hbar\varGamma/J\!=\!1$. Vertical lines denote the critical coupling $\hbar g_\text{cr}/J$ and resonances. The orange dashed lines are the scalings of the analytical approximations for $U/J\!=\!0$ [Eqs.~(\ref{eq:imbalance_universal_scaling_U0})].}
    \label{fig:cuts_U2_g4_deltavar_Gamma1_half_filling}
\end{figure}
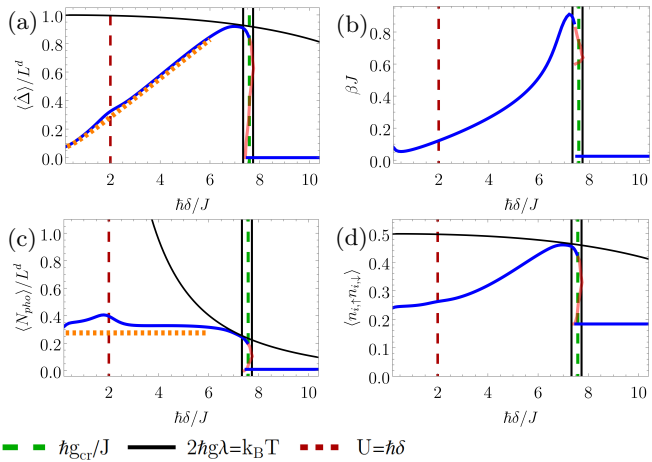

In the many-body adiabatic elimination results at coupling strengths above the critical coupling $g_\text{cr}$, a small maximum of the photon number occurs followed by a decrease and saturation at large coupling strength and strong interactions [see Fig.~\ref{fig:cuts_U8_gvar_delta4_Gamma1_half_filling}~(c)]. For smaller interactions, the maximum is slightly shifted into the self-ordered phase [see Fig.~\ref{fig:cuts_U2_gvar_delta4_Gamma1_half_filling}~(c)].
A similar behavior is evident in the density sublattice imbalance and the double occupancy which rises abruptly at the transition, shows a maximum and then decreases at even higher coupling strength. 
This is in strong contrast to the $T\!=\!0$ MF behavior, where the photon number increases with the coupling strength and the imbalance saturates at its maximal value. 
We attribute this drastic difference to the presence of the effective temperature. 
Similar to the previously discussed lower filling (Sec.~\ref{sec:self_organization_transition}), the temperature just above $g_\text{cr}$ for the solution with finite cavity field drops as the atoms order as can be seen in Fig.~\ref{fig:cuts_U8_gvar_delta4_Gamma1_half_filling} and Fig.~\ref{fig:cuts_U2_g4_deltavar_Gamma1_half_filling}~(b). 
After this drop, the effective temperature rises considerably after the onset of the bistability causing the heating and, thus, the reduction of the sublattice imbalance at larger pump strength.

\begin{figure}[h]
    \begin{tikzpicture}
    \node[anchor=north west,inner sep=0pt] at (0,0){\includegraphics[width=0.225\textwidth]{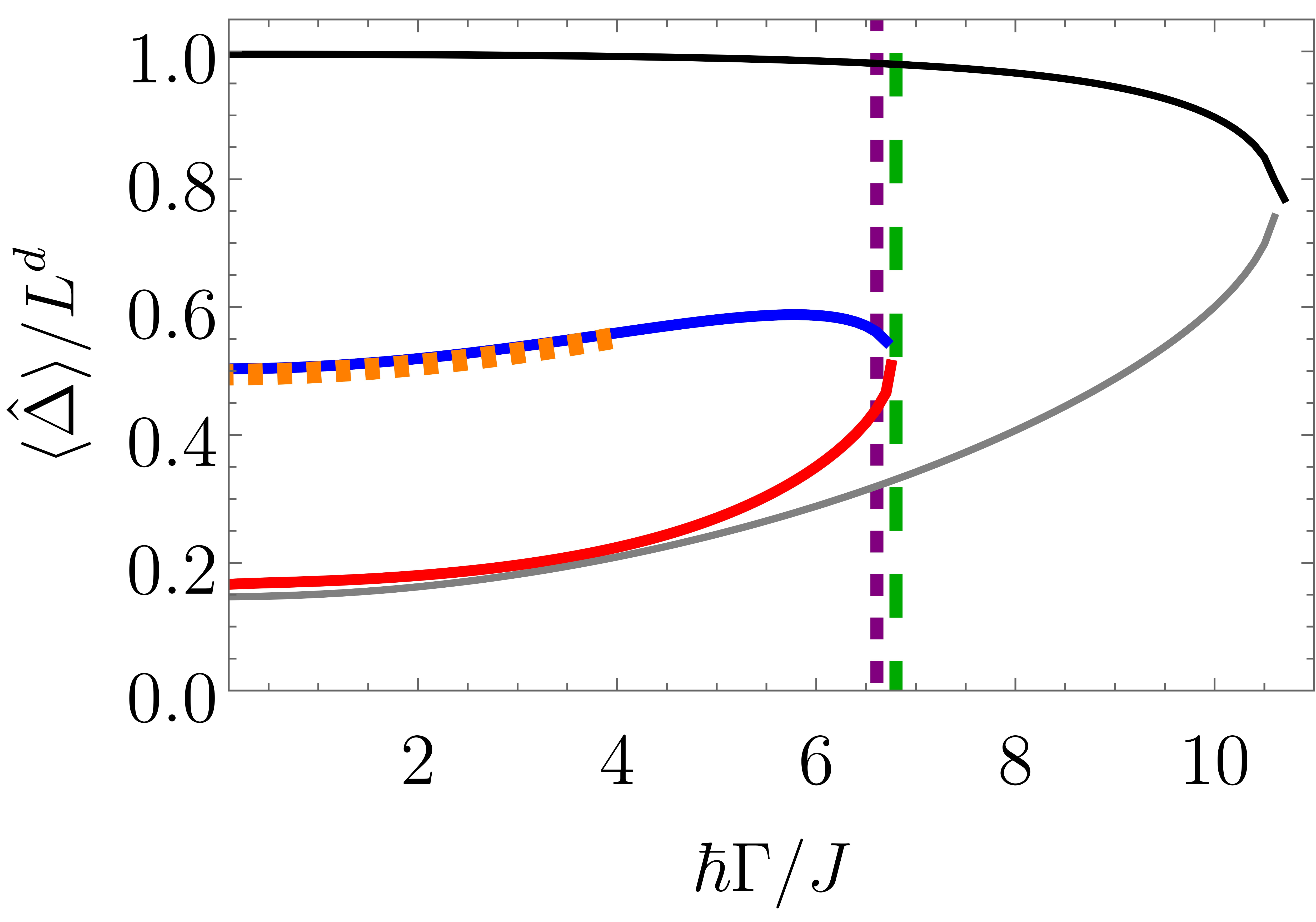}};
    \node[font=\normalfont] at (1ex,-2ex) {(a)};
    \end{tikzpicture}
    \begin{tikzpicture}
    \node[anchor=north west,inner sep=0pt] at (0,0){\includegraphics[width=0.225\textwidth]{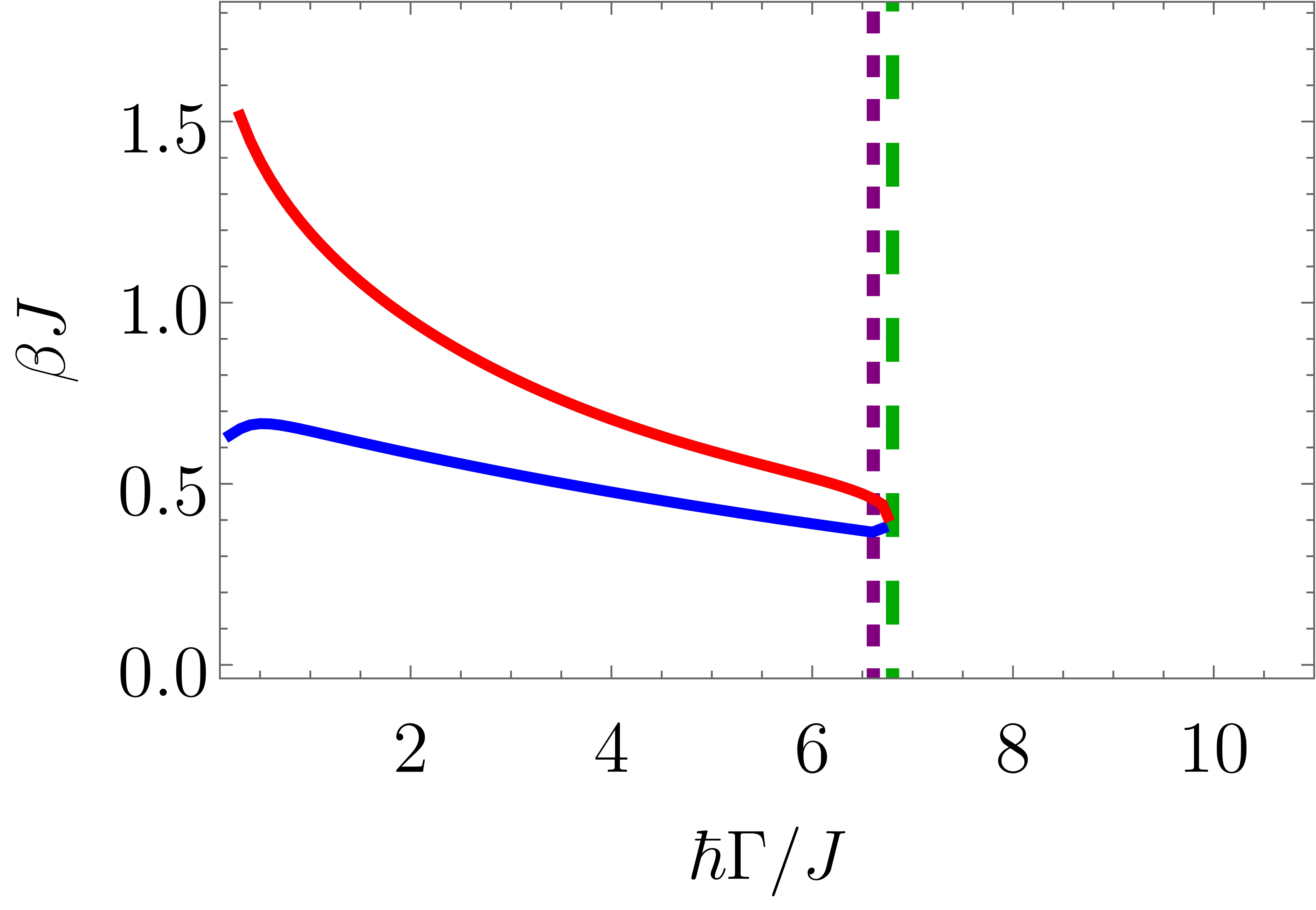}};
    \node[font=\normalfont] at (1ex,-2ex) {(b)};
    \end{tikzpicture}
    \begin{tikzpicture}
    \node[anchor=north west,inner sep=0pt] at (0,0){\includegraphics[width=0.225\textwidth]{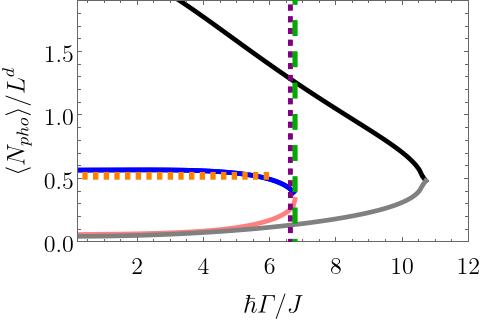}};
    \node[font=\normalfont] at (1ex,-2ex) {(c)};
    \end{tikzpicture}
    \begin{tikzpicture}
    \node[anchor=north west,inner sep=0pt] at (0,0){\includegraphics[width=0.225\textwidth]{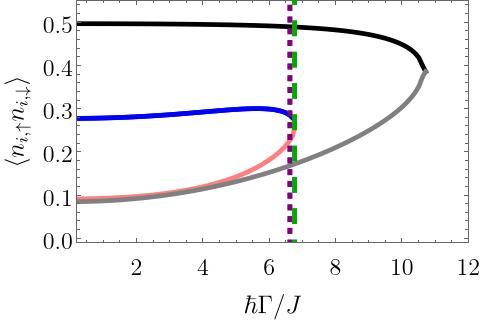}};
    \node[font=\normalfont] at (1ex,-2ex) {(d)};
    \end{tikzpicture}
    \caption[]{(a) Sublattice imbalance $\langle\hat{\Delta}\rangle/L^d$ and (b) inverse temperature $\beta J$, (c) scaled photon number and (d) average number of double occupancies as a function of $\hbar\varGamma/J$ of a finite size system $L\!=\!6$ at half filling for stable (blue) and unstable (red) results. The $T\!=\!0$ MF solution is shown as a black (stable) or gray (unstable) solid curve. The parameters used are $U/J\!=\!8$, $\hbar g/J\!=\!6$, $\hbar\delta/J\!=\!4$. The vertical lines denote $2\hbar g\lambda\!=\!U\!+\!\hbar\delta$ (purple dashed) and $\hbar g_\text{cr}/J$ (green dashed). The orange dashed line is the scaling of the analytical approximation for $U/J\!=\!0$ [Eqs.~(\ref{eq:imbalance_universal_scaling_U0})].}
    \label{fig:cuts_U8_g6_delta4_Gammavar_half_filling}
\end{figure}

The saturation of the photon number with increasing $\hbar g/J$ follows approximately the $U/J\!=\!0$ results derived in Eq.~(\ref{eq:imbalance_universal_scaling_U0}) applied for half filling [orange dashed line in Fig.~\ref{fig:cuts_U2_gvar_delta4_Gamma1_half_filling}~(a),(c) and Fig.~\ref{fig:cuts_U8_gvar_delta4_Gamma1_half_filling}~(a)]. The obtained scalings with $\hbar \delta/J$ and $\hbar\varGamma/J$ in Eq.~(\ref{eq:imbalance_universal_scaling_U0}) also agree with the results
[see Fig.~\ref{fig:cuts_U2_g4_deltavar_Gamma1_half_filling}~(a),(c) and Fig.~\ref{fig:cuts_U8_g6_delta4_Gammavar_half_filling}~(a)].
To be more specific, the finite $U/J$ leads to an offset of the curve for $U\!\gg\!\hbar\delta,\hbar\varGamma$, however the scaling seems unchanged in the presence of interaction.

In Fig.~\ref{fig:bistable_region_plots_half_filling}~(a) we show how the onset of the bistable behavior depends on the interaction strength. 
In general, the extension of the bistable region with varying $U/J$ is similar for the $T\!=\!0$ MF method and when fluctuations are included and follows approximately Eq.~(\ref{eq:gcrMFhalffilling}), which shows a scaling with $\sqrt{U}$. 
The extension of the bistable region increases considerably with the interaction strength, showing the importance of the interaction for the presence of the bistability. 
For both approaches lowering the interaction strength the bistability narrows until it can no longer be measured within a numerical accuracy of $10^{-5}$ in the coupling.
 
\begin{figure}[h]
    \begin{tikzpicture}
    \node[anchor=north west,inner sep=0pt] at (0,0){\includegraphics[width=0.235\textwidth]{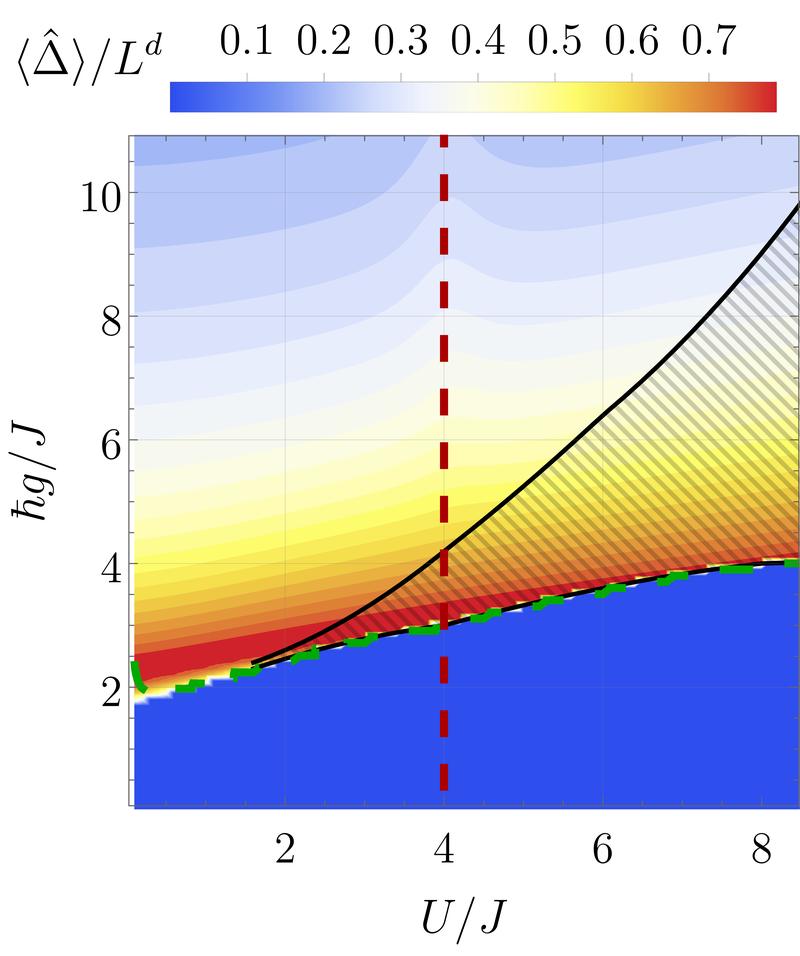}};
    \node[font=\normalfont] at (2ex,-6ex) {(a)};
    \end{tikzpicture}
    \begin{tikzpicture}
    \node[anchor=north west,inner sep=0pt] at (0,0){\includegraphics[width=0.235\textwidth]{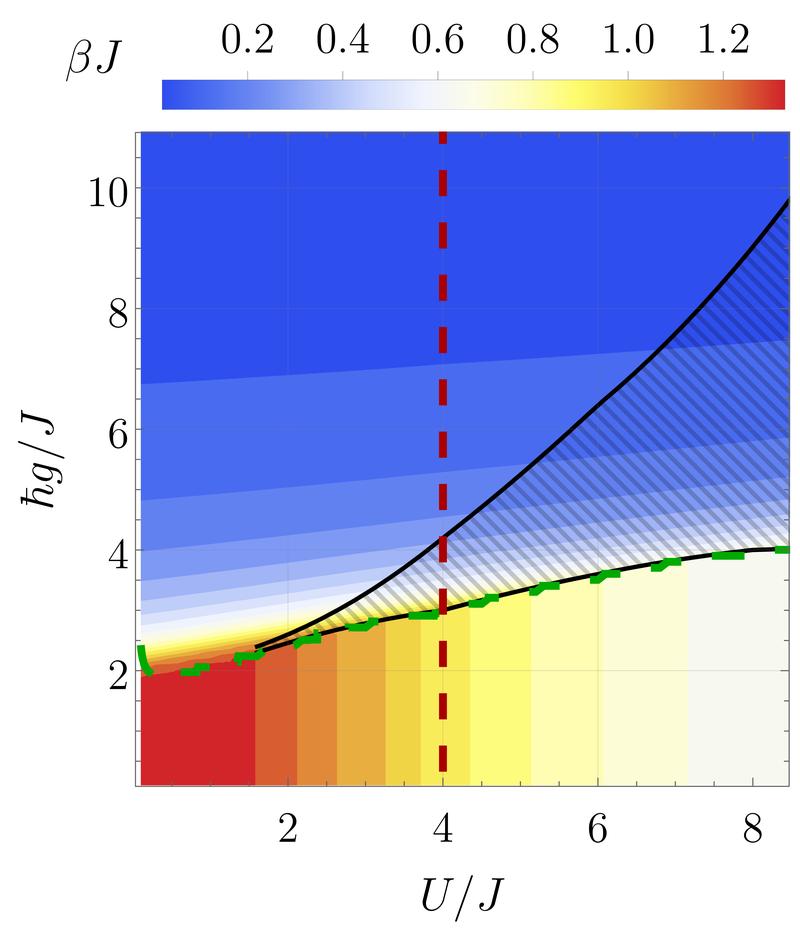}};
    \node[font=\normalfont] at (2ex,-6ex) {(b)};
    \end{tikzpicture}
    \caption[]{(a) Sublattice imbalance $\langle\hat{\Delta}\rangle/L^d$ and (b) inverse temperature $\beta J$ as a function of atoms-cavity coupling $\hbar g/J$ and on-site interaction $U/J$ of a finite size system $L\!=\!6$ at half filling. The parameters used are $\hbar\delta/J\!=\!4$, $\hbar\varGamma/J\!=\!1$. The hatched region marks the fluctuation-induced bistability region at the transition to the self-organized phase. The lines denote $U\!=\!\hbar\delta$ (red dashed) and $\hbar g_\text{cr}/J$ (green dashed).}
    \label{fig:density_plots_gvdU_delta4_G1_half_filling}
\end{figure}

The self-ordering transition at half filling involves inevitably the creation of double occupancies to form a charge density wave, which causes the strong dependence of the onset of the bistability on the interaction strength (see Fig.~\ref{fig:density_plots_gvdU_delta4_G1_half_filling}).
The increase of the region with two stable solution at commensurate filling differs from the behavior at low filling discussed in Sec.~\ref{sec:fluctuation_induced_bistability}, where an approximately linear dependence of the width $\hbar(g_\text{bi,2}\!-\!g_\text{bi,1})\!\sim\!U$ was observed. At half filling we observe a faster than linear growth of $\hbar(g_\text{bi,2}\!-\!g_\text{bi,1})$ as seen in Fig.~\ref{fig:bistable_region_plots_half_filling}~(a).

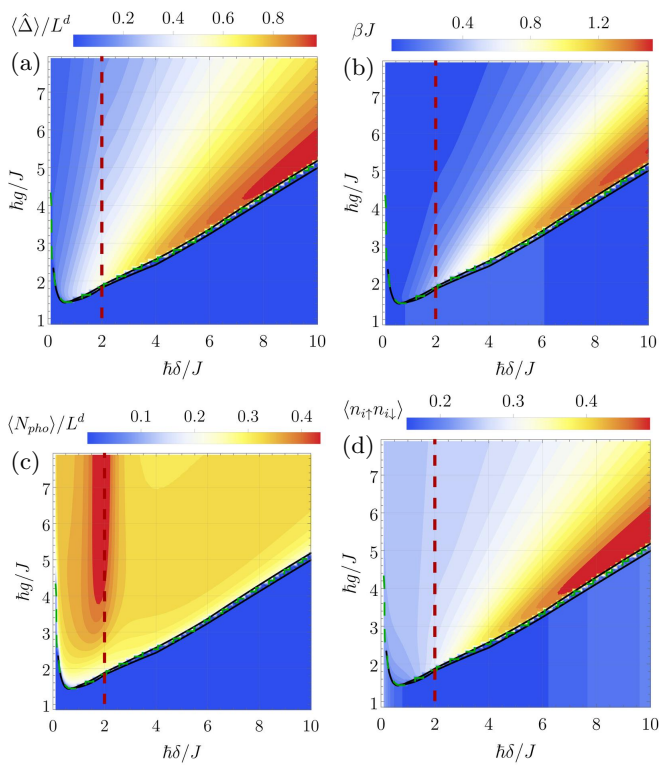
\begin{figure}[h]
    \begin{tikzpicture}
    \node[anchor=north west,inner sep=0pt] at (0,0){\includegraphics[width=0.235\textwidth]{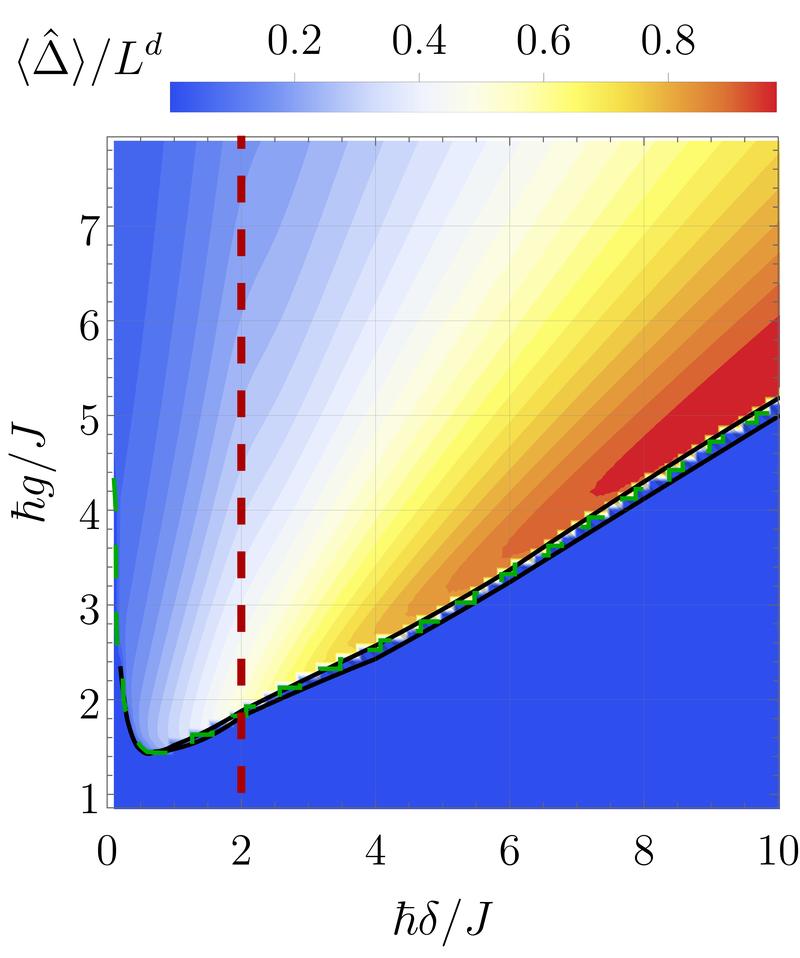}};
    \node[font=\normalfont] at (2ex,-6ex) {(a)};
    \end{tikzpicture}
    \begin{tikzpicture}
    \node[anchor=north west,inner sep=0pt] at (0,0){\includegraphics[width=0.235\textwidth]{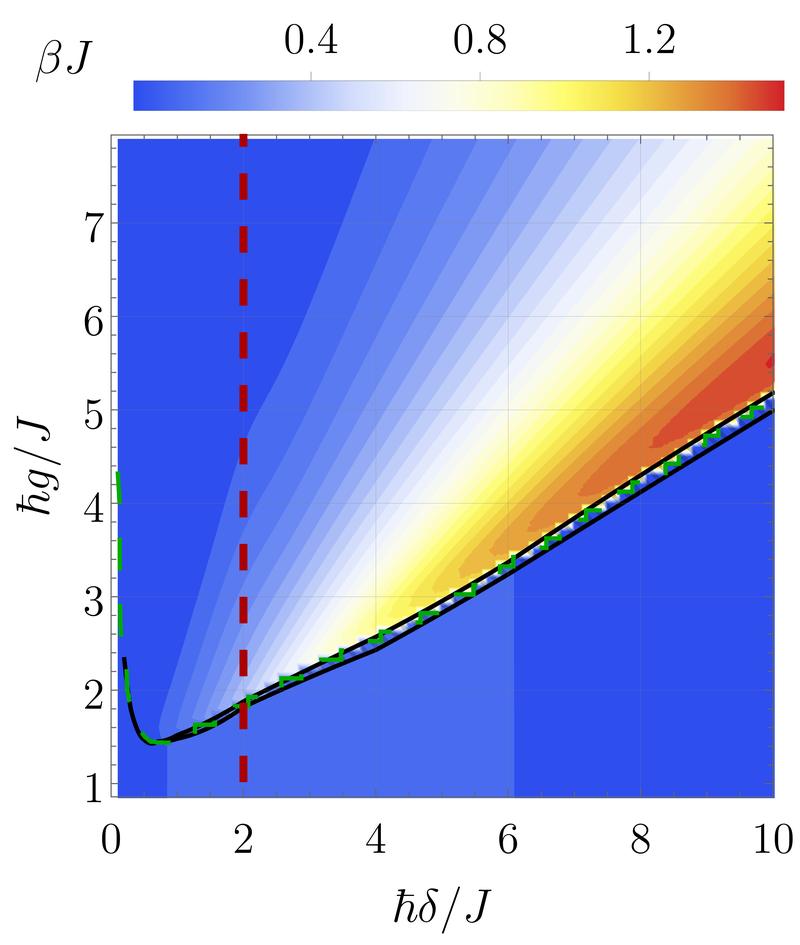}};
    \node[font=\normalfont] at (2ex,-6ex) {(b)};
    \end{tikzpicture}
     \begin{tikzpicture}
    \node[anchor=north west,inner sep=0pt] at (0,0){\includegraphics[width=0.235\textwidth]{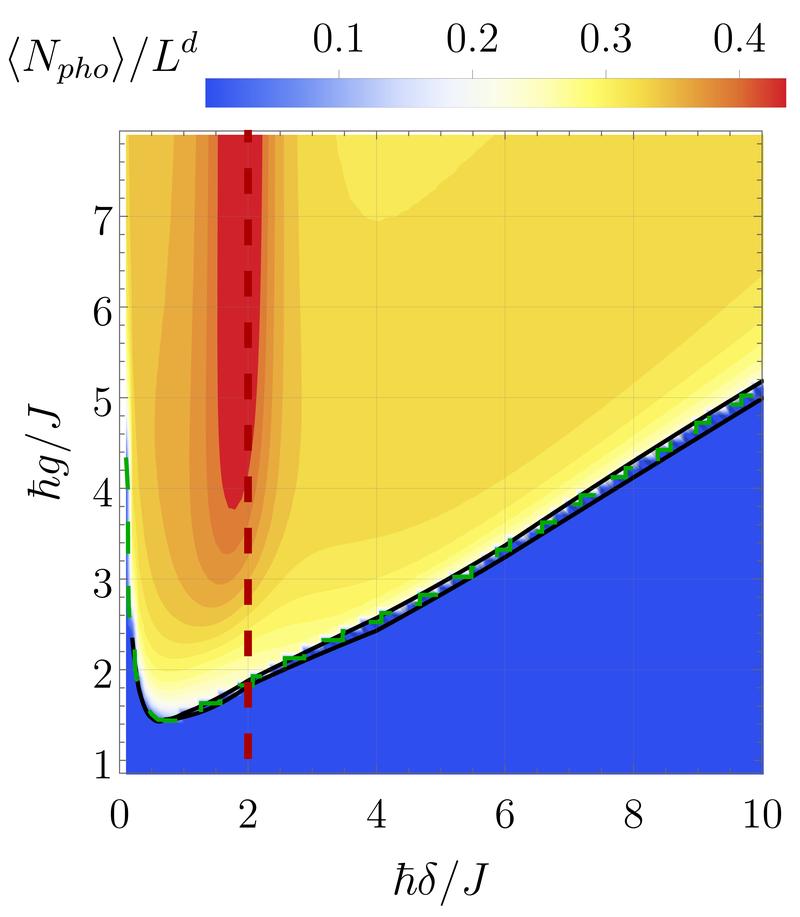}};
    \node[font=\normalfont] at (2ex,-6ex) {(c)};
    \end{tikzpicture}
    \begin{tikzpicture}
    \node[anchor=north west,inner sep=0pt] at (0,0){\includegraphics[width=0.235\textwidth]{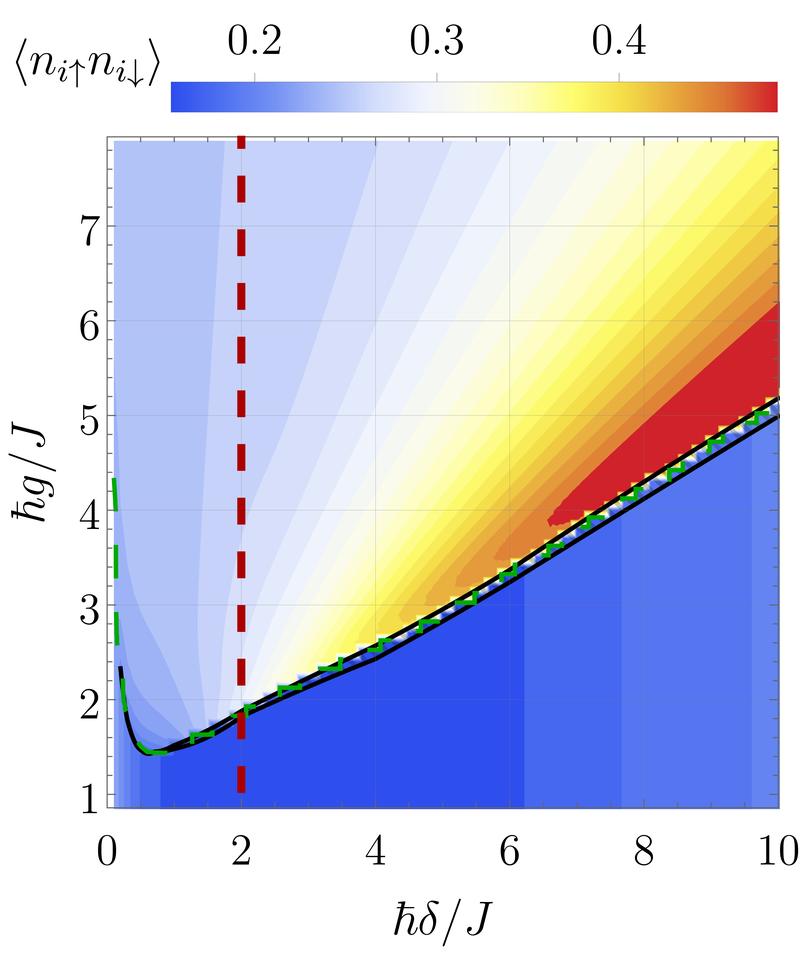}};
    \node[font=\normalfont] at (2ex,-6ex) {(d)};
    \end{tikzpicture}
    \caption[]{(a) Sublattice imbalance $\langle\hat{\Delta}\rangle/L^d$, (b) inverse temperature $\beta J$, (c) scaled photon number and (d) average number of double occupancies as a function of atoms-cavity coupling $\hbar g/J$ and detuning $\hbar\delta/J$ of a finite size system $L\!=\!6$ at half filling. The parameters used are $U/J\!=\!2$, $\hbar\varGamma/J\!=\!1$. The narrow hatched region marks the fluctuation-induced bistability region at the transition to the self-organized phase. The lines denote $U\!=\!\hbar\delta$ (red dashed) and $\hbar g_\text{cr}/J$ (green dashed).}
    \label{fig:density_plots_gvdelta_U2_G1_half_filling}
\end{figure}

In the following, we discuss the change of the critical coupling $g_\text{cr}$ and the extension of the bistability with the detuning $\hbar\delta/J$ [see Fig.~\ref{fig:bistable_region_plots_half_filling} and Fig.~\ref{fig:density_plots_gvdelta_U2_G1_half_filling}~(b)]. 
The critical coupling $g_\text{cr}$ rises with the detuning. 
In the $T\!=\!0$ MF method the rise of the critical coupling follows approximately the predictions of Eq.~(\ref{eq:gcrMFhalffilling}), i.e.~at large $\delta$ it rises as $\sqrt{\delta}$. 
In contrast to the behavior with the interaction strength, the extension of the bistable region only slightly increases with increasing $\delta$. 
Including the fluctuations, we see that the critical coupling rises even faster with the detuning. For the regime in which $\hbar\delta/J$ is the largest energy scale, the critical pump strength at which the self-ordering occurs grows almost linearly with $\hbar\delta/J$ [see Fig.~\ref{fig:bistable_region_plots_half_filling}~(b) and Fig.~\ref{fig:density_plots_gvdelta_U2_G1_half_filling}] as previously seen at lower fillings. This leads to an increase of the shift $\abs{g_\text{cr}\!-\!g_\text{cr}^\text{MF}}$ with $\hbar\delta/J$. We account this behavior partly to a larger effective temperature close to the transition at large detuning (see Fig.~\ref{fig:density_plots_gvdelta_U2_G1_half_filling}).

We turn to to the dependence with the dissipation strength $\hbar\varGamma/J$. For low values of $\hbar\varGamma/J$, the shift of the self-ordering transition between the $T\!=\!0$ MF method and the approach including fluctuations is relatively small [black and red hatched region in Fig.~\ref{fig:bistable_region_plots_half_filling}~(c)]. This is due to the fact that in this regime the self-consistently determined temperature is small compared to the excitation gap, which is proportional to $U/J$. Thus, the finite self-consistent temperature does not introduce a significant shift. 
However, increasing the dissipation $\hbar\varGamma/J$ the width of the bistability region drastically decreases when increasing $\hbar\varGamma/J$ including the fluctuations, while for the $T\!=\!0$ MF result it even increases in width [as seen in Fig.~\ref{fig:bistable_region_plots_half_filling}~(c)]. 
We attribute the shrinking of the bistability region mostly to the rise in the effective temperature when increasing $\varGamma$ [see blue points in Fig.~\ref{fig:cuts_U8_g6_delta4_Gammavar_half_filling}~(b)]. 
In the presence of the fluctuations, from $\hbar\varGamma_\text{cr}/J\!\sim\!18.3$ within the numerical accuracy no bistability can be detected anymore, hinting towards a continuous transition. 
This result resembles the finding in the model considered in \cite{FossFeigMaghrebi2017}, where, the ordering transition shows the properties of a discontinuous first-order phase transition which turns continuous in the large dissipation limit.

We comment on how some of the cooling resonances, discussed in detail in Sec.~\ref{sec:many_body_cooling}, influence the results at half filling. The effects are mostly analogous, i.e.~varying $\delta$, we note a drastic increase of the photon number around the cooling resonance $U\!=\!\hbar\delta$ [see Fig.~\ref{fig:density_plots_gvdelta_U2_G1_half_filling} and Fig.~\ref{fig:cuts_U2_g4_deltavar_Gamma1_half_filling}~(c) around the red dashed line]. Here the energy change when creating a double occupancy is resonant to the process of adding a photon to the cavity field.

In the ionic Hubbard model an interesting phase is the bond-order phase expected for the parameter regime $2\eta\!\sim\!U$ \cite{FabrizioNersesyan1999, ZhangLin2003, BatistaAligia2004, LoidaKollath2017}. 
However, in our case this corresponds to the parameter regime $2\hbar g\lambda\!\sim\!U$, which is typically within the unstable solution determined by the self-consistency condition. Furthermore, due to the finite self-consistent temperatures we obtain here, e.g.~$k_BT/J\!\sim\!3$, which broaden the features in the observables usually used for the detection of the bond-order phase, we cannot distinguish it for the system sizes we consider. A more detailed study would be required in order to settle whether this phase can be reached within the cavity-atoms system.

\section{Conclusions}

To summarize, we have investigated the steady state phase diagrams of fermionic atoms in optical lattices coupled to an optical cavity at both low filling and half filling. 
We take the quantum fluctuations in the atoms-cavity coupling into account, going beyond the $T\!=\!0$ mean-field method. 
By including the fluctuations within the many-body adiabatic elimination method we obtain mixed steady states described by an effective temperature.
We investigate the self-organization transition, where the finite value of the effective temperature shifts the transition between the normal and self-organized phases with respect to the ground-state ($T\!=\!0$) approximation. 
We analyze both the low filling scenario, in which the transition is continuous, and the commensurate half-filled case where a bistability occurs at the phase transition. We show that in the limit of strong dissipation the quantum fluctuations can reduce the width of the bistability arising at half-filling with respect to the predictions of the $T\!=\!0$ mean-field method.
We substantiate our results around the phase transition with approximate analytical expressions in various limits. 
Deep in the self-organized phase, a saturation in the cavity field occupation is observed and by employing an analytical approximation we show that the photon number becomes independent of the atoms-cavity coupling strength.

Our approach can capture the dynamical energy exchange between the atoms and the cavity field mediated by the fluctuations. Thus, we identified cavity-cooling processes as well as many-body cooling processes and associated them with resonances between the photon energy and gap in the effective atomic model. 
In the vicinity of these resonances, the energy transfer between excited eigenstates of the atomic system to the photonic mode is facilitated. 
At low filling, as exemplarily shown with results at quarter filling, we find the occurrence of a fluctuation-induced bistability region \cite{TolleHalati2024}, which does not exist in the $T\!=\!0$ MF solution. 
The emergence of the bistability relies on the many-body cooling processes and can only be capture due to the self-consistent determination of the effective temperature present in the many-body adiabatic elimination approach.

In the many-body adiabatic elimination approach we employ to include the fluctuations in the atoms-cavity coupling we use exact diagonalization to compute the eigenstates of the effective atomic model, the one-dimensional ionic Hubbard model. However, we complement these results with analytic perturbative calculations in the limit of weak kinetic energy.
Since the analytic derivations are obtained independent of specific filling or the underlying lattice dimensionality and geometry, we expect our results to be generic features of hybrid atoms-cavity systems. The cooling processes we described in our work will emerge in parameter regimes in which the cavity can probe and resolve atomic many-body energy gaps.

\section*{DATA AVAILABILITY}

The supporting data for this article are openly available at Zenodo \cite{tolle_2025_16962567}.

\section*{ACKNOWLEDGMENTS}

We thank A.~Bezvershenko, J.~P.~Brantut, T.~Donner, S.~B.~J\"ager, F.~Mivehvar, G.~Morigi, H.~Ritsch, A.~Rosch for fruitful discussions. 
We acknowledge support by the Deutsche Forschungsgemeinschaft (DFG, German Research Foundation) under project number 277625399 - TRR 185 (B4), project number 277146847 - CRC 1238 (C05), and project number 511713970 - CRC 1639 NuMeriQS (“Numerical Methods for Dynamics and Structure Formation in Quantum Systems”) – and under Germany’s Excellence Strategy – Cluster of Excellence Matter and Light for Quantum Computing (ML4Q) EXC 2004/1 – 390534769.
We further acknowledge support by the Swiss National Science Foundation under Division II grant 200020-219400. 
This research was supported in part by the National Science Foundation under Grants No.~NSF PHY-1748958 and PHY-2309135.

 \appendix

 \section{Approximations for the critical coupling\label{app:gcr_approximations}}

In this appendix, we give details on the derivation of the approximate analytical expressions [Eq.~(\ref{eq:gcrMF}) and Eq.~(\ref{eq:gcr_approx_highT}) in the main text] for the critical coupling at both $T\!=\!0$ MF level and in the many-body adiabatic elimination at finite temperature. 
We restrict to the two limits of the on-site interactions $U/J\!=\!0$ and $U/J\!\to\!\infty$.
The idea is to find the smallest coupling $g$ at which one obtains a small but non-zero solution for the cavity order parameter $\lambda$.
For non-interacting fermions, we can use the expressions derived in Sec.~\ref{sec:ionic_hubbard_hamiltonian} on the ionic Hubbard model for the sublattice imbalance, Eq.~(\ref{eq:density_imbalance_momentum_space}). Inserting this equation into the self-consistency equation Eq.~(\ref{eq:cav_MF}) we obtain an implicit equation given by
\begin{align}
\label{eq:Delta_discrete_sum}
   \frac{\delta^2+(\varGamma/2)^2}{2g_\text{cr}^2\delta/L}\!&=\!\sum_{k}\!\frac{\hbar }{E(k,\lambda=0)}\\
   &\times(\langle\hat{\gamma}^\dagger_{-,k,\sigma}\hat{\gamma}_{-,k,\sigma}\rangle+\langle\hat{\gamma}^\dagger_{+,k,\sigma}\hat{\gamma}_{+,k,\sigma}\rangle)|_{\lambda=0}.\nonumber
\end{align}

The expectation value is calculated with respect to the ground state for the $T\!=\!0$ MF method and using the thermal density matrix in the adiabatic elimination method. We first consider the $T\!=\!0$ MF approach. The equation simplifies to
\begin{align}
\label{eq:DeltaMF_discrete_sum}
 \frac{\delta^2+(\varGamma/2)^2}{2g_\text{cr}^2\delta/L}\!=\!\sum_{k}\!\frac{\hbar }{E(k,\lambda\!=\!0)}.
\end{align}
The momentum sum has to be taken over all occupied momenta. Thus we obtain Eq.~(\ref{eq:gcrMF}) in the main text with $\text{F}^{\text{MF},L}$ given by Eq.~(\ref{eq:F_MF_discrete}).

Assuming the thermodynamic limit, we can approximate the sum by an integral over continuous momenta, i.e.~$\sum_{k}\to\frac{L}{2\pi}\int_{-n\pi/2}^{n\pi/2}\!dk$. This leads to the analytic expression [Eq.~(\ref{eq:gcrMF}) with Eq.~(\ref{eq:F_MF_Linf})]
\begin{equation*}    g^\text{MF}_\text{cr}=\sqrt{\frac{J\pi\big(\delta^2+(\varGamma/2)^2\big)}{\hbar\delta\text{F}\!\left(\left.n \pi/2\right|1\right)}}
\end{equation*}
where $\text{F}(\phi|m)\!=\!\int_0^{\phi}(1\!-\!m\sin^2(\theta ))^{-1/2}d\theta$ is the elliptic integral of the first kind. When considering $U/J\!\to\!\infty$, the ground state is approximated by spinless fermions filling the lower band completely at quarter filling.

At finite temperature, however, the occupation of the momentum states in Eq.~(\ref{eq:density_imbalance_momentum_space}) is captured by the Fermi-distribution 
  \begin{align*}
     \langle\hat{\gamma}^\dagger_{\pm,k,\sigma}\hat{\gamma}^{\phantom{\dagger}}_{\pm,k,\sigma}\rangle_T&=\!\frac{1}{e^{\beta(\mp E(k,\lambda)-\mu)}+1},
 \end{align*}
 corresponding to Eq.~(\ref{eq:gcrMF}) with Eq.~(\ref{eq:F_AE_discrete}).
The chemical potential is fixed by the particle density Eq.~(\ref{eq:n_U0_momentum}), i.e.
\begin{align}
    n\!=\!\sum_{k,\sigma}\!\Big(1\!-\!\frac{\hbar g\lambda}{E(k,\lambda)}\Big)\!\Big[\frac{1}{e^{\beta(E(k,\lambda)-\mu)}\!+\!1}\!+\!\frac{1}{e^{\beta(-E(k,\lambda)-\mu)}\!+\!1}\Big] \nonumber
\end{align}
The condition for the critical coupling within the many-body adiabatic elimination approach Eq.~(\ref{eq:Delta_discrete_sum}) becomes
\begin{align}
\label{eq:DeltaAE_discrete_sum}
    \!\frac{\delta^2+(\varGamma/2)^2}{2 g_\text{cr}^2\delta/L}&=\sum_{k}\!\frac{\hbar}{E(k,\lambda\!=\!0)}\\&\times\Big[\frac{1}{e^{\beta(E(k,\lambda\!=\!0)-\mu)}\!+\!1}\!+\!\frac{1}{e^{\beta(-E(k,\lambda\!=\!0)-\mu)}\!+\!1}\Big]\nonumber
\end{align}
We evaluate the discrete sum expression using the self-consistently determined values of $\beta$.

In the thermodynamic limit, the sum can be replaced by an integral over the Brillouin-zone, i.e. $\sum_{k}\!\to\!\frac{L}{2\pi}\int_{-\pi/2}^{\pi/2}\!dk$. For this expression $\beta$ still depends on the system parameters and needs to be solved together with
Eq.~(\ref{eq:energy_transfer}) ($\partial\langle\hat{H}_\text{eff}\rangle/\partial t\!=\!0$). Thus, further approximations need to be performed to recover simple analytical expressions.

In the \emph{large dissipation limit} $\varGamma,\varGamma^2/\delta\!\gg\! J/\hbar,g\lambda$ we use the expression Eq.~(\ref{eq:beta_approx_large_dissipation})
\begin{equation}
    \beta\!\approx\!\frac{4\delta/\hbar}{\delta^2\!+\!(\varGamma/2)^2}\nonumber
\end{equation}
If the large-temperature approximation $\langle\hat{\Delta}\rangle/L\!\propto\!\hbar g\lambda\beta$, which will be derived in the paragraph following Eq.~(\ref{eq:beta_approx_large_glambda}), is valid close to the self-ordering transition, using the self-consistency condition Eq.~(\ref{eq:cav_MF}) the critical coupling is given by [Eq.~(\ref{eq:gcr_approx_highT})]

\begin{equation*}
   g_\text{cr}(\beta\!\ll\!1)\propto\frac{\delta^2\!+\!(\varGamma/2)^2}{\delta}
\end{equation*}

\section{Universal scaled photon number\label{app:universal_scaled_photon_number}}

Deep in the self-ordered regime, $\hbar g\lambda\!\gg\! U,\hbar\delta,J$, we observe a rise in the effective temperature that ultimately leads to an increasing weight of states with density imbalance smaller than the maximal value in the steady state and a saturation in the cavity field independent of specific system parameters. 
In this appendix, we give details on the derivation of the approximate expressions for the universal scaled photon number as well as the corresponding temperature and density imbalance [Eq.~(\ref{eq:imbalance_universal_scaling_U0}) and Eq.~(\ref{eq:imbalance_universal_scaling_Uinf}) in the main text].

At low filling the lowest energy state of the effective Hamiltonian $\hat{H}_\text{eff}$, Eq.~(\ref{eq:Heff_atoms}), consists of low potential sublattice sites being singly occupied. 
The relevant processes in this case are the first-order hopping events, which give an energy change of $\hbar\omega\!=\!\pm2g\lambda$. All other processes with energy changes $\hbar\omega\!\sim\!\pm U$ are suppressed by a factor of $\sim\!J^2/\hbar g\lambda$. 
Using this assumptions we further simplify the equations derived in Sec.~\ref{sec:pertubation_J} and obtain a steady state condition of the energy transfer, Eq.~(\ref{eq:EOM_1D}), for both the non-interacting $U/J\!=\!0$ and $U/J\!\to\!\infty$ limit as
\begin{align}
\frac{-e^{\beta\hbar g\lambda}}{(2g\lambda+\delta)^2+(\varGamma/2)^2}+\frac{e^{-\beta\hbar g\lambda}}{(2g\lambda-\delta)^2+(\varGamma/2)^2}\nonumber=0,
\end{align}
from which we obtain
\begin{equation}
\label{eq:expbetaglambda_exact}
   e^{\beta\hbar g\lambda}=\sqrt{\frac{(\varGamma/2)^2+(\delta+2g\lambda)^2}{(\varGamma/2)^2+(\delta-2g\lambda )^2}}.
\end{equation}
Note that one arrives at the same equation by assuming the chemical potential to be independent of $\lambda$ and taking the large-interaction limit $e^{\beta U}\!\gg\!1$ (see Supplemental Material of Ref.~\cite{TolleHalati2024}).
For $\beta \hbar g\lambda\! \ll\! 1$ and $\hbar g\lambda\! \to\! \infty$ we get an approximation for the scaling of $\beta$ with the effective potential imbalance
\begin{equation}
    \label{eq:beta_approx_large_glambda}
    \beta\approx\frac{\delta}{\hbar(g\lambda)^2},
\end{equation}
leading to the temperature rising quadratically with $g\lambda$.

In the following, we first consider the non-interacting limit ($U/J\!=\!0$).
Here, since the spin species occupy the states independently, one can block diagonalize the Hamiltonian and solve the system of equations for each spin independently.
This results in
\begin{align}
\label{eq:lambda_universal_scaling_U0}
    \lambda(U\!=\!0)\!=\!\sqrt{\frac{\delta^2(2-n)n}{(\varGamma/2)^2+\delta^2}}.
\end{align}
From the expression of the cavity field we can obtain the inverse temperature, imbalance and photon number in the strong coupling regime [Eq.~(\ref{eq:imbalance_universal_scaling_U0})]
\begin{align*}
    \beta J&\approx\frac{(\varGamma/2)^2+\delta^2}{g^2n(2-n)}, \\
     \frac{\langle\hat{\Delta}\rangle}{L}&=\frac{\sqrt{(2-n)n((\varGamma/2)^2+\delta ^2)}}{2g}\nonumber, \\
    \frac{\langle\hat{a}^\dagger\hat{a}\rangle}{L}&\approx\frac{n}{4}(2-n)\nonumber.
\end{align*}

The results slightly change when considering the $U/J\!\to\!\infty$ limit. 
The particle conservation condition Eq.~(\ref{eq:pc_1D}) results in the expression
\begin{align}
\textstyle
    e^{\beta\mu}\!=\!\frac{(2-4n)\cosh(\beta \hbar g\lambda)+\sqrt{2(1-2n)^2\cosh(2\beta \hbar g\lambda)-8(n-1)n+2}}{8(n-1)}.\nonumber
\end{align}
We only confirmed the validity of this result for $N/L\!\leq\!0.5$, since the approximation breaks down as one approaches half filling due to the fact that the chemical potential diverges with $U$ and can no longer be assumed finite.
Inserting these results into the self-consistency condition Eq.~(\ref{eq:sc_1D}) and using $g\lambda\gg \delta,\varGamma$ yields
\begin{align}
\label{eq:lambda_universal_scaling_Uinf}
    \lambda(U\!\to\!\infty)\!=\!\sqrt{\frac{2\delta^2(1-n)n}{(\varGamma/2)^2+\delta^2}}.
\end{align}
From the expression of the cavity field we can obtain the inverse temperature, imbalance and photon number in the strong coupling regime [Eq.~(\ref{eq:imbalance_universal_scaling_Uinf})]
\begin{align*}
    \beta&\approx\frac{(\varGamma/2)^2+\delta^2}{2\hbar\delta g^2n(1-n)}, \\
     \frac{\langle\hat{\Delta}\rangle}{L}&=\frac{\sqrt{(1-n)n((\varGamma/2)^2+\delta^2)}}{\sqrt{2} g}\nonumber \\
    \frac{\langle\hat{a}^\dagger\hat{a}\rangle}{L}&\approx\frac{n}{2}(1-n) \nonumber.
\end{align*}

\end{document}